\documentclass[10pt]{article}
\usepackage{amsmath,amssymb,amsthm,bm,bbm,graphicx,natbib,hyperref,caption,algorithm,adjustbox}
\usepackage{subcaption}

\usepackage{color}

\usepackage{natbib}

\usepackage{float}  % load float package first!
\usepackage{hyperref}

\newtheorem{dfn}{Definition}

\long\def\comment#1{}

\DeclareMathOperator{\diag}{diag}

\newtheorem{case}{Case}

\title{Bayesian sequential parameter estimation with a Laplace type approximation}
\author{Tiep Mai\thanks{Bell Laboratories, Ireland. \texttt{tiep.mai@alcatel-lucent.com}} \and Simon P.~Wilson\thanks{School of Computer Science and Statistics, Trinity College, Dublin 2, Ireland. \texttt{swilson@tcd.ie}}}
\date{}

\begin{document}

\maketitle

\begin{abstract}
A method for sequential inference of the fixed parameters of a dynamic latent Gaussian models is proposed and evaluated that is based on the iterated Laplace approximation.  The method provides a useful trade-off between computational performance and the accuracy of the approximation to the true posterior distribution. Approximation corrections are shown to improve the accuracy of the approximation in simulation studies.  A population-based approach is also shown to provide a more robust inference method.
\end{abstract}

\section{Introduction}
This paper explores inference for the parameters of the latent linear Gaussian model:
\begin{align}
y_{t} &\sim p(y_{t} \, | \, x_{t},\varphi), \label{eqn:ssm_obs} \\
x_{t} &= A x_{t-1} + u_{t}, \label{eqn:ssm_state}
\end{align}
where $y_t$ are conditionally independent noisy observations of $x_t$, $u_t \sim N(0,\Sigma_u)$ are independent Gaussian innovations and $\varphi$ is the set of model parameters: $A$, $\Sigma_u$ and any others associated with $p(y_{t} \, | \, x_{t},\varphi)$. They are an important class of dynamic state space models and have seen extensive applications in signal processing, epidemiology, environmental statistics, to name but a few fields. Statistical inference tasks for these models focus on learning about the latent process, and in some cases the model parameters, from observation of the $y_t$. 

Nowadays, many statistical applications concern the analysis of streaming data for which fast algorithms are required e.g.\ dynamic classification \citep{ref:Mccormick2011} or online analysis of sensor data \citep{ref:Hill2009, ref:Yang2015}. A desirable property in these applications is that the inference result can be efficiently re-used or modified when another observation arrives. As a result, traditional batch processing methods such as MCMC \citep{ref:Roberts2004}, variational Bayes \citep{ref:Beal2003}, expectation propagation \citep{ref:Minka2001} and INLA \citep{ref:Rue2009} are not ideal solutions as the approximation must be re-computed 'from scratch' when the next datum is observed. Furthermore, in the field of sequential inference, there have been quite many well-established works on the state variable inference such as variants of Kalman filter \citep{ref:Julier1997,ref:Wan2000} and particle filter \citep{ref:Pitt1999,ref:Doucet2000b}. However, these algorithms do not completely fit in most practical applications which require not only the estimation of the state variable but also the unknown fixed parameters.

Sequential parameter learning is non-trivial for several reasons: the non-regeneration characteristic of fixed parameters, the assumption of one-datum-at-a-time, the strong coupling between state variable and the unknown parameters, and the degeneracy issue of particle-based solutions. A comprehensive review of sequential parameter estimation is in \citet{ref:Kantas2014}. There are broadly two types of approach: maximum-likelihood-based methods try to obtain a point estimate of $\varphi$, and Bayesian methods that approximate the whole distribution $p(\varphi \, | \, y_{1:t})$; most solutions are based on particle filters.

For Bayesian sequential parameter learning, which is the focus of this paper, \citet{ref:Liu2001} uses kernel artificial dynamics for unknown parameters and maintains an invariant parameter variance with a shrinkage factor. \citet{ref:Fearnhead2002, ref:Storvik2002} propose a method to use sufficient statistics to re-sample the unknown parameters, lessening the degeneracy effect of particle filters. \citet{ref:Carvalho2010} further extends the idea by combining the sufficient statistics with auxiliary particle filters, with explicit inference for conditional dynamic linear models. There are also other Bayesian works such as practical filtering \citep{ref:Polson2008} and SMC$^2$ \citep{ref:Chopin2013}. However, as practical filtering uses lagged observations in fixed lagged approximations and SMC$^2$ refreshes particle samples by MCMC with past data upon degeneracy encounter, they are not suitable for truly online scenario.

All above methods are based on particle filters, which are susceptible to outlier and path degeneracy \citep{ref:Andrieu2005,ref:Kantas2014}. Hence, in this paper, we propose an alternative approach for sequential inference, based on a functional approximation, in the expectation that a smoothed approximation of both state variable and unknown parameters is more robust to outliers than a discrete particle approximation. Like \citet{ref:Liu2001}, the proposed method does not require the existence of sufficient statistics and hence can be applied in various cases \footnote{The proposed method is less generalised than \citet{ref:Liu2001} due to the assumption of Gaussian linear state equation. Logically, with an extra computational cost, this constraint can be overcome either by a joint functional approximation $\widetilde{p}_(x_{t+1}, x_t, \varphi \, | \, y_{1:t})$, or double functional approximations $\widetilde{p}_(x_t, \varphi \, | \, y_{1:t})$, $\widetilde{p}_(x_{t+1}, \varphi \, | \, y_{1:t})$ at each time point. However, this extension is not covered in this paper.}.

The paper focuses on approximation of $p(x_t, \varphi \, | \, y_{1:t})$ and $p(x_{t+1}, \varphi \, | \, y_{1:t})$, where $y_{1:t} = \{ y_1,\ldots,y_t\}$. These approximations are derived as a generalisation of iterLap, an iterative Laplace approximation described in \cite{ref:Bornkamp2011}, to a sequential setting. As the approximations are based on Gaussian mixtures, $\varphi$ is easily marginalised to yield approximations to $p(x_t \, | \, y_{1:t})$ and $p(x_{t+1} \, | \, y_{1:t})$. Some approximation corrections by importance sampling and a population-based strategy of selective exploration are also developed, and shown to produce a more robust approximation.

The paper is organised as follows.  A brief description of the iterLap approximation is given in Section \ref{sec:iterlap}. In Section \ref{sec:sequential_iterlap}, the method is adapted for use in a sequential setting. The performance of the algorithm and the approximation corrections is evaluated in Section \ref{sec:examples}. In Section \ref{sec:pop_based_strategies}, a population-based approach is developed that adds further robustness to the method. There are some concluding remarks in Section \ref{sec:conclusion}.

\section{The iterLap approximation}
\label{sec:iterlap}

The Laplace approximation \citep{ref:Tierney1986} is one of the most commonly used functional approximations in statistical methods. For an non-normalised density $q(x)$, it uses a second order Taylor expansion of $\log(q(x))$ about a local maximum to obtain a Gaussian approximation $\widetilde{q}(x) = N(x \, | \, \mu=\widehat{x},\Sigma=\widehat{Q}^{-1})$, where the mean $\widehat{x}$ is a local maximum of $q$ and the precision matrix $\widehat{Q}$ is the Hessian of $\log(q(x))$ evaluated at $x = \widehat{x}$. When a posterior distribution converges to the Gaussian asymptotically, the Laplace approximation can be very efficient. However, for non-Gaussian distributions, this approximation suffers from several shortcomings, of which we mention: it is a uni-modal approximation and so ignores other modes if the target density is multi-modal, it does not approximate highly skewed or heavy-tailed distributions well, and the Gaussian distribution implies a linear correlation between components of $x$ and so cannot account for non-linear dependencies.

The Laplace approximation can be generalised to a mixture of Gaussians approximation
\[ \widetilde{p}_m(x) \: = \: \sum_{i=1}^m w_i N(x \, | \, \mu_i, Q_i^{-1}), \]
with component weights $w_i$. \citet[chap. 12]{ref:Gelman2003} obtain the $\mu_i$ and $Q_i$ by simultaneous optimisations with different starting points, from which the $w_i$ are evaluated by equating the target density $q(x)$ to $\widetilde{q}_m(x)$. The difficulties with this approach are that it is often inefficient when $q(x)$ is a skewed uni-modal target density, as the optimisation solutions cluster around the unique global maximum of $q(x)$ and produce a mixture density that is unimodal and almost symmetric. Another difficulty is the inefficiency in identifying local modes through random starting points. 

The iterated Laplace (iterLap) approximation, introduced in \citet{ref:Bornkamp2011}, tries to overcome these difficulties by constructing the Gaussian mixture iteratively.  Given an approximation with $m$ components, another component is added to the mixture by constructing a Laplace approximation for the difference between the target density and the approximation so far. This yields a new component mean $\mu_{m+1}$ and variance $\mathit{\Sigma_{m+1}}$. Weights for the new and already-existing components are then re-evaluated by quadratic programming. Further components are added until one of the stopping criteria for the algorithm is met.

\subsection{Modifying the iterLap approximation for sequential inference}
\label{subsec:mod-iterlap}

Many modifications to the original iterLap approximation are possible. \citet{mref:Mai_arxiv2015B} describes several that are implemented in this work for different parts of the algorithm, and have been shown empirically to give better performance when applied to latent Gaussian models. These include: modifying the stopping criteria, selecting initial optimisation points, scaling the Hessian matrix, and proposing alternative methods for re-assigning component weights.

\begin{figure}[ht]
\centering
\begin{subfigure}{0.48\textwidth}
  \centering
    \includegraphics[scale=0.6]{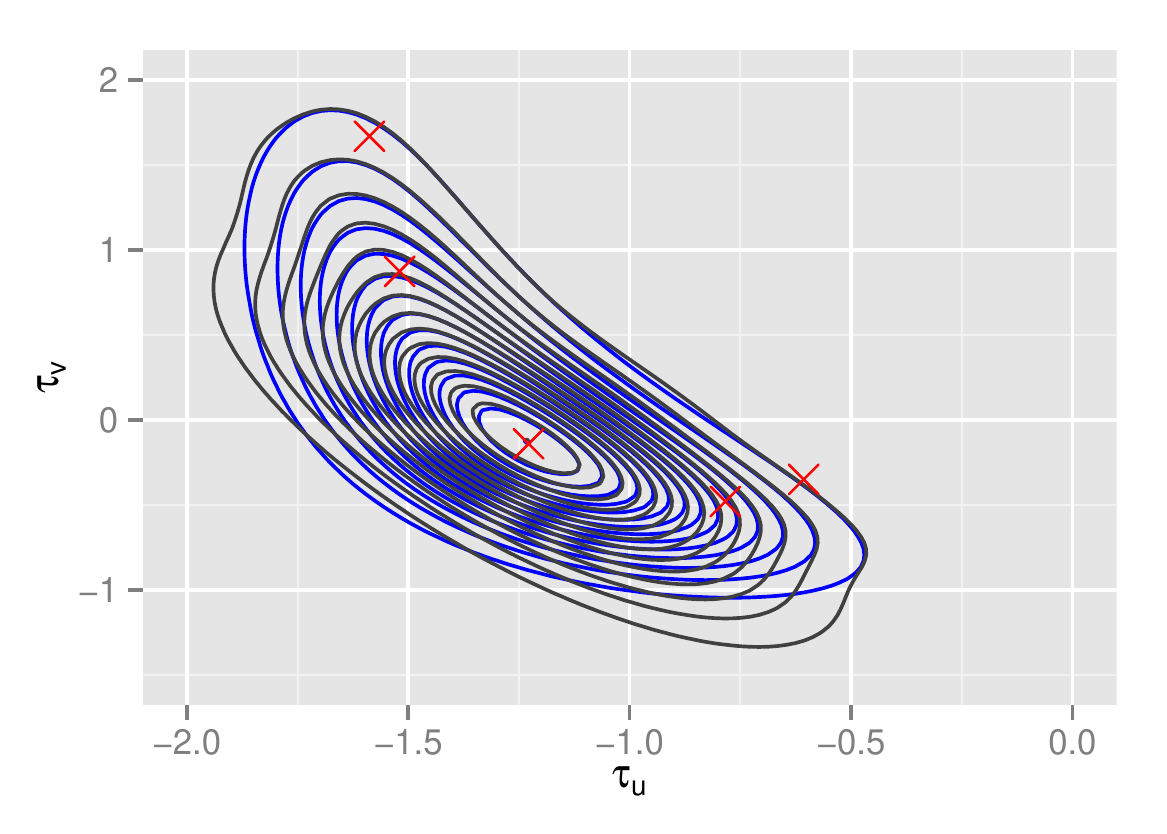}
  \caption{original iterLap: $(\tau_u,\tau_v)$ ($0.498$ seconds)}
    \label{fig_iterlap:ex2_iterlap_tau}
\end{subfigure}
\begin{subfigure}{0.48\textwidth}
    \centering
    \includegraphics[scale=0.6]{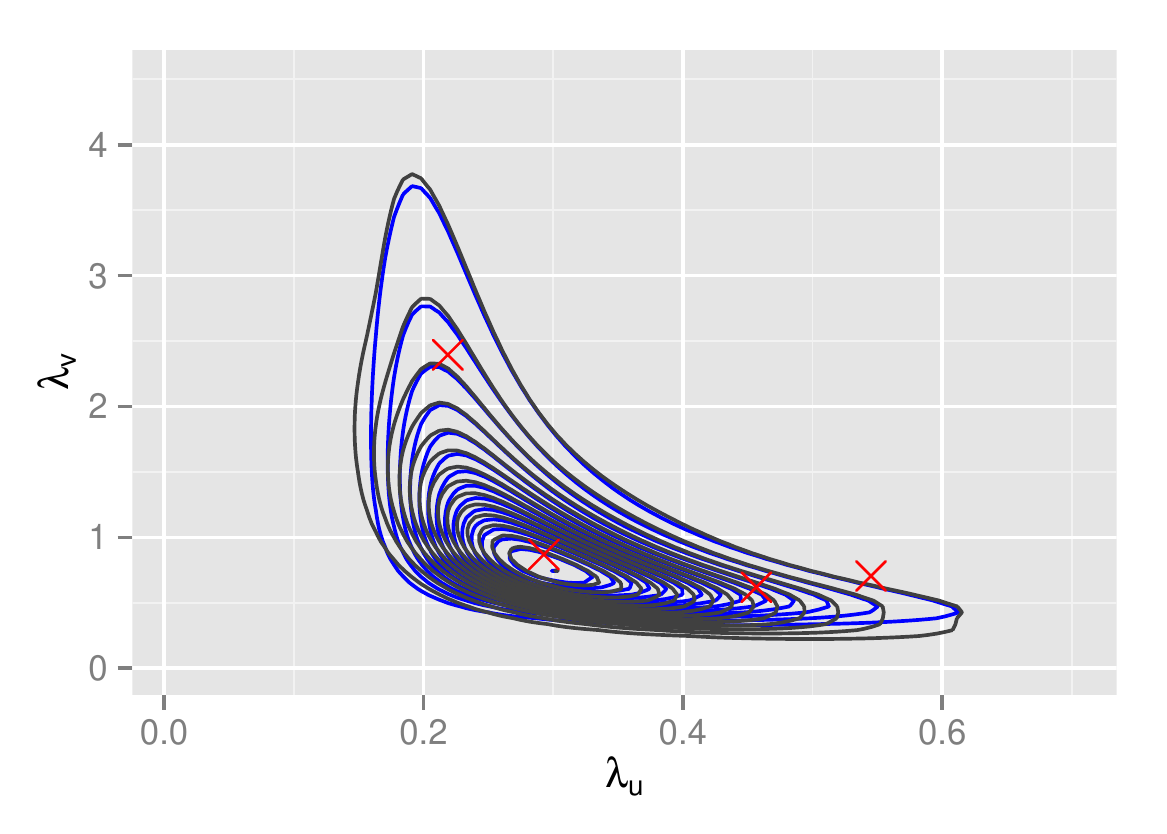}
    \caption{original iterLap: $(\lambda_u,\lambda_v)$ ($0.498$ seconds)}
    \label{fig_iterlap:ex2_iterlap_lambda}
\end{subfigure}

\begin{subfigure}{0.48\textwidth}
    \centering
    \includegraphics[scale=0.6]{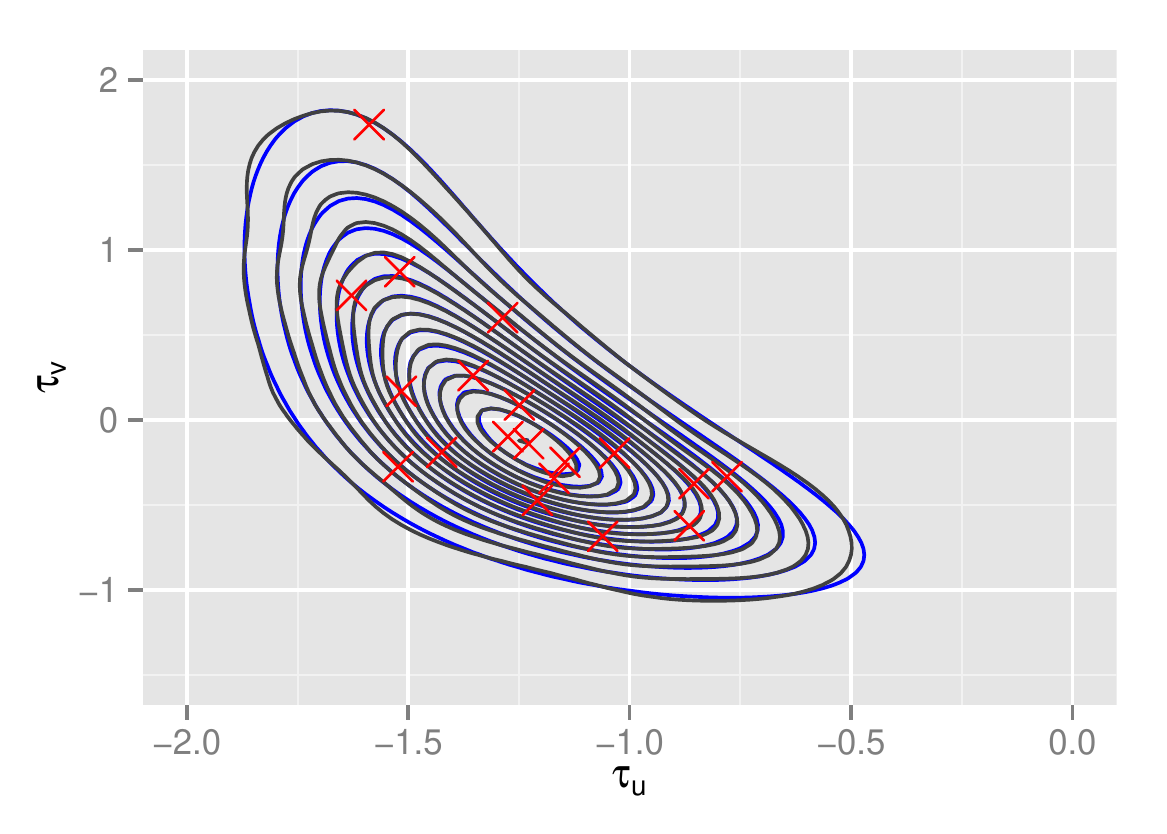}
    \caption{modified iterLap: $(\tau_u,\tau_v)$ ($5.024$ seconds)}
    \label{fig_iterlap:ex2_mod_iterlap_tau}
\end{subfigure}
\begin{subfigure}{0.48\textwidth}
    \centering
    \includegraphics[scale=0.6]{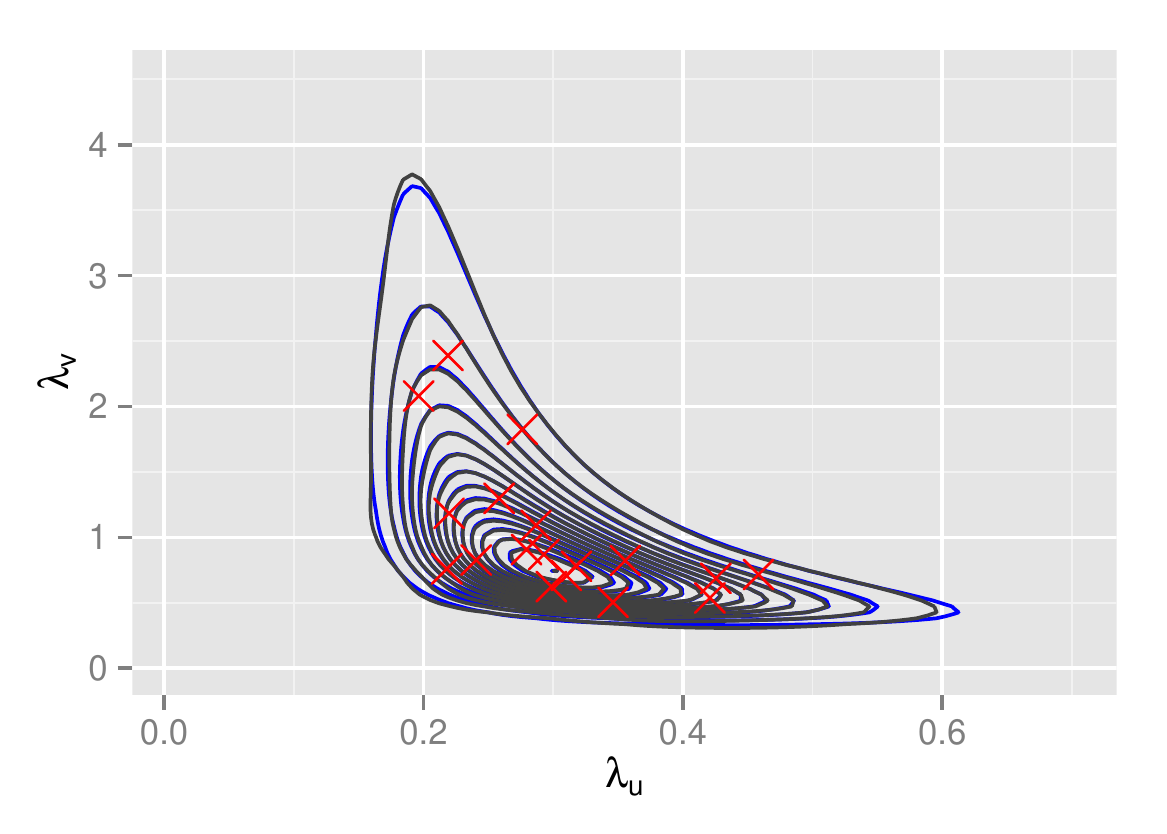}
    \caption{modified iterLap: $(\lambda_u,\lambda_v)$ ($5.024$ seconds)}
    \label{fig_iterlap:ex2_mod_iterlap_lambda}
\end{subfigure}
\caption{Left: the target posterior density contours (blue) and iterLap approximation (black) contours for $p(\tau_u,\tau_v \, | \, y)$.  Right: the approximation transformed back to $(\lambda_u, \lambda_v)$. The red crosses are the iterLap component means.}
\label{fig_iterlap:ex2}
\end{figure}

As an example, Figure \ref{fig_iterlap:ex2} shows the iterLap approximation of the posterior density of the precision parameters $(\lambda_v, \lambda_u)$ in the Gaussian dynamic linear model :
\begin{align*}
y_t &= x_t + v_t,\\
x_t &= x_{t-1} + u_t, 
\end{align*}
where $u_{t} \sim N(0,\sigma_u^2=\lambda_u^{-1})$, $v_{t} \sim N(0,\sigma_v^2=\lambda_v^{-1})$. The priors for precision parameters $\lambda_u$ and $\lambda_v$ are exponential with mean 2. One hundred observations are generated from the model with $\lambda_u^{\star} = 0.25$, $\lambda_v^{\star} = 1$. The posterior $p(\lambda_v,\lambda_u \, | \, y_{1:100})$ can be evaluated up to a constant in closed form by marginalising out $x_{1:100}$, allowing comparison between the approximations and the target. The original iterLap approximation of \cite{ref:Bornkamp2011} and that of \cite{mref:Mai_arxiv2015B} are applied to $p(\tau_u,\tau_v \, | \, y)$ where $\tau_u = \log(\lambda_u)$ and $\tau_v = \log(\lambda_v)$. The maximum number of mixture components $m_{max}$ is set to $30$.

The optimal iterLap approximation stops with $6$ components and the modified iterLap with $19$ components. More comparison examples of two R implementations are given in \citep{mref:Mai_arxiv2015B}, illustrating the trade-off between approximation accuracy and running time; in general the modified algorithm offers improved accuracy at a modest increase in computation time. The modified iterLap performs quite well in low-dimensional space but suffers from the curse of dimensionality and so performance deteriorates with the dimension of the target density. Still, the iterative and non-sampling nature of the algorithm means that it is faster than Monte Carlo based solutions to density approximation. Furthermore, these functional approximations can complement Monte Carlo methods by serving as non-linear multi-modal sampling proposals, providing an efficient way to explore complex parameter spaces.

%%%%%%%%%%%%%%%%%%%%%%%%%%%%%%%%%%%%%%%%%%%%%%%%%%%%%%%%%%%%%%%%%%%%%
%%%%%%%%%%%%%%%%%%%%%%%%%%%%%%%%%%%%%%%%%%%%%%%%%%%%%%%%%%%%%%%%%%%%%%

\section{Extending iterLap for sequential learning}
\label{sec:sequential_iterlap}

\subsection{Base algorithm}
\label{subsec:baseSI}

Given a Gaussian mixture approximation to $p(x_{t-1},\varphi \, | \, y_{1:(t-1)})$, the properties of the Gaussian and the linearity of the state equation permit a closed-form expression for the prediction distribution $p(x_t,\varphi \, | \, y_{1:(t-1)})$. Using this, iterLap can be applied to obtain an approximation to the next filtering density $p(x_t,\varphi \, | \, y_{1:t})$.

Assume that iterLap has provided a Gaussian mixture approximation to $p(x_{t-1},\varphi \, | \, y_{1:(t-1)})$:
\begin{equation*}
\widetilde{p}(x_{t-1},\varphi \, | \, y_{1:(t-1)}) \: = \:  \sum_{i=1}^{m_{t-1}} w_i^{(t-1)} N(x_{t-1},\varphi \, | \, \mu_i^{(t-1)}, (Q_i^{(t-1)})^{-1}).
%\label{alg_baseSI:SIBS_fdent_iterlap}
\end{equation*}

Each Gaussian component of $\widetilde{p}(x_{t-1}, \varphi \, | \, y_{1:(t-1)})$ is decomposed into a Gaussian marginal for $\varphi$ and a conditional Gaussian for $x_{t-1}$ given $\varphi$:
\begin{equation*}
N(x_{t-1}, \varphi \, | \, \mu_i^{(t-1)}, \Sigma_i^{(t-1)} = (Q_i^{(t-1)})^{-1}) = N(\varphi \, | \, \mu_{i,\varphi}^{(t-1)}, (Q_{i,\varphi}^{(t-1)})^{-1}) \: N(x_{t-1} \, | \, \mu_{i,x_{t-1}|\varphi}^{(t-1)}, (Q_{i,x_{t-1}|\varphi}^{(t-1)})^{-1}),
\end{equation*}
where:
\begin{itemize}
\item $\mu_{i,x_{t-1}|\varphi}^{(t-1)} = \mu_{i,x_{t-1}}^{(t-1)} - (Q_{i,x_{t-1} x_{t-1}}^{(t-1)})^{-1} Q_{i,x_{t-1} \varphi}^{(t-1)} (\varphi - \mu_{i,\varphi}^{(t-1)})$;
\item $Q_{i,\varphi}^{(t-1)}=(\Sigma_{i,\varphi \varphi}^{(t-1)})^{-1}$;
%\item $Q_{i,x_{t-1}|\varphi}^{(t-1)}=Q_{i,x_{t-1} x_{t-1}}^{(t-1)}$ ($Q_{i,x_{t-1} x_{t-1}}^{(t-1)}$, $Q_{i,x_{t-1} \varphi}^{(t-1)}$,
\item $Q_{i,x_{t-1}|\varphi}^{(t-1)}=Q_{i,x_{t-1} x_{t-1}}^{(t-1)}$,
\end{itemize}
where $Q_{i,x_{t-1} x_{t-1}}^{(t-1)}$, $Q_{i,x_{t-1} \varphi}^{(t-1)}$, $Q_{i,\varphi \varphi}^{(t-1)}$ are relevant sub-matrices of $Q_{i}^{(t-1)}$ and $\Sigma_{i,\varphi \varphi}^{(t-1)}$  is a sub-matrix of $\Sigma_{i}^{(t-1)}$.

Since the state equation is linear and $v_t$ is Gaussian, so an approximation to the joint density of $x_{t}$, $x_{t-1}$ and $\varphi$ can be obtained:
\begin{align}
\nonumber &~~~~ \widetilde{p}(x_{t-1},x_{t},\varphi \, | \, y_{1:(t-1)}) \; = \; \widetilde{p}(x_{t-1},\varphi \, | \, y_{1:(t-1)}) \: p(x_{t} \, | \, x_{t-1}, \varphi) \\ 
&= \sum_{i=1}^{m_{t-1}} w_i^{(t-1)} N(\varphi \, | \, \mu_{i,\varphi}^{(t-1)}, (Q_{i,\varphi}^{(t-1)})^{-1}) \: N(x_{t-1} \, | \, \mu_{i,x_{t-1}|\varphi}^{(t-1)}, (Q_{i,x_{t-1}|\varphi}^{(t-1)})^{-1}) \: N(x_{t} \, | \, A x_{t-1} ,Q_u^{-1}).
\label{eqn_baseSI_t_and_t+1}
\end{align}

Then $x_{t-1}$ can be marginalised out to get the approximation to $p(x_{t}, \varphi \, | \, y_{1:(t-1)})$ of Equation \ref{eqn_baseSI:SIBS_mgndent}:
\begin{align}
\nonumber &~~~~ \widetilde{p}(x_{t}, \varphi \, | \, y_{1:(t-1)}) \; = \; \int \widetilde{p}(x_{t-1},x_{t},\varphi \, | \, y_{1:(t-1)}) \: \mathrm{d}x_{t-1}\\
\nonumber &= \sum_{i=1}^{m_{t-1}} w_i^{(t-1)} N(\varphi \, | \, \mu_{i,\varphi}^{(t-1)}, (Q_{i,\varphi}^{(t-1)})^{-1}) \: \int  N(x_{t-1} \, | \, \mu_{i,x_{t-1}|\varphi}^{(t-1)}, (Q_{i,x_{t-1}|\varphi}^{(t-1)})^{-1}) \: N(x_{t} \, | \, A x_{t-1} ,Q_u^{-1}) \: \mathrm{d}x_{t-1}\\
\nonumber &= (2 \pi)^{-d_x/2} |Q_u|^{1/2} \exp \left( -\frac{1}{2} x_t^{T} Q_u x_t\right) 
\: \sum_{i=1}^{m_{t-1}} \left[ \vphantom{\frac{1}{2}}  w_i^{(t-1)} N(\varphi \, | \, \mu_{i,\varphi}^{(t-1)}, (Q_{i,\varphi}^{(t-1)})^{-1})  \right. \\
 &~~~~~~~~~~~ \left.
\times \:  |Q_{i,x_{t-1}|\varphi}^{(t-1)}|^{1/2} |\overline{Q}_{i,\varphi}|^{-1/2} \:\exp \left( -\frac{1}{2}\left( (\mu_{i,x_{t-1}|\varphi}^{(t-1)})^{T} Q_{i,x_{t-1}|\varphi}^{(t-1)} \mu_{i,x_{t-1}|\varphi}^{(t-1)} - \overline{\mu}_{i,\varphi}^{T}\overline{Q}_{i,\varphi}\overline{\mu}_{i,\varphi} \right) \right) \right],
\label{eqn_baseSI:derived_mgn01}
\end{align}
with:
\begin{eqnarray*}
\overline{Q}_{i,\varphi} &=& Q_{i,x_{t-1}|\varphi}^{(t-1)} + A^T Q_u A,\\
\overline{\mu}_{i,\varphi} &=& \overline{Q}_{i,\varphi}^{-1} (Q_{i,x_{t-1}|\varphi}^{(t-1)} \mu_{i,x_{t-1}|\varphi}^{(t-1)} + A^T Q_u x_t).
\end{eqnarray*}
Notice that even though it is more compact to write Equation \ref{eqn_baseSI:derived_mgn01} using the variance representation, evaluating $\widetilde{p}(x_{t}, \varphi \, | \, y_{1:(t-1)})$ with the precision matrix is more efficient computationally.

\begin{algorithm}[!ht]
\caption{SIBS: Base algorithm of sequential inference with iterLap}
\label{alg_baseSI:SIBS}
\begin{enumerate}
\item For $t=1$, derive a Gaussian mixture approximation $\widetilde{p}(x_1,\varphi \, | \, y_1)$ to $p(x_1,\varphi \, | \, y_1)$ by modified iterLap:
\begin{align}
\label{alg_baseSI:SIBS_fden1}
\widetilde{p}(x_1,\varphi) & \approx  p(y_1 \, | \, x_1,\varphi) \: p(x_1,\varphi) \propto p(x_1,\varphi \, | \, y_1).
\end{align}
\item For $t=2:n$, assuming a Gaussian mixture approximation $\widetilde{p}(x_{t-1},\varphi \, | \,y_{1:t-1})$  exists, then derive
\begin{align}
\label{eqn_baseSI:SIBS_mgndent}
\widetilde{p}(x_t,\varphi \, | \, y_{1:(t-1)}) &= \int p(x_t \, | \, x_{t-1},\varphi) \: \widetilde{p}(x_{t-1},\varphi \, | \, y_{t-1}) \: \mathrm{d}x_{t-1},
\end{align}
which is a non-Gaussian mixture and an approximation to $p(x_t, \varphi \, | \, y_{1:(t-1)})$. Then derive a Gaussian mixture approximation $\widetilde{p}(x_t,\varphi \, | \, y_{1:t})$ to $p(x_t,\varphi \, | \, y_{1:t})$ by modified iterLap:
\begin{align}
\label{alg_baseSI:SIBS_fdent}
\widetilde{p}(x_t,\varphi \, | \, y_{1:t}) &\approx  p(y_t \, | \, x_t,\varphi) \: \widetilde{p}(x_t,\varphi \, | \, y_{1:(t-1)}) \approx p(y_t \, | \, x_t,\varphi) \: p(x_t,\varphi \, | \, y_{1:(t-1)}) \propto p(x_t, \varphi \, | \, y_{1:t}).
\end{align}
\end{enumerate}
\end{algorithm}

An approximation (up to a constant) for $p(x_t, \varphi \, | \, y_{1:t})$ is $\widetilde{p}(x_{t}, \varphi \, | \, y_{1:(t-1)}) \: p(y_t \, | \, x_t, \varphi)$.  IterLap is applied to this approximation to derive the Gaussian mixture approximation $\widetilde{p}(x_{t}, \varphi \, | \, y_{1:t})$. This completes one cycle of the algorithm. A summary is given in Algorithm \ref{alg_baseSI:SIBS}.

Since $\widetilde{p}(x_t, \varphi \, | \, y_{1:t})$ is a Gaussian mixture, so are the marginal densities $\widetilde{p}(\varphi \, | \, y_{1:t})$ and $\widetilde{p}(x_{t} \, | \, y_{1:t})$ and easily derived. Also, $\log(\widetilde{p}(x_{t}, \varphi \, | \, y_{1:(t-1)}))$ is quadratic in $x_{t}$, hence the conditional density $\widetilde{p}(x_{t} \, | \, y_{1:(t-1)}, \varphi)$ is also a Gaussian mixture. However, $\log(\widetilde{p}(x_{t}, \varphi \, | \, y_{1:(t-1)}))$ is not quadratic in $\varphi$ due to the presence of the determinant term and so $\widetilde{p}(\varphi\, | \, y_{1:(t-1)})$ is of more complicated form.

\subsection{Error correction using EM}
\label{subsec:bias_correction}

To improve the sequential approximation, density correction is done by combining importance sampling with expectation maximization to obtain Algorithm \ref{alg_baseSI:bias_em}. Importance sampling is applied to the target density $p(x_t,\varphi \, | \, y_{1:t})$ with  $\widetilde{p}(x_t,\varphi \, | \, y_{1:t})$ as the proposal function. Then, expectation maximization is used to fit a Gaussian mixture to the re-sampled points, following \citet[chap. 9]{ref:Bishop2006}, using $\widetilde{p}(x_t,\varphi \, | \, y_{1:t})$ as the initial solution. This algorithm is denoted SIEM.

\begin{algorithm}[ht]
\caption{SIEM: Bias correction with expectation maximization}
\label{alg_baseSI:bias_em}
For each $t$:
\begin{enumerate}
\item Derive a Gaussian mixture approximation $\widetilde{p}_{IL}(x_{t},\varphi \, | \,y_{1:t})$  by modified iterLap as in Algorithm \ref{alg_baseSI:SIBS};
\item Sample $(x_t^{(1)},\varphi^{(1)}),\ldots,(x_t^{(M)},\varphi^{(M)})$ from $\widetilde{p}_{IL}(x_t,\varphi \, | \, y_{1:t})$;
\item \label{substep3}Calculate weights $\omega^{(i)} \propto p(x_t^{(i)},\varphi^{(i)} \, | \, y_{1:t}) / \widetilde{p}_{IL}(x_t^{(i)},\varphi^{(i)} \, | \, y_{1:t})$;
\item \label{substep4}Re-sample $M^*$ values from $(x_t^{(1)},\varphi^{(1)}),\ldots,(x_t^{(M)},\varphi^{(M)})$ with weights $\omega^{(i)}$ to get re-weighted samples  $(x_t^{(*,1)},\varphi^{(*,1)}),\ldots,(x_t^{(*,M^*)},\varphi^{(*,M^*)})$;
\item \label{substep:fitGMM} Apply the EM algorithm to find a Gaussian mixture fit $\widetilde{p}_{EM}(x_t,\varphi \, | \, y_{1:t})$ to the $(x_t^{(*j)},\varphi^{(*j)})$ with the same number of mixture components as $\widetilde{p}(x_t,\varphi \, | \, y_{1:t})$.
\item Use the EM approximation $\widetilde{p}_{EM}(x_t,\varphi \, | \, y_{1:t})$ as $\widetilde{p}(x_t,\varphi \, | \, y_{1:t})$ in Algorithm \ref{alg_baseSI:SIBS} and derive the next iterLap approximation (back to step 1).
\end{enumerate}
\end{algorithm}

A simpler variant of Algorithm \ref{alg_baseSI:bias_em} is just to fit a single Gaussian to replace Steps 3 to 5, with weighted mean and variance calculated from sampled values $(x_t^{(1)},\varphi^{(1)}),\ldots,(x_t^{(M)},\varphi^{(M)})$. We denote this approach as SIG (Algorithm 3).  It is noted that SIG is not the simple Laplace approximation as it produces a single Gaussian that cover multiple components of iterLap approximation, creating a "smoothed" version of the iterLap density.

%%%%%%%%%%%%%%%%%%%%%%%%%%%%%%%%%%%%%%%%%%%%%%%%%%%%%%%%%%%%%%%%%%%%%%
%%%%%%%%%%%%%%%%%%%%%%%%%%%%%%%%%%%%%%%%%%%%%%%%%%%%%%%%%%%%%%%%%%%%%%
\subsection{Example model and non-identifiability}
\label{sec:ssm}

An example of the model defined in Equations \ref{eqn:ssm_obs} and \ref{eqn:ssm_state}, which will be used in this paper, is:
\begin{align}
y_{t} &= \alpha_t^2 + v_t, \label{eqn_ssm:ex1_ssm_obs} \\
\alpha_t &= \mathbbm{1}(c_1 x_t + c_2 z_t + 5 \geq 0) \: (c_1 x_t + c_2 z_t + 5), \label{eqn_ssm:ex1_ssm_alpha} \\
x_{t} &= a x_{t-1} + u_{t}, \label{eqn_ssm:ex1_ssm_state} 
\end{align}
where $\mathbbm{1}(\cdot)$ is the indicator function; $z_t$ is a known exogenous time series generated by an iid Gaussian distribution $z_{t} \sim N(0,\sigma_z^2=0.5^2)$ ; $u_{t}$, $v_{t}$ are univariate iid Gaussian variables with $u_{t} \sim N(0,\sigma_u^2=\lambda_u^{-1})$, $v_{t} \sim N(0,\sigma_v^2=\lambda_v^{-1})$; $\tau_u = \log(\lambda_u)$ and $\tau_v = \log(\lambda_v)$. The full parameter vector is $(a,c_{1:2},\tau_u,\tau_v)$ or $(a,c_{1:2},\sigma_u,\sigma_v)$. Like many latent models, inference for the parameters and latent process in this case can suffer from non-identifiability in the parameters and latent process, and this has an impact on inference performance and how one assesses the performance of the approximation. In Bayesian methods, this is often seen in a posterior distribution as a posterior with multiple modes of the same weight, or a path of values whose posterior probabilities are very near the modal value. Formally, the type of  non-identifiability under consideration is defined as follows.
\begin{dfn}[Non-identifiability]
A state space model has a non-identifiability $\mathcal{NID}(\varrho_A)$ when there is a transformation $(\varrho_A^{\prime},x_{1}^{\prime})=f(\varrho_A,x_{1})$ such that
$p(y_{1:n} \, | \, x_{1}^{\prime},\varrho_A^{\prime},\varrho_B)=p(y_{1:n} \, | \, x_1,\varrho_A,\varrho_B)$ where $\varrho_A \subset \varphi$ and $\varrho_B=\varphi \setminus \varrho_A$.
\end{dfn}

Under this definition, the model of Equations \ref{eqn_ssm:ex1_ssm_obs} to \ref{eqn_ssm:ex1_ssm_state} has 2 non-identifiability issues: $\mathcal{NID}(c_1,\sigma_u)$ and $\mathcal{NID}(c_1)$ (see Appendix \ref{app:nonident} for details). We do not address this issue here, but reduce its impact for this model in two ways: place a prior on $c_1$ that favours one of the identifiable modes, or simply assume that one of parameters in the non-identifiable set is known.

\section{Examples}
\label{sec:examples}

All the examples in this section are based on the model of Equations \ref{eqn_ssm:ex1_ssm_obs} to \ref{eqn_ssm:ex1_ssm_state}, with $n=5000$ data points and parameters: $a^{\star}=0.8$, $c_1^{\star}=1.5$, $c_2^{\star}=-1$, $\sigma_u^{\star}=0.3$ ($\tau_u^{\star} \approx 2.407$) and $\sigma_v^{\star}=10$ ($\tau_v^{\star} \approx -4.605$).

\subsection{Example 1}
\label{subsec:example_1}

\begin{figure}[!ht]
\centering
\begin{adjustbox}{center}
\begin{subfigure}[b]{0.55\textwidth}
    \centering
    \includegraphics[scale=0.40]{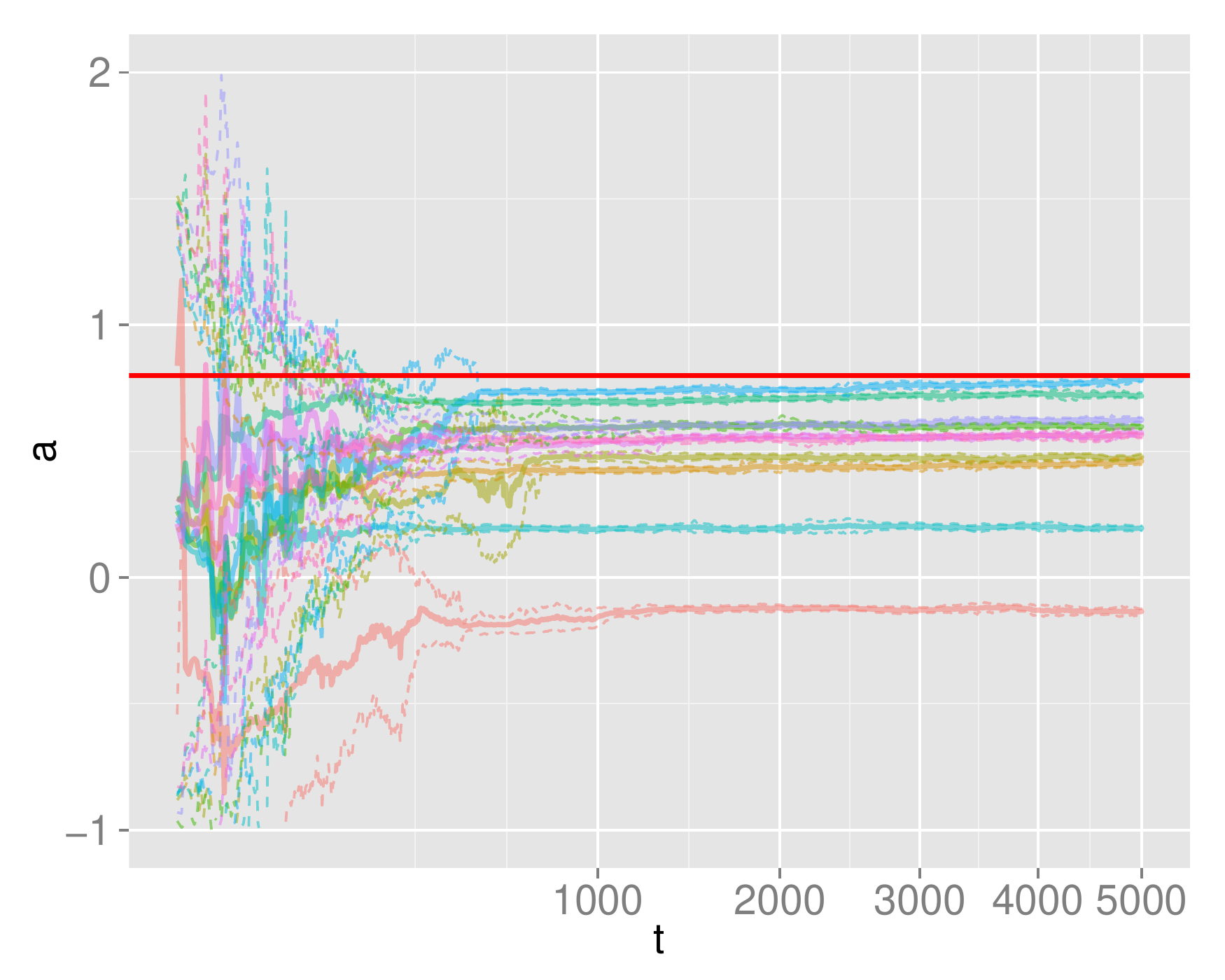}
    \subcaption{$a$}
    \label{fig_si:ex01_SIEM_mfd_a}
\end{subfigure}
\begin{subfigure}[b]{0.55\textwidth}
    \centering
    \includegraphics[scale=0.40]{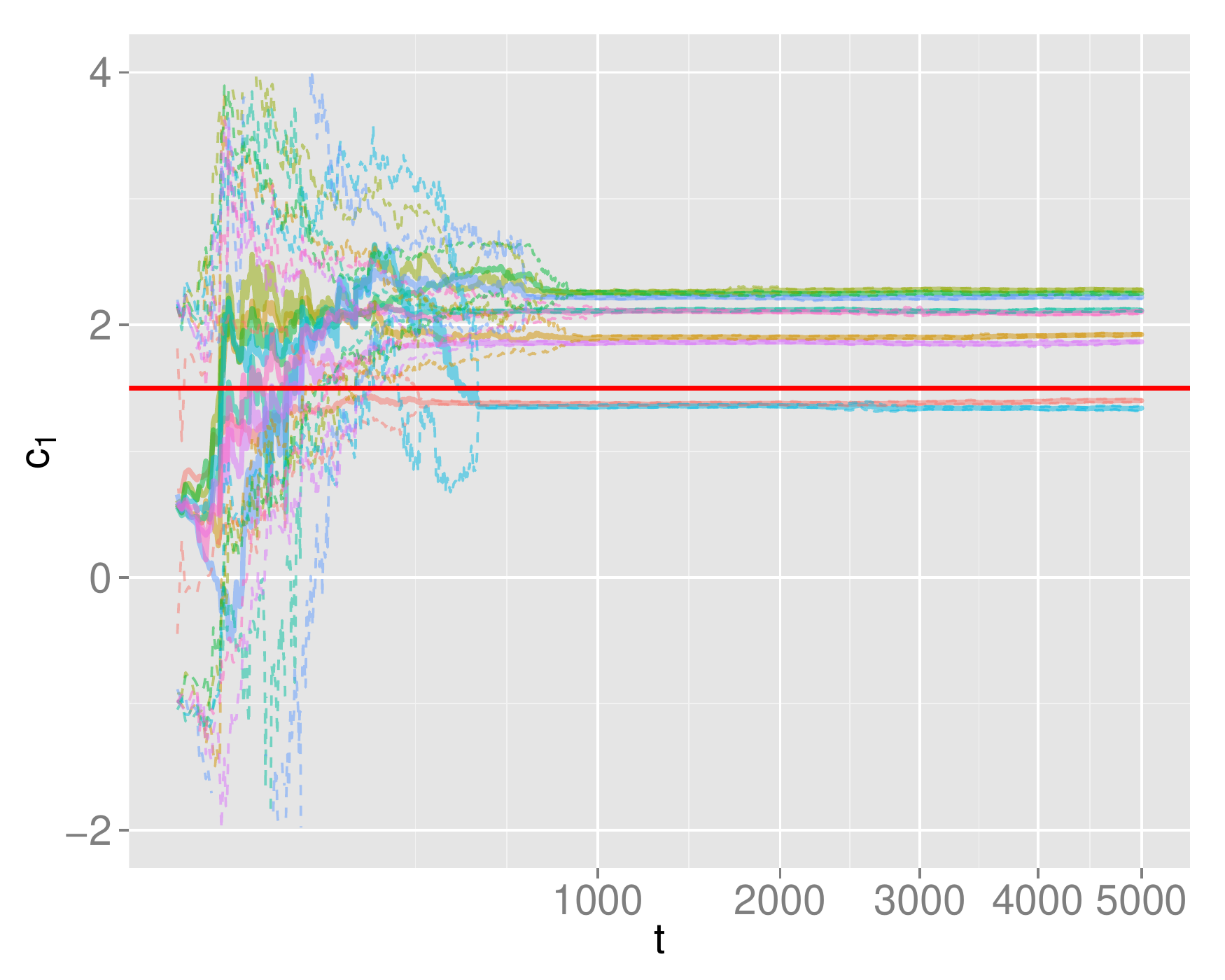}
    \subcaption{$c_1$}
    \label{fig_si:ex01_SIEM_mfd_b1}
\end{subfigure}
\end{adjustbox}

\begin{adjustbox}{center}
\begin{subfigure}[b]{0.55\textwidth}
    \centering
    \includegraphics[scale=0.40]{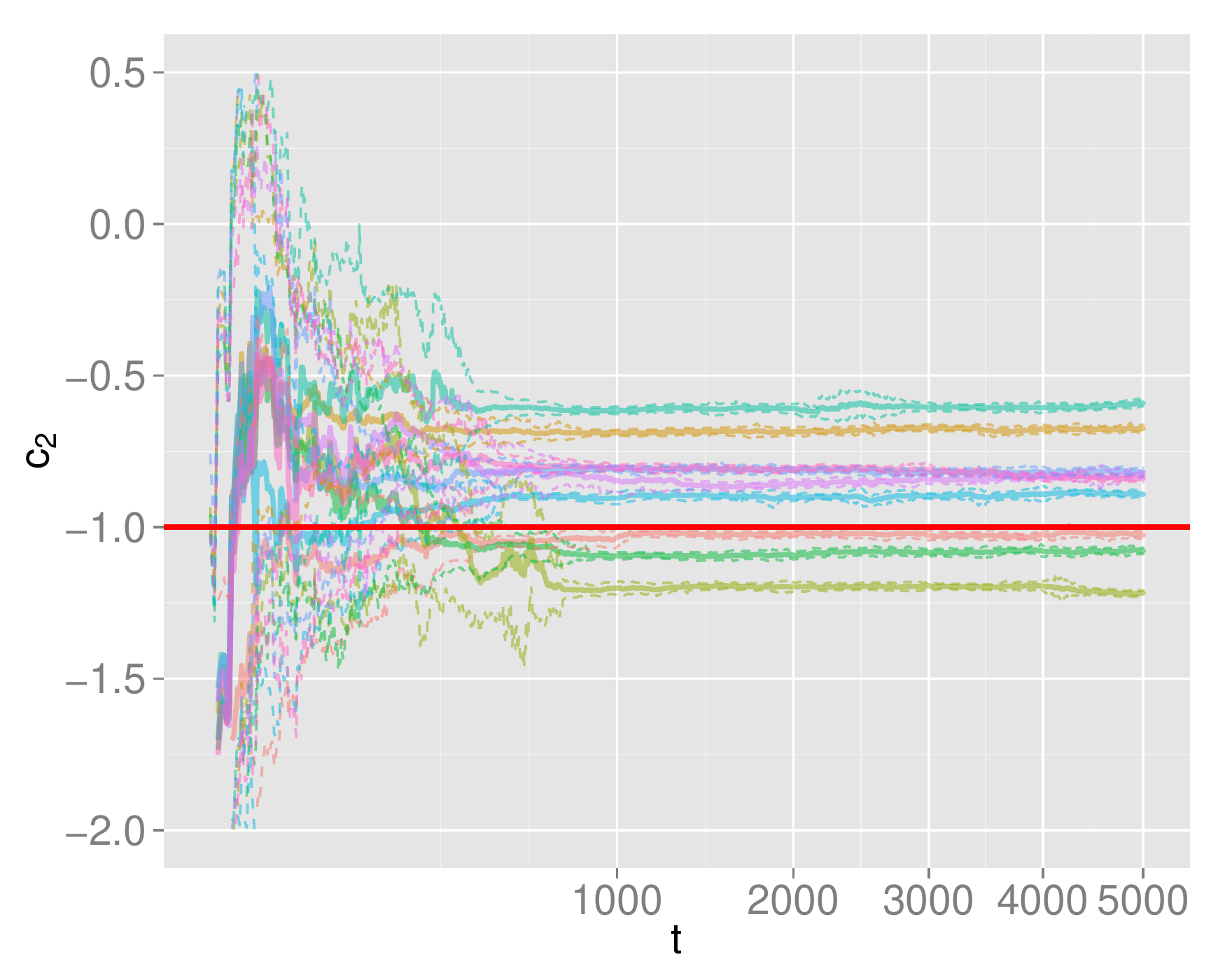}
    \subcaption{$c_2$}
    \label{fig_si:ex01_SIEM_mfd_b2}
\end{subfigure}
\begin{subfigure}[b]{0.55\textwidth}
    \centering
    \includegraphics[scale=0.40]{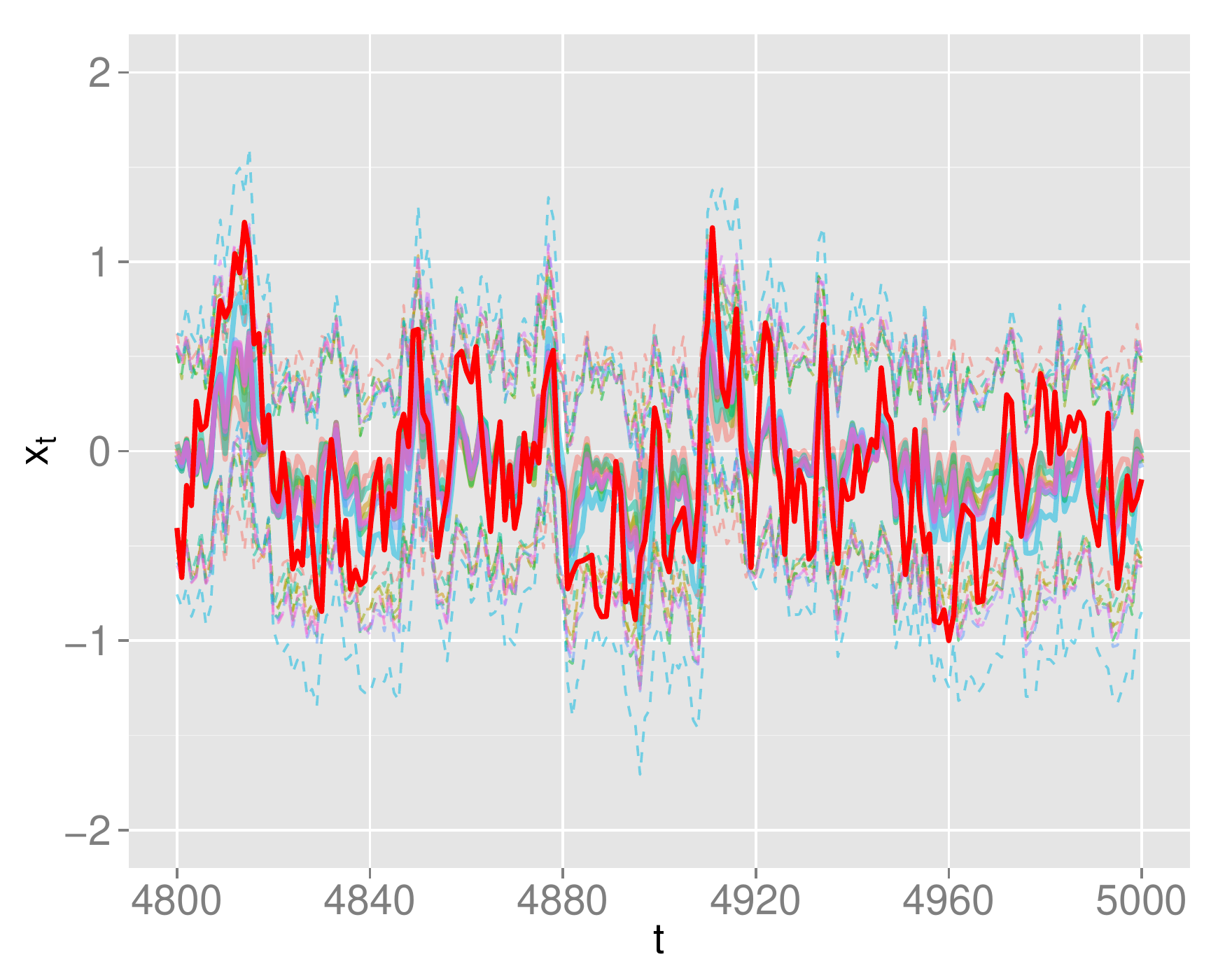}
    \subcaption{$x_t$}
    \label{fig_si:ex01_SIEM_mfd_xt}
\end{subfigure}
\end{adjustbox}
\caption{SIEM approach: the marginal filtering distributions of unknown parameters and state variable for the example in Section \ref{subsec:example_1} over 10 runs. The bright red curves represent the true values of unknown parameters and state variable.}
\label{fig_si:ex01_SIEM_mfd}
\end{figure}

\begin{figure}[!ht]
\centering
\begin{adjustbox}{center}
\begin{subfigure}[b]{0.55\textwidth}
    \centering
    \includegraphics[scale=0.40]{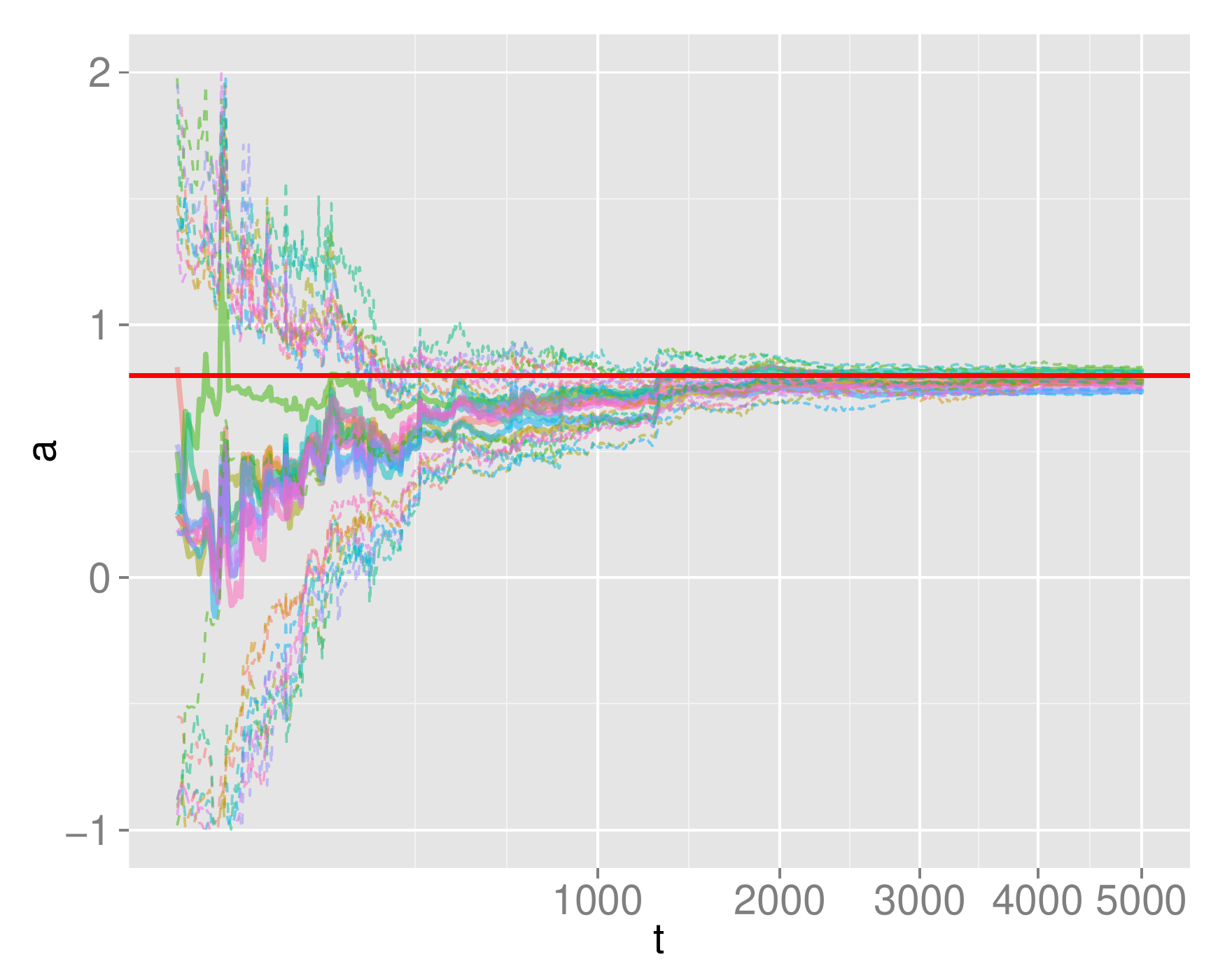}
    \subcaption{$a$}
    \label{fig_si:ex01_SIG_mfd_a}
\end{subfigure}
\begin{subfigure}[b]{0.55\textwidth}
    \centering
    \includegraphics[scale=0.40]{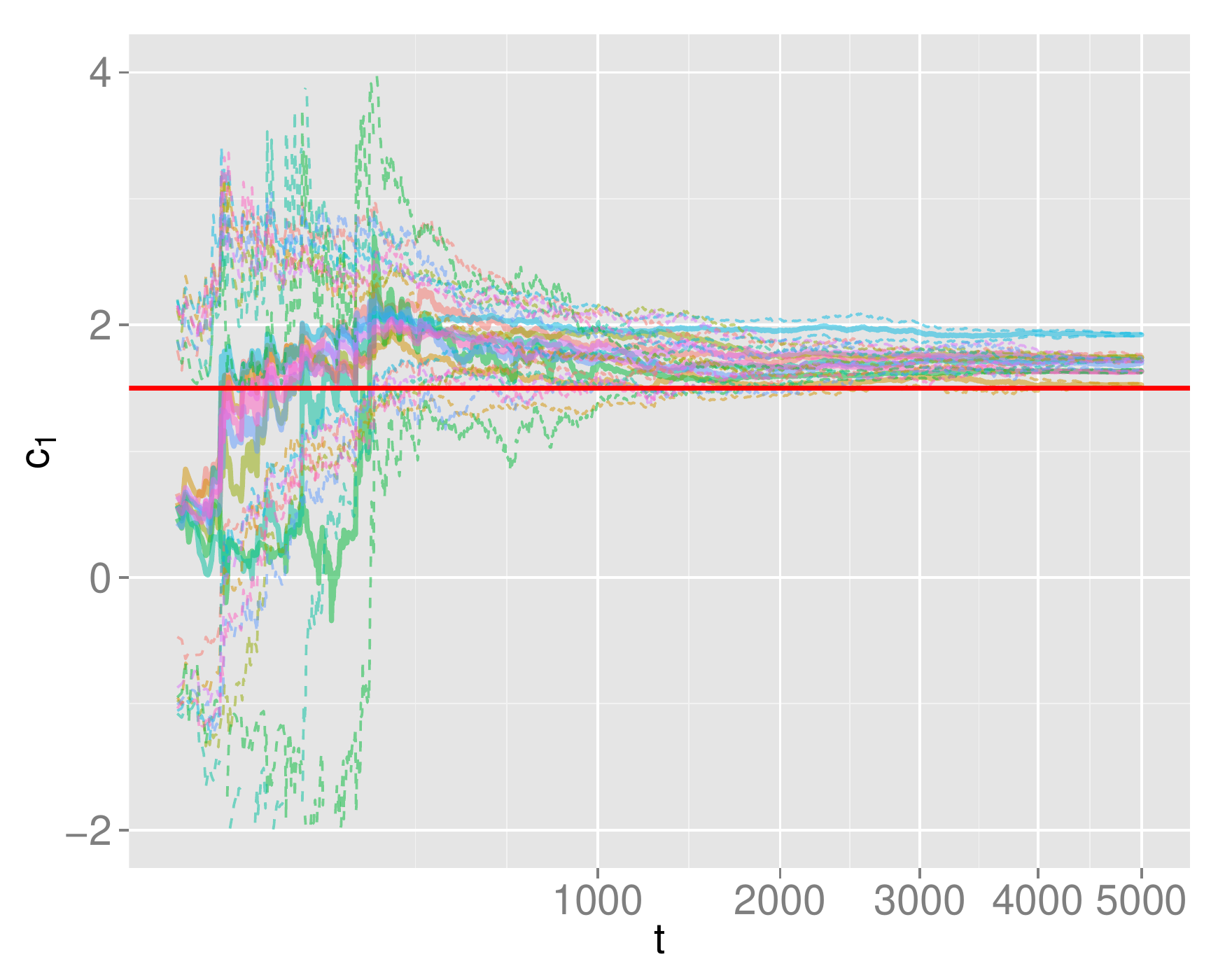}
    \subcaption{$c_1$}
    \label{fig_si:ex01_SIG_mfd_b1}
\end{subfigure}
\end{adjustbox}

\begin{adjustbox}{center}
\begin{subfigure}[b]{0.55\textwidth}
    \centering
    \includegraphics[scale=0.40]{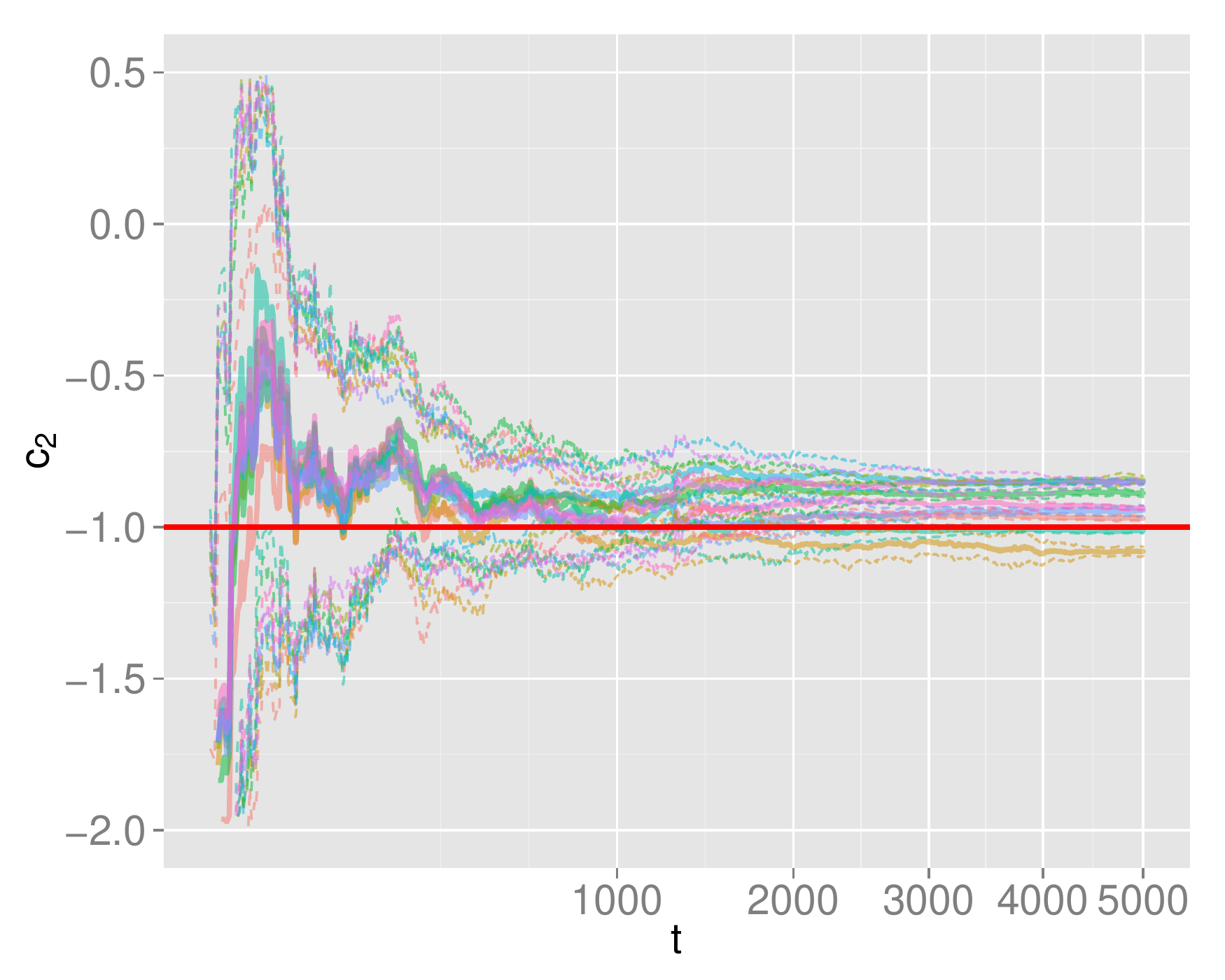}
    \subcaption{$c_2$}
    \label{fig_si:ex01_SIG_mfd_b2}
\end{subfigure}
\begin{subfigure}[b]{0.55\textwidth}
    \centering
    \includegraphics[scale=0.40]{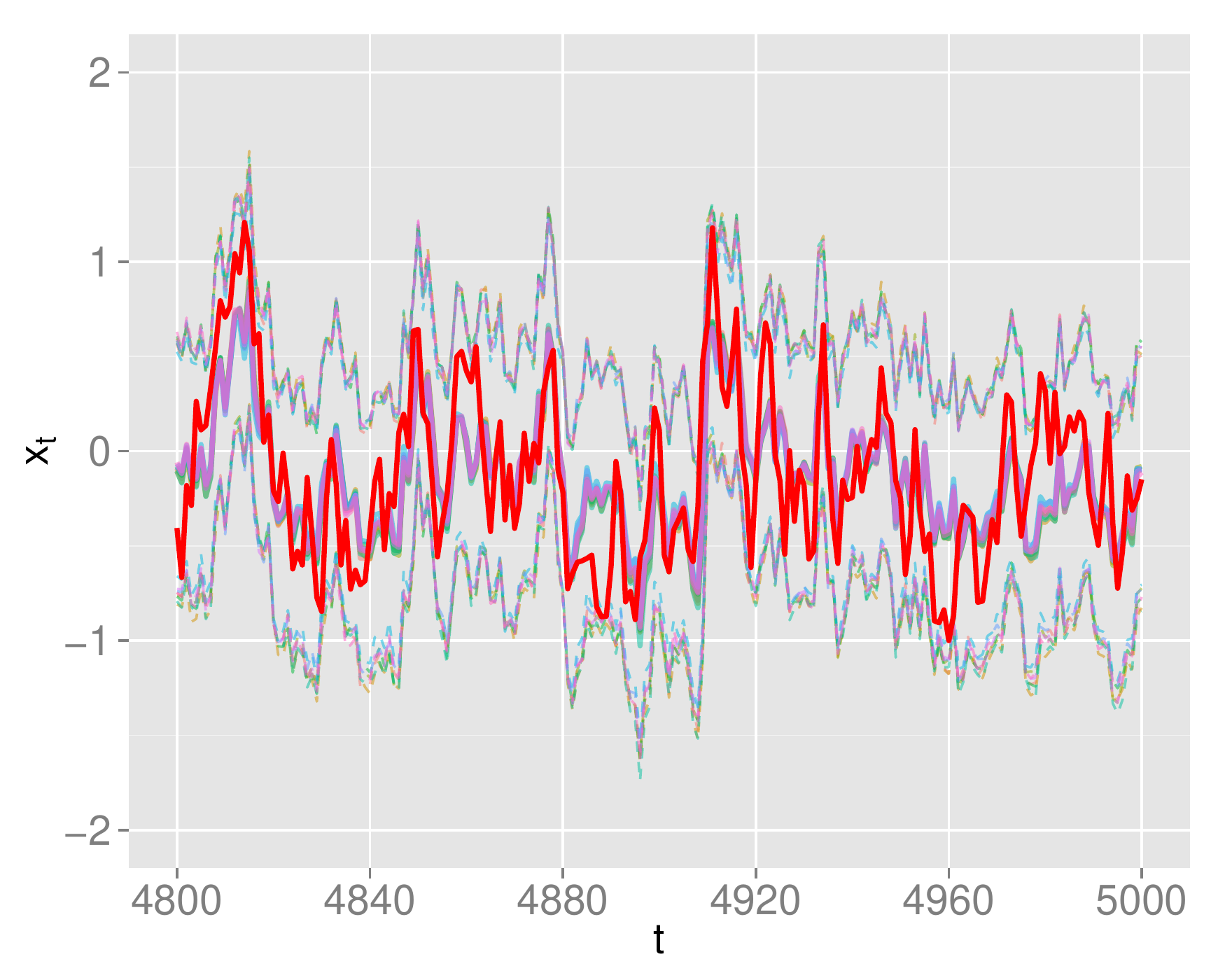}
    \subcaption{$x_t$}
    \label{fig_si:ex01_SIG_mfd_xt}
\end{subfigure}
\end{adjustbox}
\caption{SIG approach: the marginal filtering distributions of unknown parameters and state variable for the example in Section \ref{subsec:example_1}.}
\label{fig_si:ex01_SIG_mfd}
\end{figure}

It is assumed that $\sigma_u^{\star}$ and $\sigma_v^{\star}$ are known or can be estimated offline; and the prior for the other unknown parameters $\varphi=(a,c_1,c_2)$ and the initial state $x_1$ is:
\begin{align}
p(x_1, \varphi,) &= N(x_1, \varphi \, | \, \mu=(0, 0.5, 1, -3), \Sigma=\diag(2^2, 0.9^2, 1, 1)).
\end{align}
The SIEM and SIG algorithms are run ten times on the same data so that differences between the runs are due to the importance sampling step of the algorithm. The iterLap approximation is used with $5$ mixture components. The marginal filtering distributions of unknown parameters and state variable are plotted in Figures \ref{fig_si:ex01_SIEM_mfd} and \ref{fig_si:ex01_SIG_mfd} as a function of $t$ for the SIEM and SIG approaches. The means with 2 standard deviation limits are shown. For clarity, time is plotted on a square root scale and truncated to the last 200 observations for $x_t$.  These plots show that Monte Carlo variation between runs is considerably larger for the Gaussian mixture approximation (SIEM) than for the single component approximation (SIG). Both appear to produce approximations with good marginal fit to the data, although with some possible bias in parameter estimation.

\begin{figure}[!ht]
\centering
\begin{adjustbox}{center}
\begin{subfigure}[b]{0.55\textwidth}
    \centering
    \includegraphics[scale=0.40]{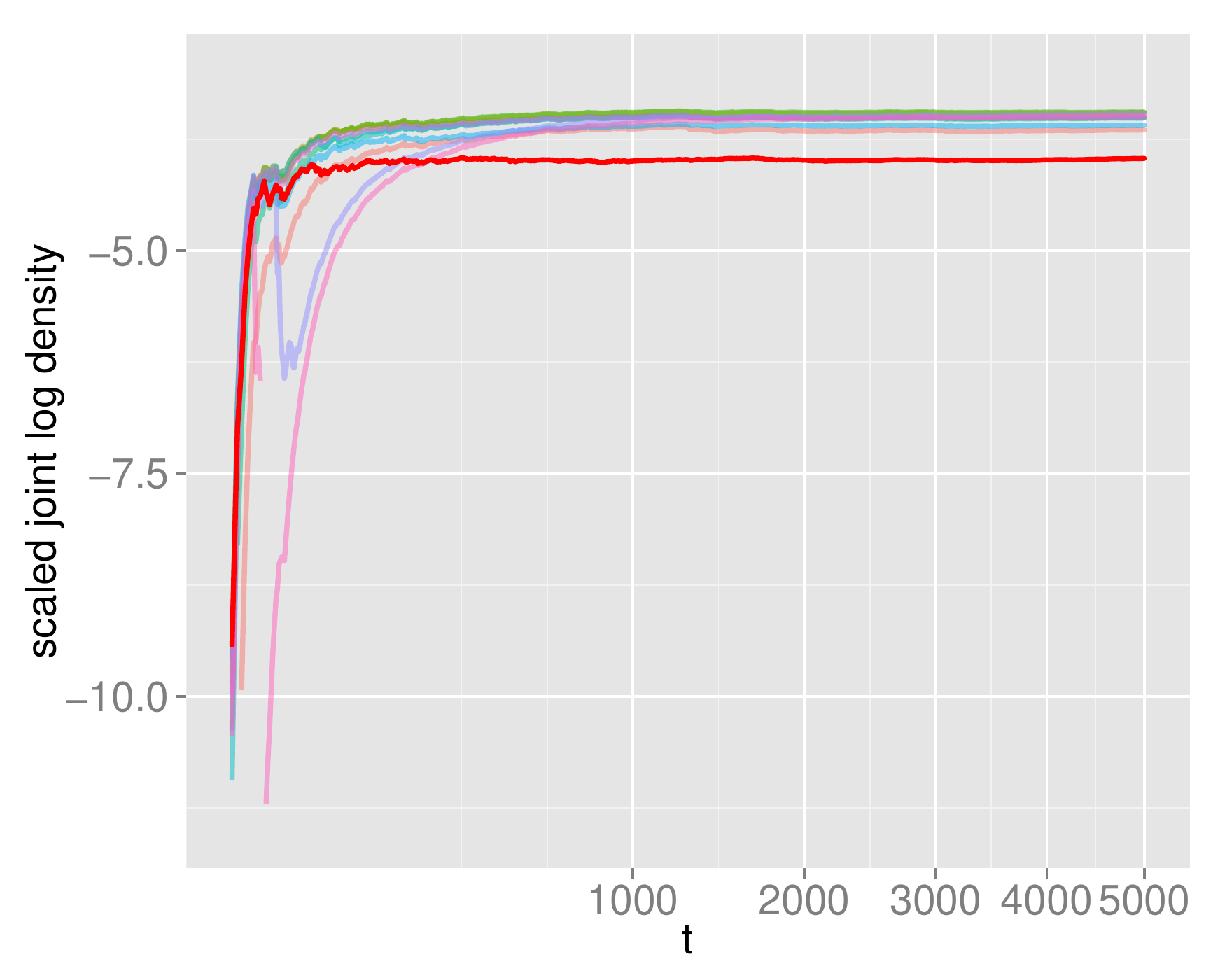}
    \subcaption{SIEM}
    \label{fig_si:ex01_sjd_SIEM}
\end{subfigure}
\begin{subfigure}[b]{0.55\textwidth}
    \centering
    \includegraphics[scale=0.40]{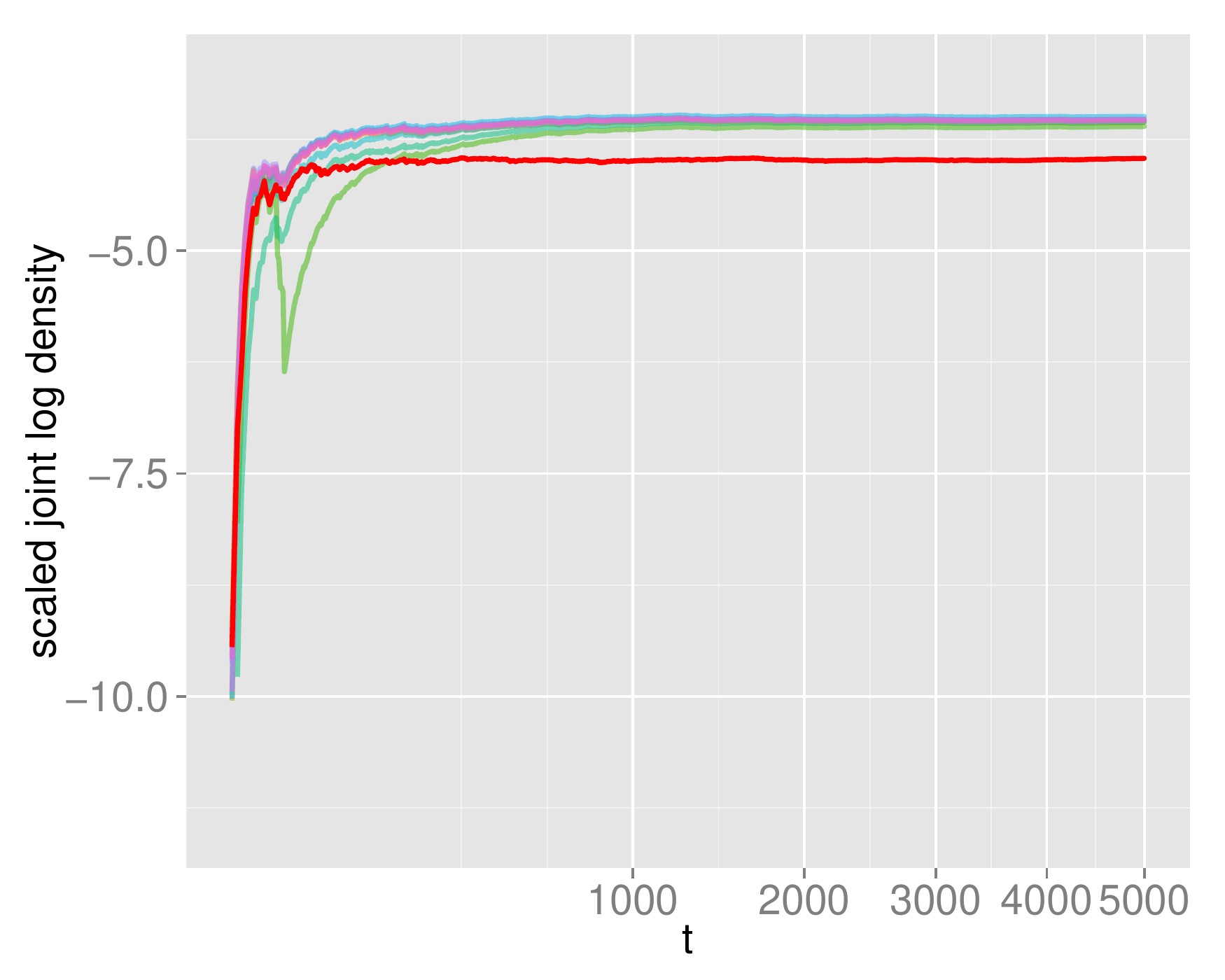}
    \subcaption{SIG}
    \label{fig_si:ex01_sjd_SIG}
\end{subfigure}
\end{adjustbox}
\caption{The scaled joint log density $\log(p(y_{1:t}, \widetilde{x}_{1:t}, \widetilde{\varphi}))/t$ versus time stamp in square root scale for the example in Section \ref{subsec:example_1}. The bright red curve corresponds with the evaluation at the true values of $\varphi$ and $x_1$.}
\label{fig_si:ex01_sjd}
\end{figure}

Figure \ref{fig_si:ex01_sjd} explores the quality of the approximation in more details.  It shows the scaled joint model log probability $\log(p(y_{1:t}, \widetilde{x}_{1:t}, \widetilde{\varphi}))/t$ as a function of $t$, where $\widetilde{x}_i$ is the marginal mean with respect to $\widetilde{p}_{EM}(x_i,\varphi \, | \, y_{1:i})$ and $\widetilde{\varphi}$ is the marginal mean with respect to  $\widetilde{p}_{EM}(x_{5000},\varphi \, | \, y_{1:5000})$. The figure shows that in many cases the SIEM and SIG algorithms infer a solution that is a better fit, in terms of model probability, than the true parameter values but that the algorithm tends not to move between possible solutions. This is a numerical counterpart to the non-identifiability issue mentioned in the previous section. 

\begin{figure}[!ht]
\centering
\includegraphics[scale=0.40]{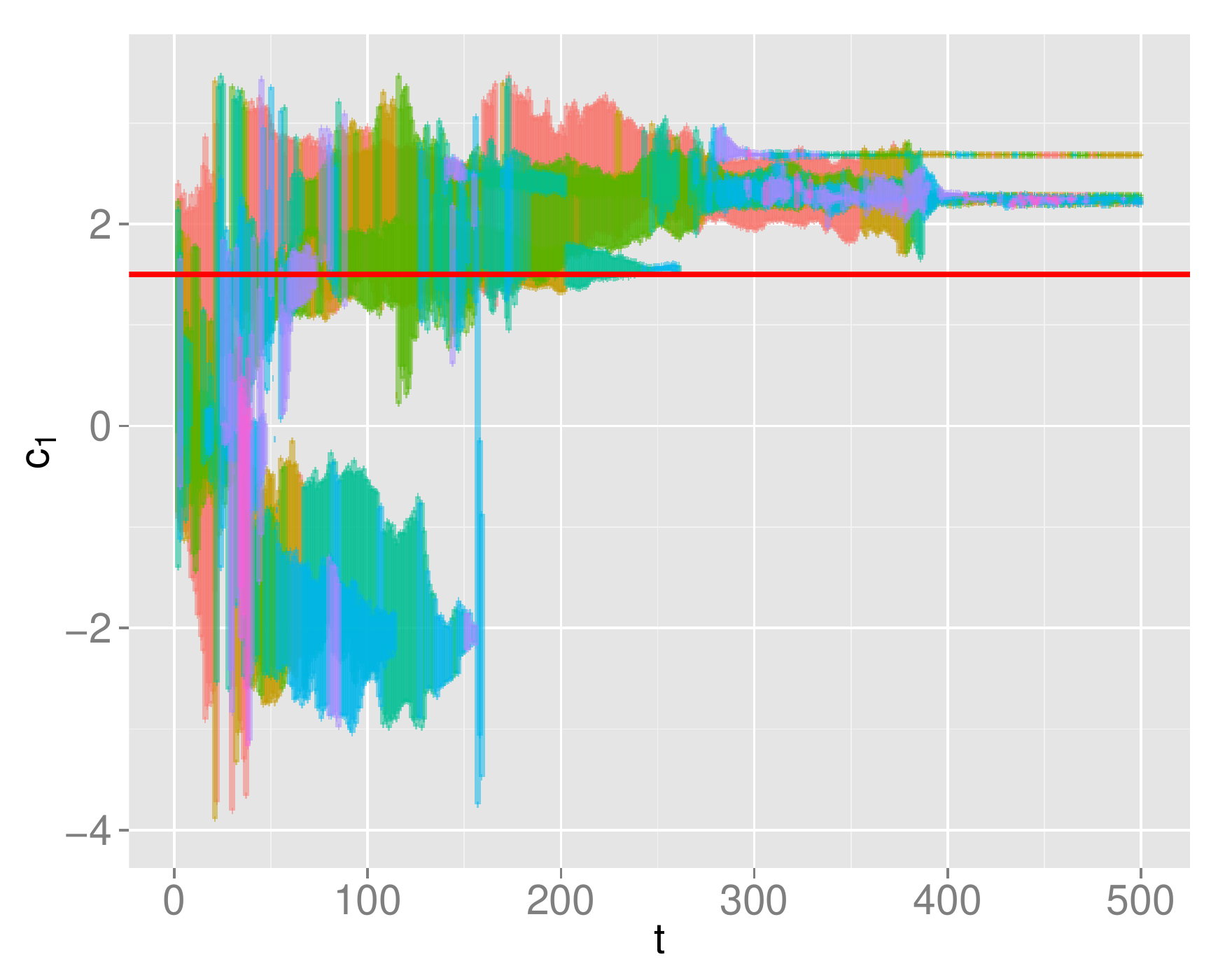}
\caption{Mixture components of $\widetilde{p}_{EM}(c_1 \, | \, y_{1:t})$ by the SIEM approach for the example in Section \ref{subsec:example_1}. Each shaded colour corresponds to the two standard deviations around the mean of the different mixture components.}
\label{fig_si:ex01_mc_SIEM_b18}
\end{figure}

To understand the difference between SIEM and SIG, and the "smoothing" effect of SIG, Figure \ref{fig_si:ex01_mc_SIEM_b18} plots the mixture components of the marginal density of $c_1$ from $\widetilde{p}_{EM}(x_t,\varphi \, | \, y_{1:t})$ of a single SIEM run. It can be seen that the approximation SIEM eliminates a component that is very close to the true value of $c_1$ around $t=260$.  This happens because the approximation will try to best accommodate the latest observation, and so is vulnerable to outlier observations that leave large weight on components that do not fit past or future observations very well. On the other hand, if a single Gaussian component is used to cover multiple components in Figure \ref{fig_si:ex01_mc_SIEM_b18} around $t=200$, the approximation will be more conservative and more robust to outliers. This explains the more consistent performance of the single component approximation SIG; this single component smooths out any local large change in the filtering distributions and so is more robust to outlier observations.

Our experience is that the SIG approach produces more consistent and robust results across a whole set of examples that we have tried. Hence, the remainder of the paper will focus on SIG and its extensions.

\subsection{Example 2}
\label{subsec:example_3}

The previous example showed differences in the approximation across separate runs due to Monte Carlo error in the importance sampling. In this example, the sensitivity of the inference to the prior is explored. The state and observation precisions are also assumed unknown; the non-identifiability issue between them makes the inference very sensitive to the choice of prior, and this can increase the differences between runs.  As mentioned earlier, the SIG approach is used as it has smaller Monte Carlo variation and more robust results.

\begin{figure}[!htb]
\centering
\begin{adjustbox}{center}
\begin{subfigure}[b]{0.55\textwidth}
    \centering
    \includegraphics[scale=0.36]{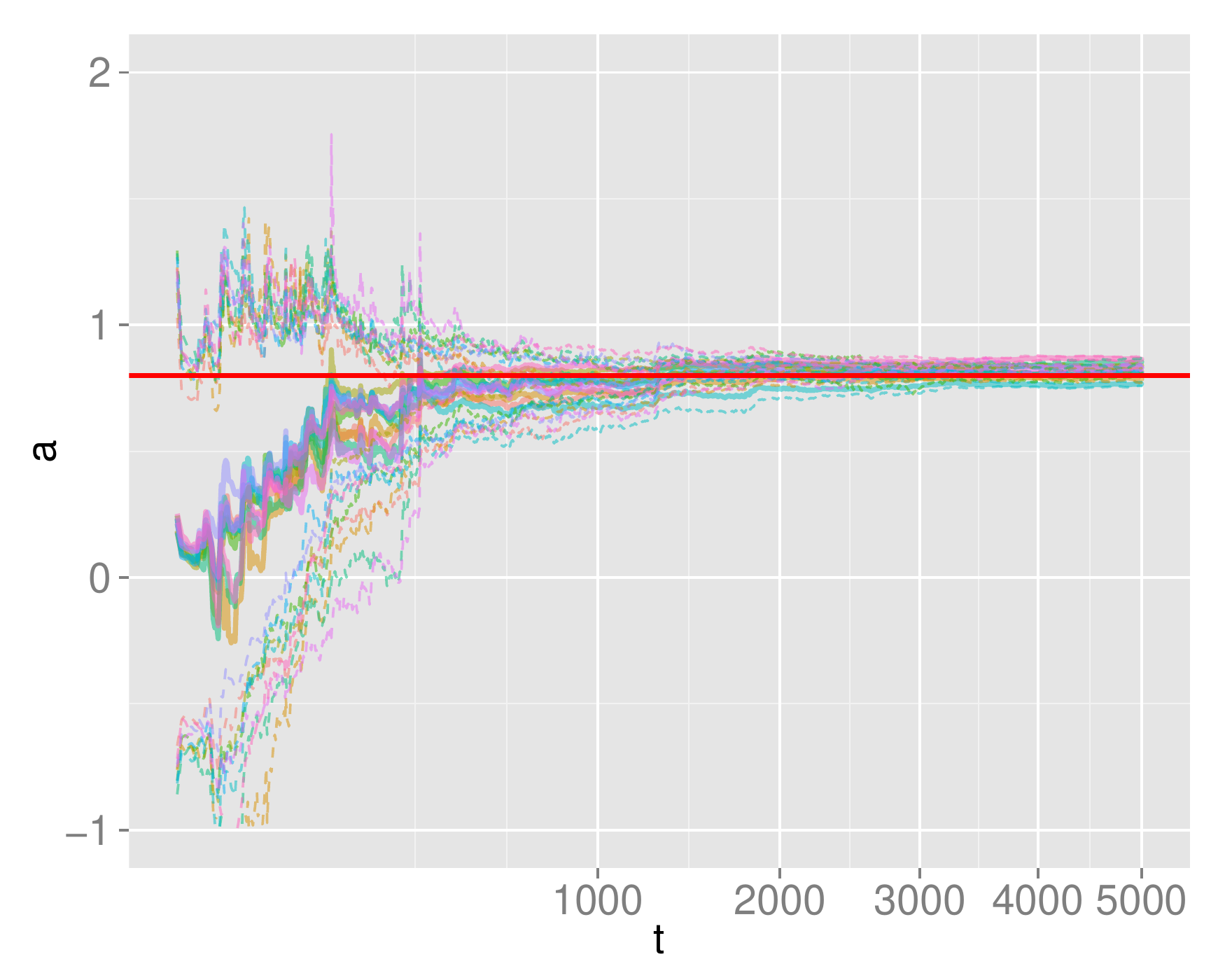}
    \subcaption{$a$}
    \label{fig_si:ex03_prior01_SIG_mfd_a}
\end{subfigure}
\begin{subfigure}[b]{0.55\textwidth}
    \centering
    \includegraphics[scale=0.36]{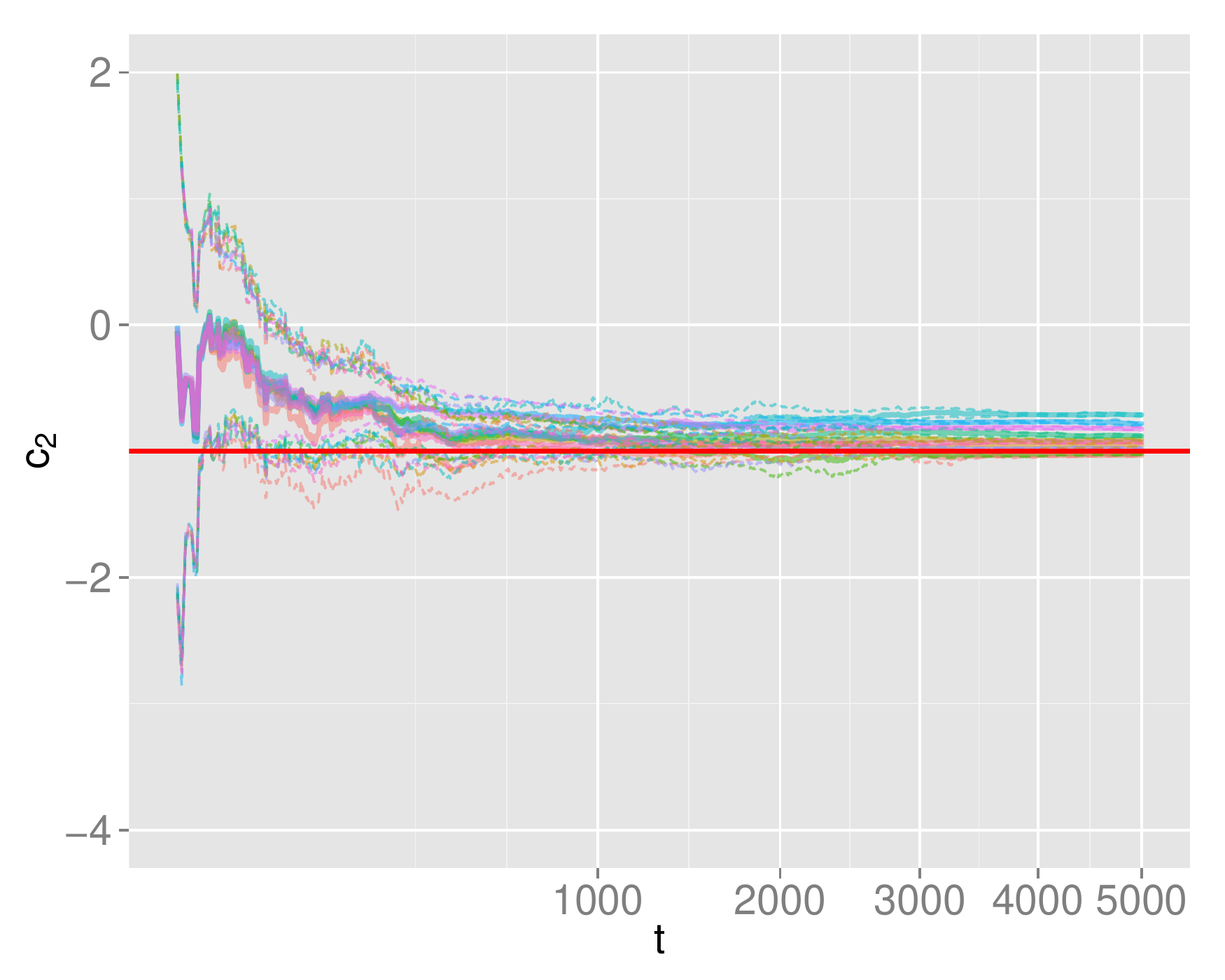}
    \subcaption{$c_2$}
    \label{fig_si:ex03_prior01_SIG_mfd_b2}
\end{subfigure}
\end{adjustbox}

\begin{adjustbox}{center}
\begin{subfigure}[b]{0.55\textwidth}
    \centering
    \includegraphics[scale=0.36]{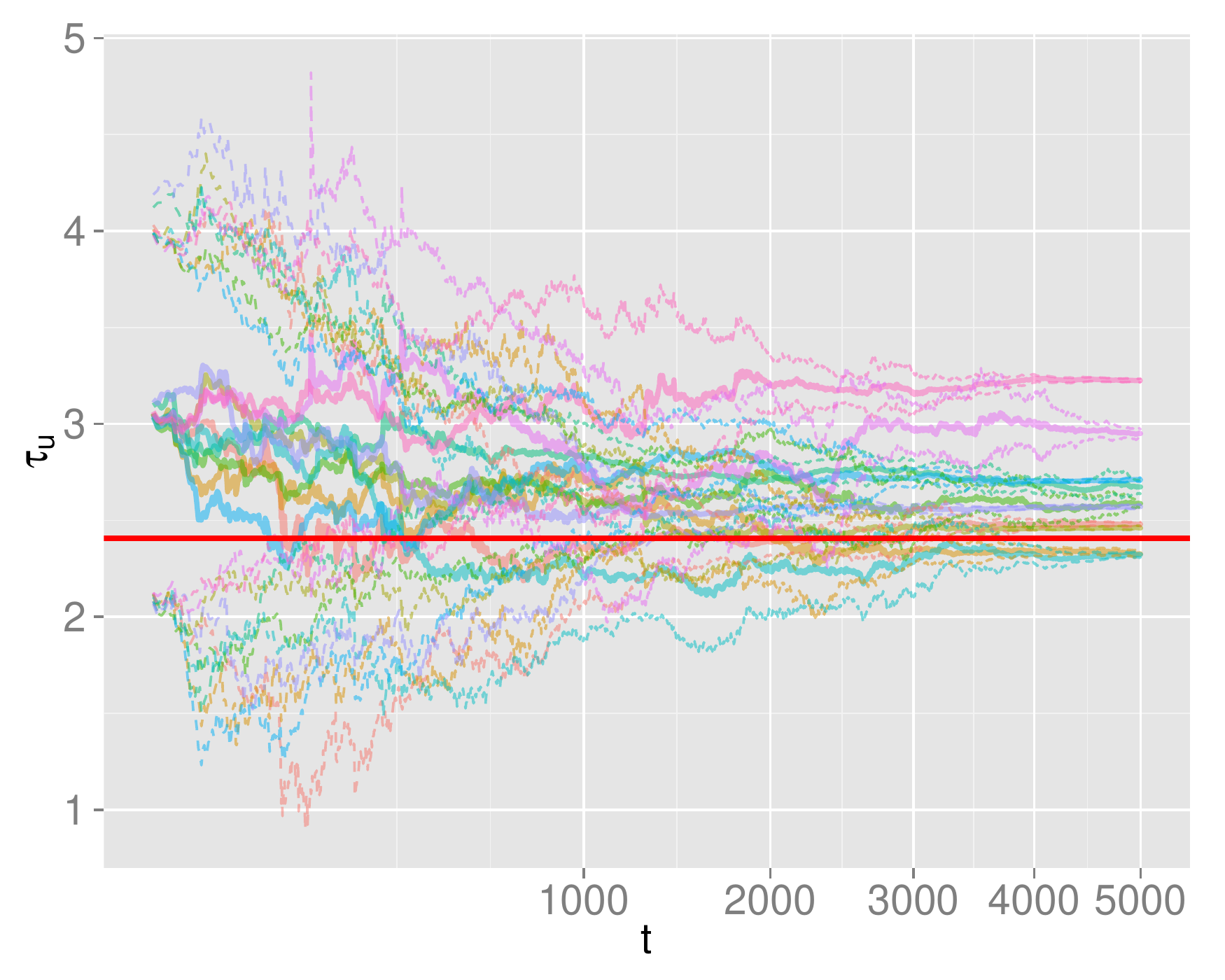}
    \subcaption{$\tau_u$}
    \label{fig_si:ex03_prior01_SIG_mfd_tauu}
\end{subfigure}
\begin{subfigure}[b]{0.55\textwidth}
    \centering
    \includegraphics[scale=0.36]{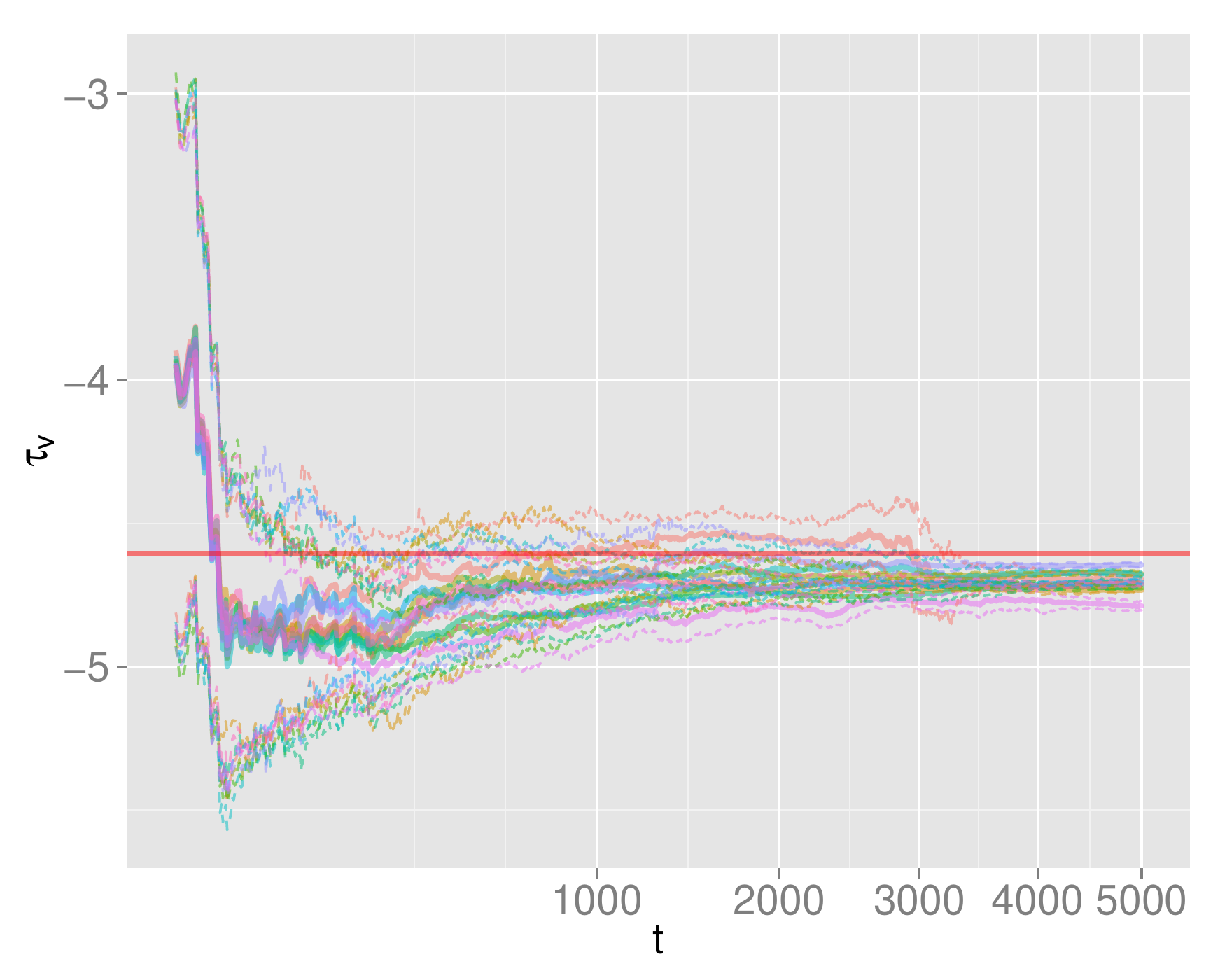}
    \subcaption{$\tau_v$}
    \label{fig_si:ex03_prior01_SIG_mfd_tauv}
\end{subfigure}
\end{adjustbox}

\begin{adjustbox}{center}
\begin{subfigure}[b]{0.55\textwidth}
    \centering
    \includegraphics[scale=0.36]{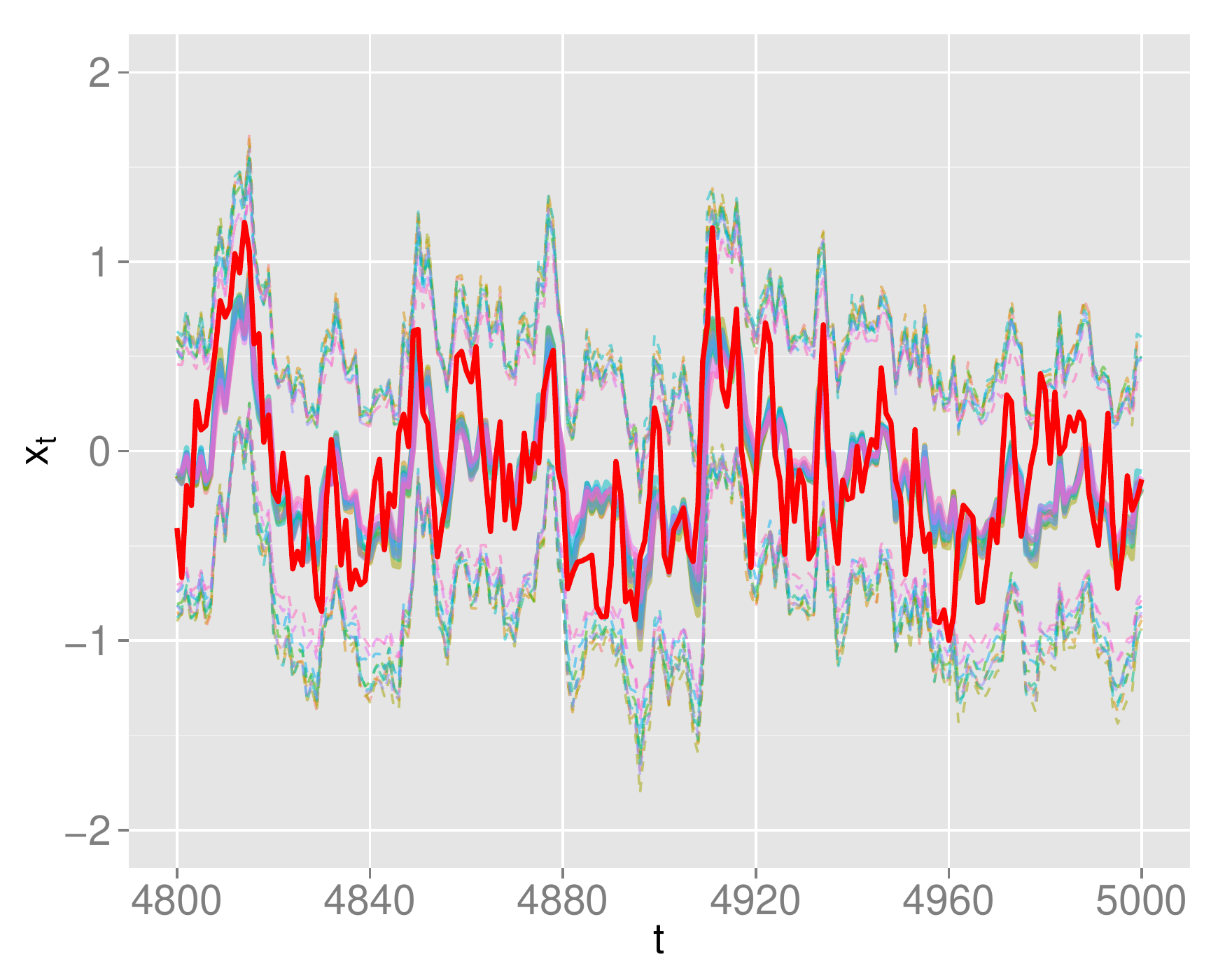}
    \subcaption{$x_t$}
    \label{fig_si:ex03_prior01_SIG_mfd_xt}
\end{subfigure}
\begin{subfigure}[b]{0.55\textwidth}
    \centering
    \includegraphics[scale=0.36]{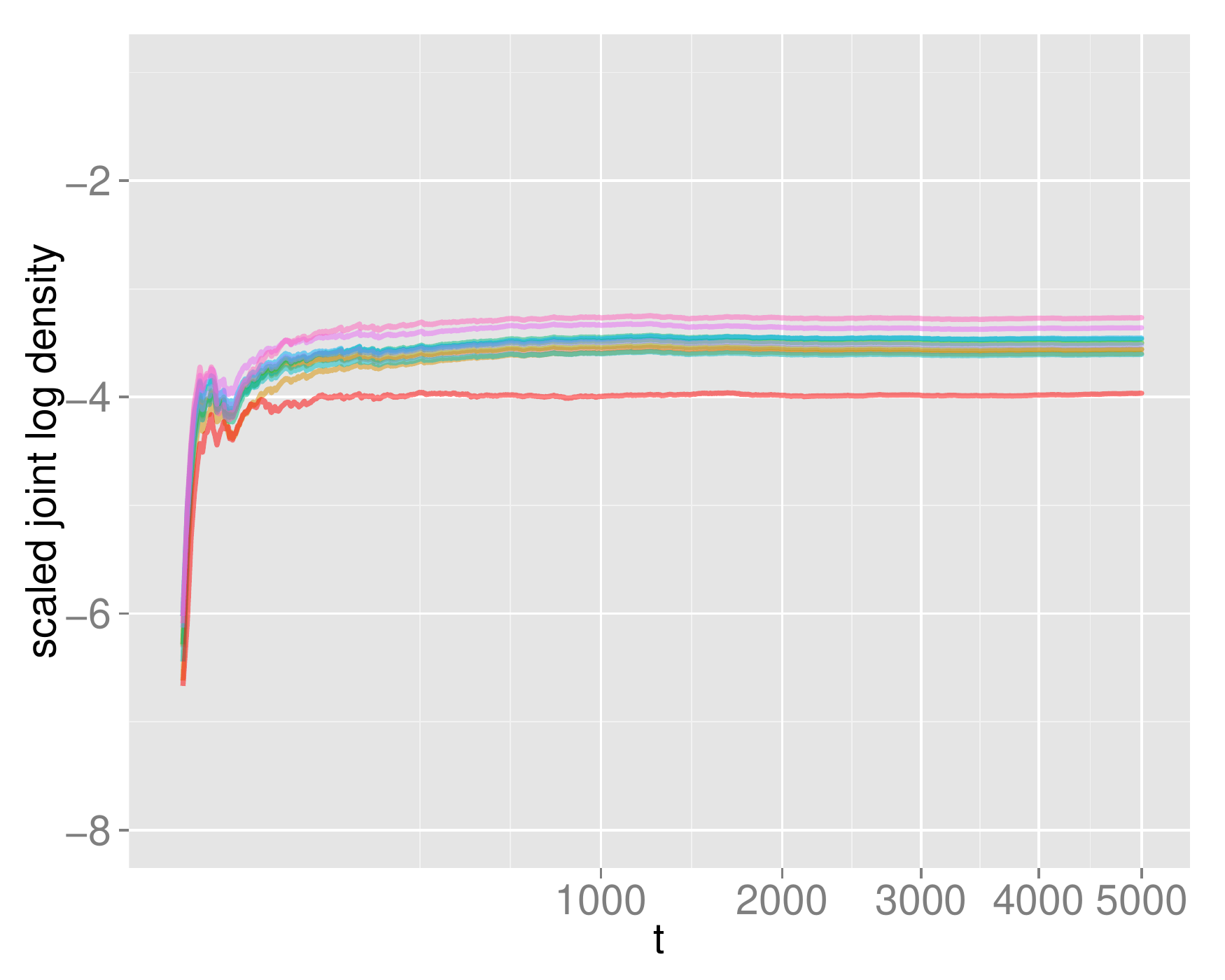}
    \subcaption{$\log(p(y_{1:t}, x_{1:t}, \varphi))/t$}
    \label{fig_si:ex03_prior01_sjd}
\end{subfigure}
\end{adjustbox}
%\Cref{fig_si:ex03_prior01_SIG_mfd_a,fig_si:ex03_prior01_SIG_mfd_b2,fig_si:ex03_prior01_SIG_mfd_tauu,fig_si:ex03_prior01_SIG_mfd_tauv,fig_si:ex03_prior01_SIG_mfd_xt}
\caption{SIG approach. Figures \ref{fig_si:ex03_prior01_SIG_mfd_a} to \ref{fig_si:ex03_prior01_SIG_mfd_xt} show the marginal filtering distributions of unknown parameters and state variable for the example in Section \ref{subsec:example_3} with prior in Equation \ref{eqn:ex03_prior01}. Figure \ref{fig_si:ex03_prior01_sjd} plots the scaled joint log density with the bright red curve corresponds with the evaluation at the true values.}
\label{fig_si:ex03_prior01_SIG_mfd}
\end{figure}

\begin{figure}[!htb]
\centering
\begin{adjustbox}{center}
\begin{subfigure}[b]{0.55\textwidth}
    \centering
    \includegraphics[scale=0.36]{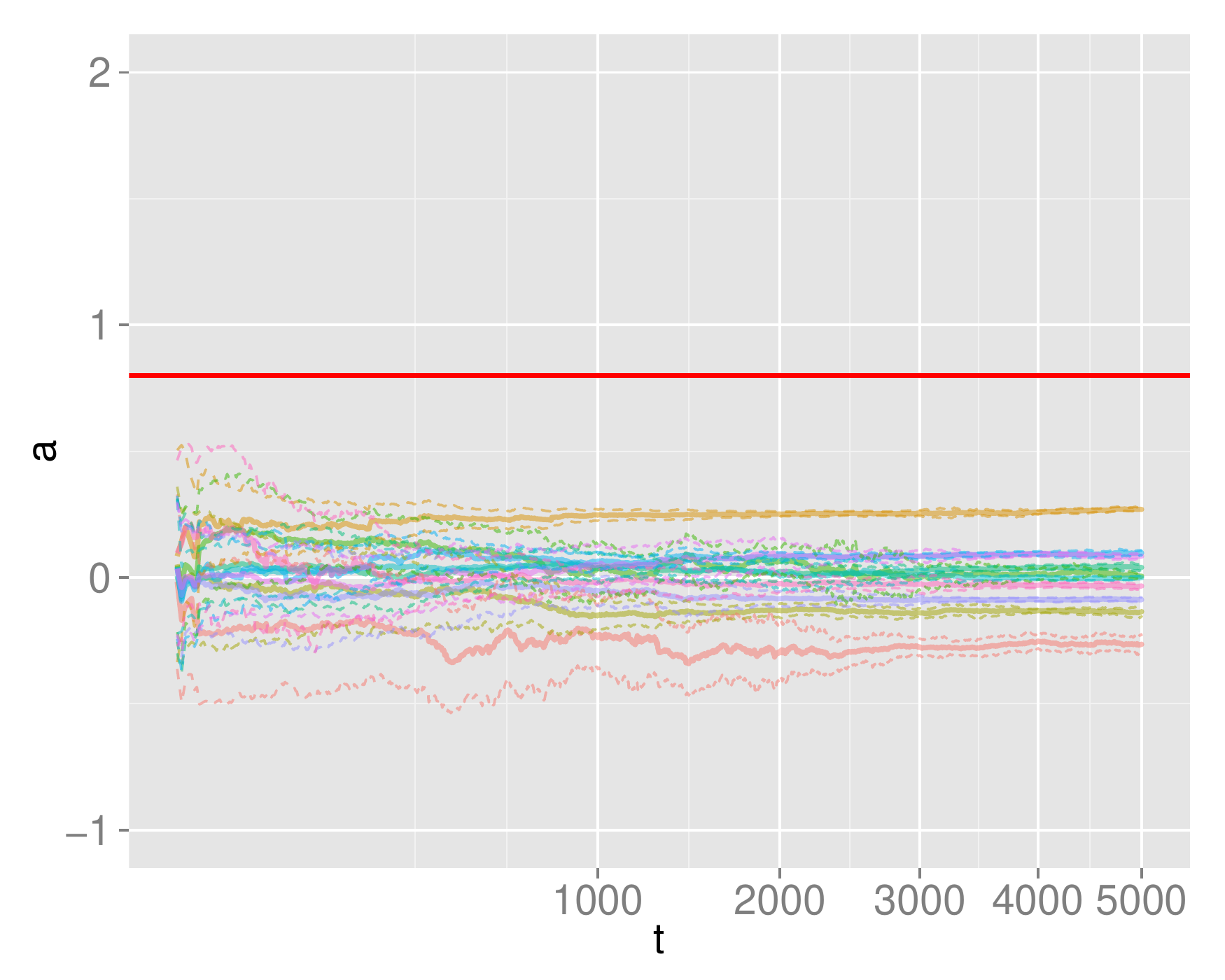}
    \subcaption{$a$}
    \label{fig_si:ex03_prior02_SIG_mfd_a}
\end{subfigure}
\begin{subfigure}[b]{0.55\textwidth}
    \centering
    \includegraphics[scale=0.36]{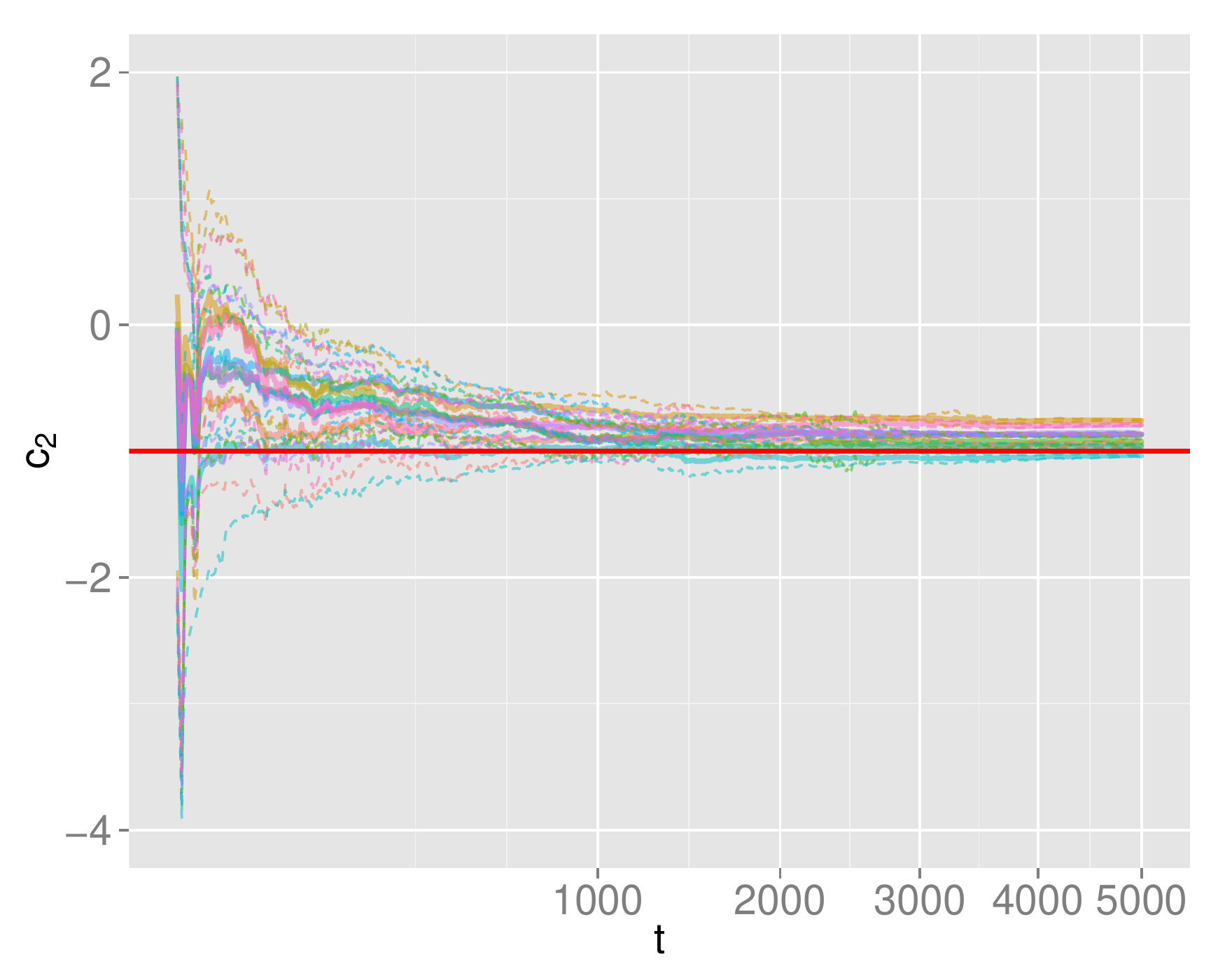}
    \subcaption{$c_2$}
    \label{fig_si:ex03_prior02_SIG_mfd_b2}
\end{subfigure}
\end{adjustbox}

\begin{adjustbox}{center}
\begin{subfigure}[b]{0.55\textwidth}
    \centering
    \includegraphics[scale=0.36]{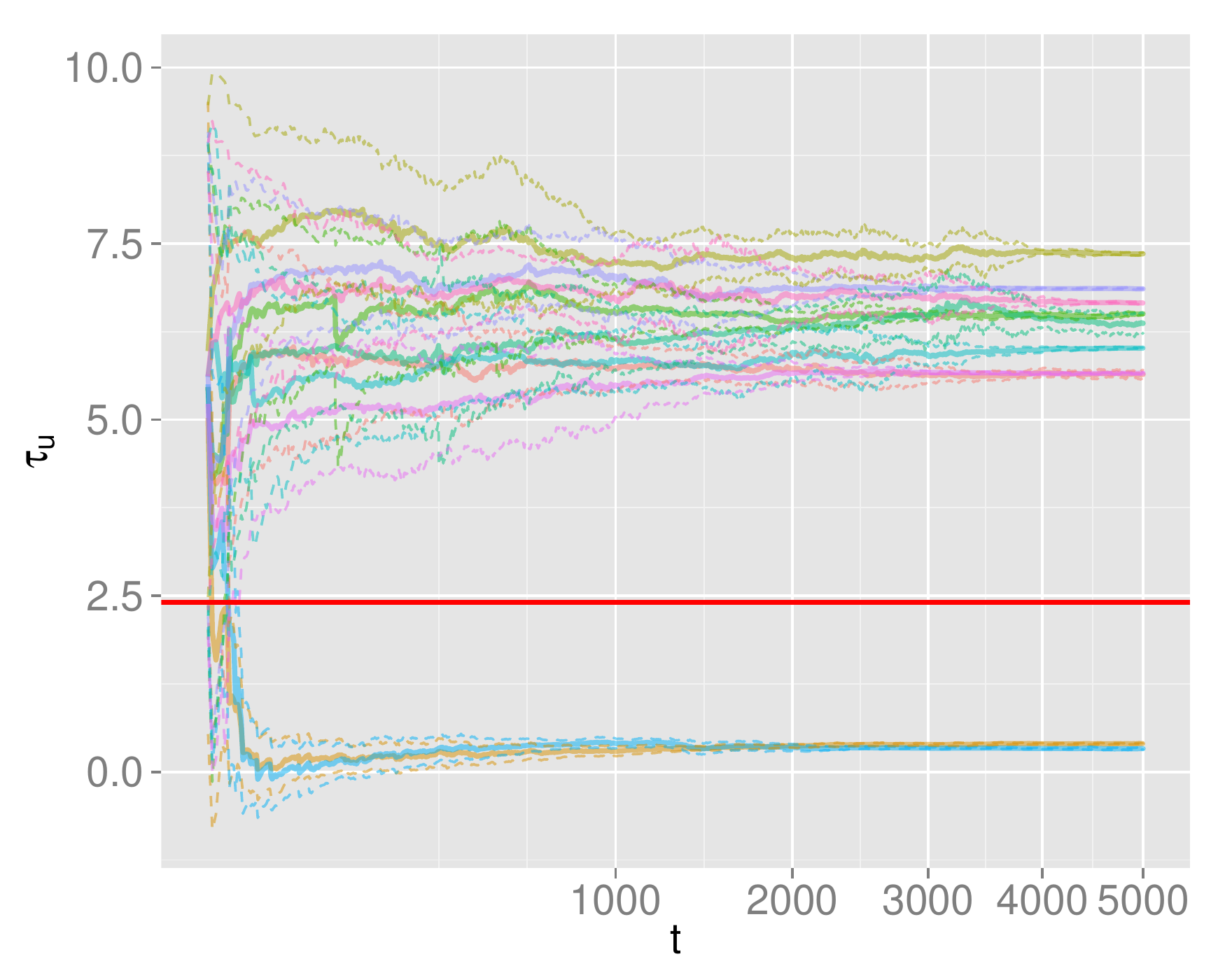}
    \subcaption{$\tau_u$}
    \label{fig_si:ex03_prior02_SIG_mfd_tauu}
\end{subfigure}
\begin{subfigure}[b]{0.55\textwidth}
    \centering
    \includegraphics[scale=0.36]{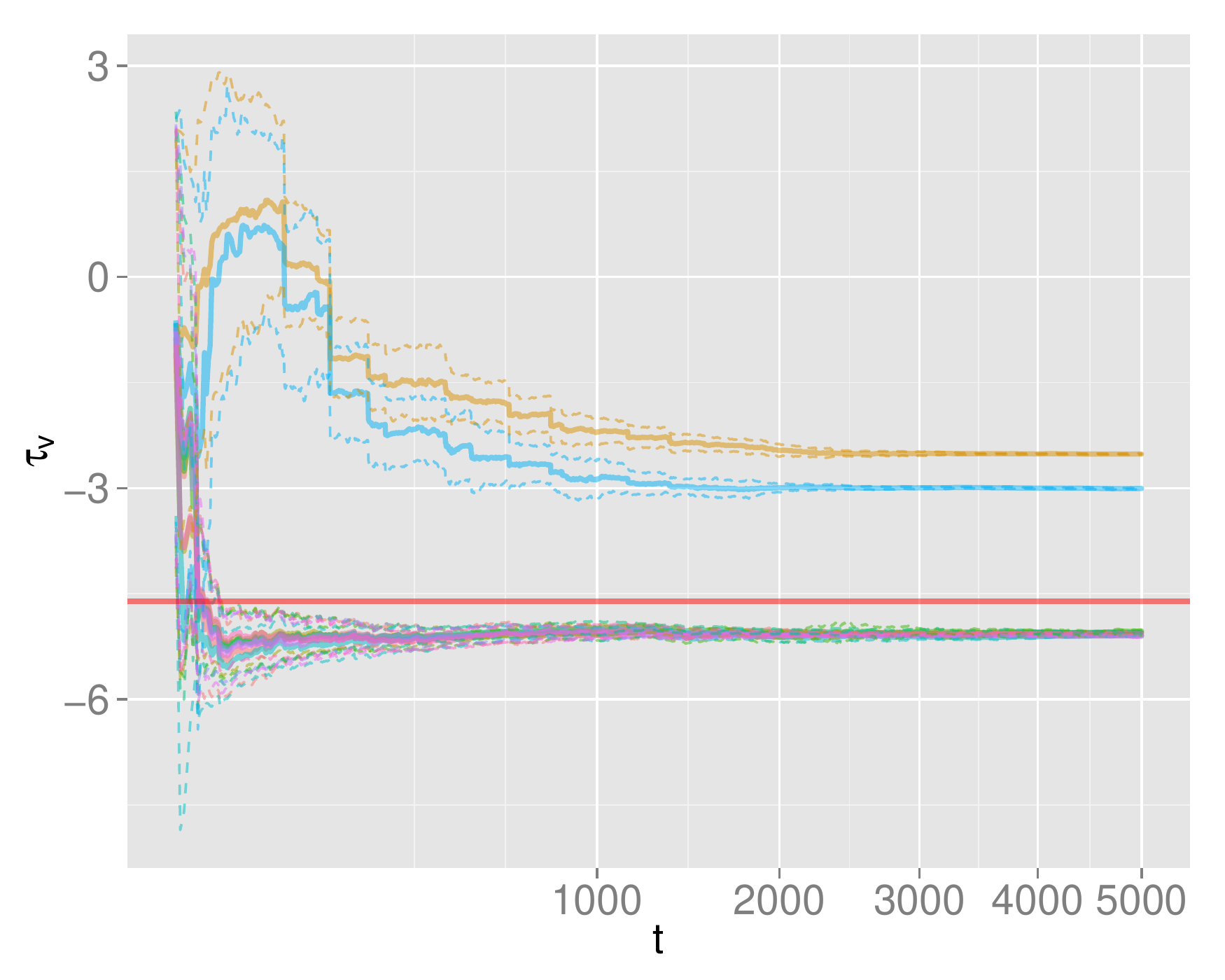}
    \subcaption{$\tau_v$}
    \label{fig_si:ex03_prior02_SIG_mfd_tauv}
\end{subfigure}
\end{adjustbox}

\begin{adjustbox}{center}
\begin{subfigure}[b]{0.55\textwidth}
    \centering
    \includegraphics[scale=0.36]{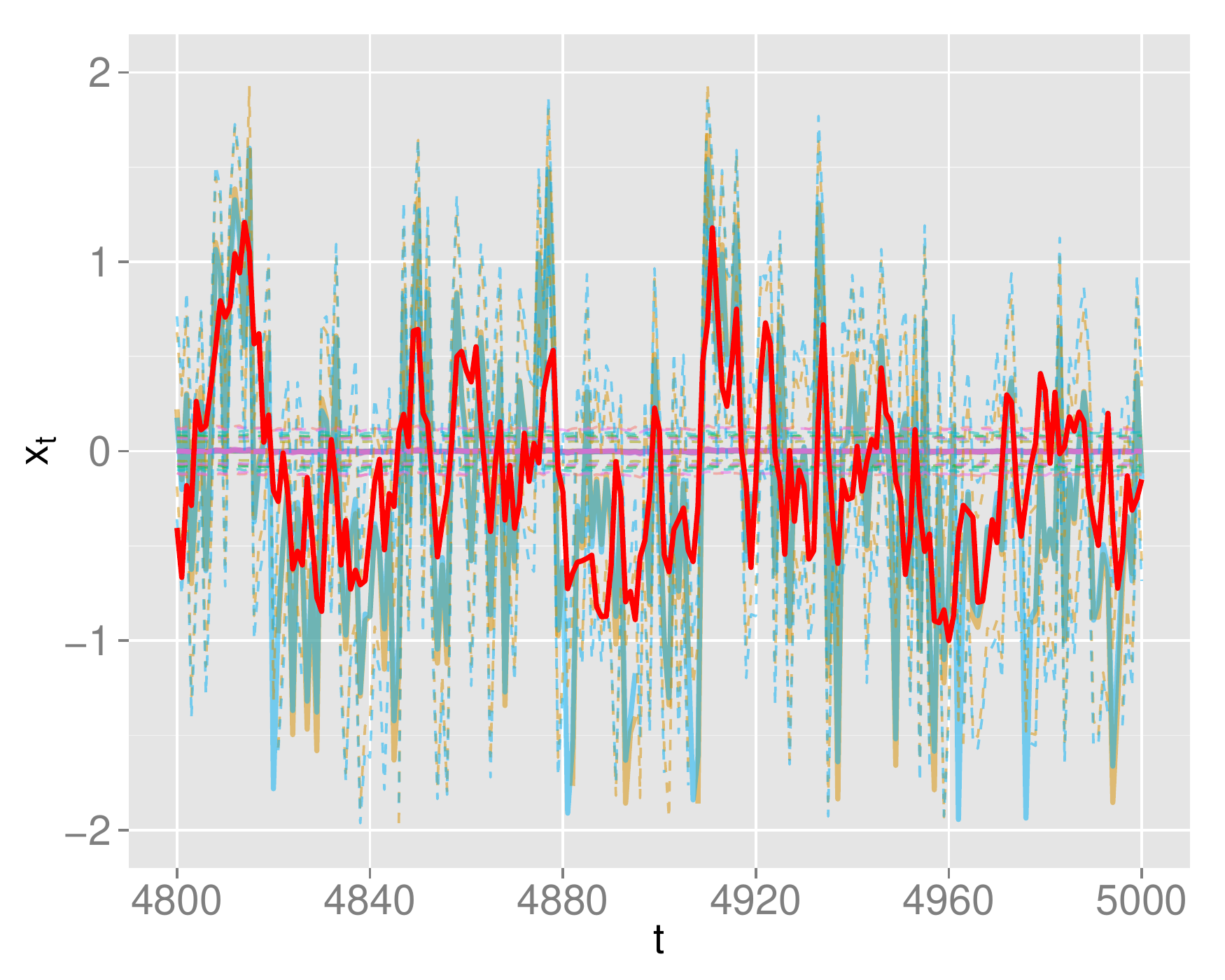}
    \subcaption{$x_t$}
    \label{fig_si:ex03_prior02_SIG_mfd_xt}
\end{subfigure}
\begin{subfigure}[b]{0.55\textwidth}
    \centering
    \includegraphics[scale=0.36]{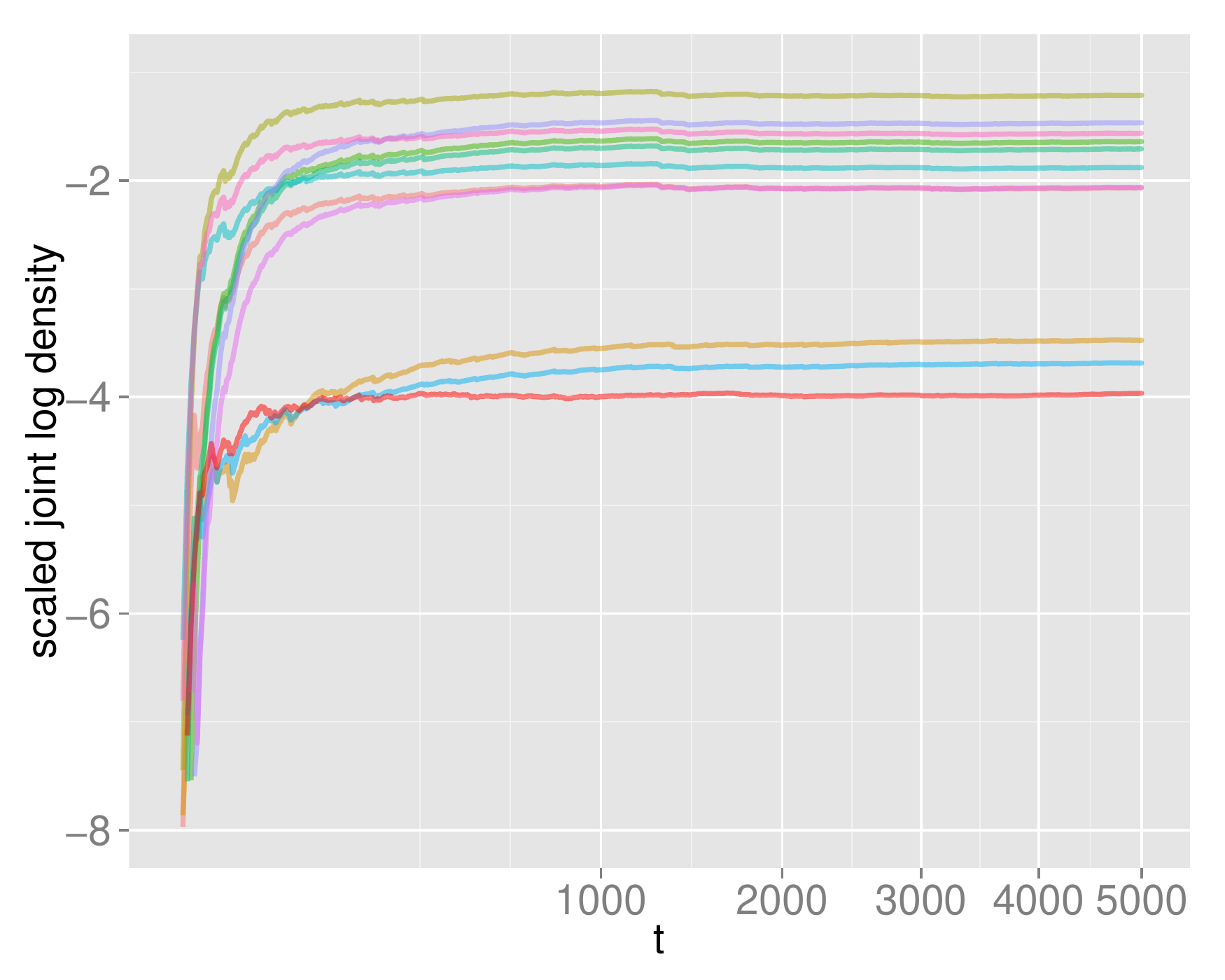}
    \subcaption{$\log(p(y_{1:t}, x_{1:t}, \varphi))/t$}
    \label{fig_si:ex03_prior02_sjd}
\end{subfigure}
\end{adjustbox}
%\Cref{fig_si:ex03_prior02_SIG_mfd_a,fig_si:ex03_prior02_SIG_mfd_b2,fig_si:ex03_prior02_SIG_mfd_tauu,fig_si:ex03_prior02_SIG_mfd_tauv,fig_si:ex03_prior02_SIG_mfd_xt}
\caption{SIG approach. Figures \ref{fig_si:ex03_prior02_SIG_mfd_a} to \ref{fig_si:ex03_prior02_SIG_mfd_xt} show the marginal filtering distributions of unknown parameters and state variable for the example in Section \ref{subsec:example_3} with prior in Equation \ref{eqn:ex03_prior02}. Figure \ref{fig_si:ex03_prior02_sjd} plots the scaled joint log density with the bright red curve corresponds with the evaluation at the true parameter values.}
\label{fig_si:ex03_prior02_SIG_mfd}
\end{figure}

The unknown parameter vector is $\varphi=(a,c_2,\tau_u,\tau_v)$, with $c_1$ assumed known in order to avoid the non-identifiability mentioned in Section \ref{sec:ssm}. Data are simulated with the same parameter values as Example 1 ($a^{\star}=0.8$, $c_1^{\star}=1.5$, $c_2^{\star}=-1$, $\tau_u^{\star} = 2.407$ and $\tau_v^{\star} = -4.605$).  Two priors are used.  The first is quite informative and is close to the true parameter values:
\begin{align}
\label{eqn:ex03_prior01}
p(x_1, \varphi) &= N(x_1, \varphi \, | \, \mu_{x_1, \varphi}^{(p)}=(0, 0.5, 0, 3, -4), \Sigma_{x_1, \varphi}^{(p)}=\diag(2^2, 0.9^2, 1, 0.5^2, 0.5^2)).
\end{align}
The second is less informative with a higher variance and a support further away from the true values $\varphi^{\star}$:
\begin{align}
\label{eqn:ex03_prior02}
p(x_1, \varphi) &= N(x_1, \varphi \, | \, \mu_{x_1, \varphi}^{(p)}=(0, 0.2, 0, 5, -2), \Sigma_{x_1, \varphi}^{(p)}=\diag(2^2, 0.9^2, 1, 2^2, 2^2)).
\end{align}
The marginal filtering densities from the SIG approach with priors by Equations \ref{eqn:ex03_prior01} and \ref{eqn:ex03_prior02} are shown in Figures \ref{fig_si:ex03_prior01_SIG_mfd} and \ref{fig_si:ex03_prior02_SIG_mfd} respectively.

Figure \ref{fig_si:ex03_prior01_SIG_mfd} shows that SIG runs do converge to the true parameter values with the first informative prior. However, inference results with the prior by Equation \ref{eqn:ex03_prior02} show significant differences, especially in Figures \ref{fig_si:ex03_prior02_SIG_mfd_tauu} to \ref{fig_si:ex03_prior02_SIG_mfd_xt}. The runs can be divided into two groups. In one group, the state equation precision $\tau_u$ is inferred to be smaller and the observation precision $\tau_v$ to be higher than the true values, which induces the inferred value of $x_t$ to strongly track the observations. In the other group, $\tau_u$ is inferred to be larger and $\tau_v$ smaller, which leaves the inferred value $x_t$ to stay constant. Although both are reasonable local approximations, the first group is more interesting than the second in terms of interpretability.

The scaled joint log densities for results of two priors are plotted in Figures \ref{fig_si:ex03_prior01_sjd} and \ref{fig_si:ex03_prior02_sjd}. For the case of the second prior, Figure \ref{fig_si:ex03_prior02_sjd} shows that the densities values of two discussed groups above are still higher than the one of true parameters values, hinting again at the problem of numerical non-identifiability for these local approximations.

\subsection{Comparison with pomp}
\label{subsec:pomp}

In this section, the SIG approach is compared with the bsmc function of R package pomp \citep{ref:King2015}, which is an implementation of Liu and West's method \citep{ref:Liu2001}. We choose this method as it has similar settings with our proposed method, a Bayesian solution with no assumption of sufficient statistics.

\begin{figure}[!htb]
\centering
\begin{adjustbox}{center}
\begin{subfigure}[b]{0.55\textwidth}
    \centering
    \includegraphics[scale=0.37]{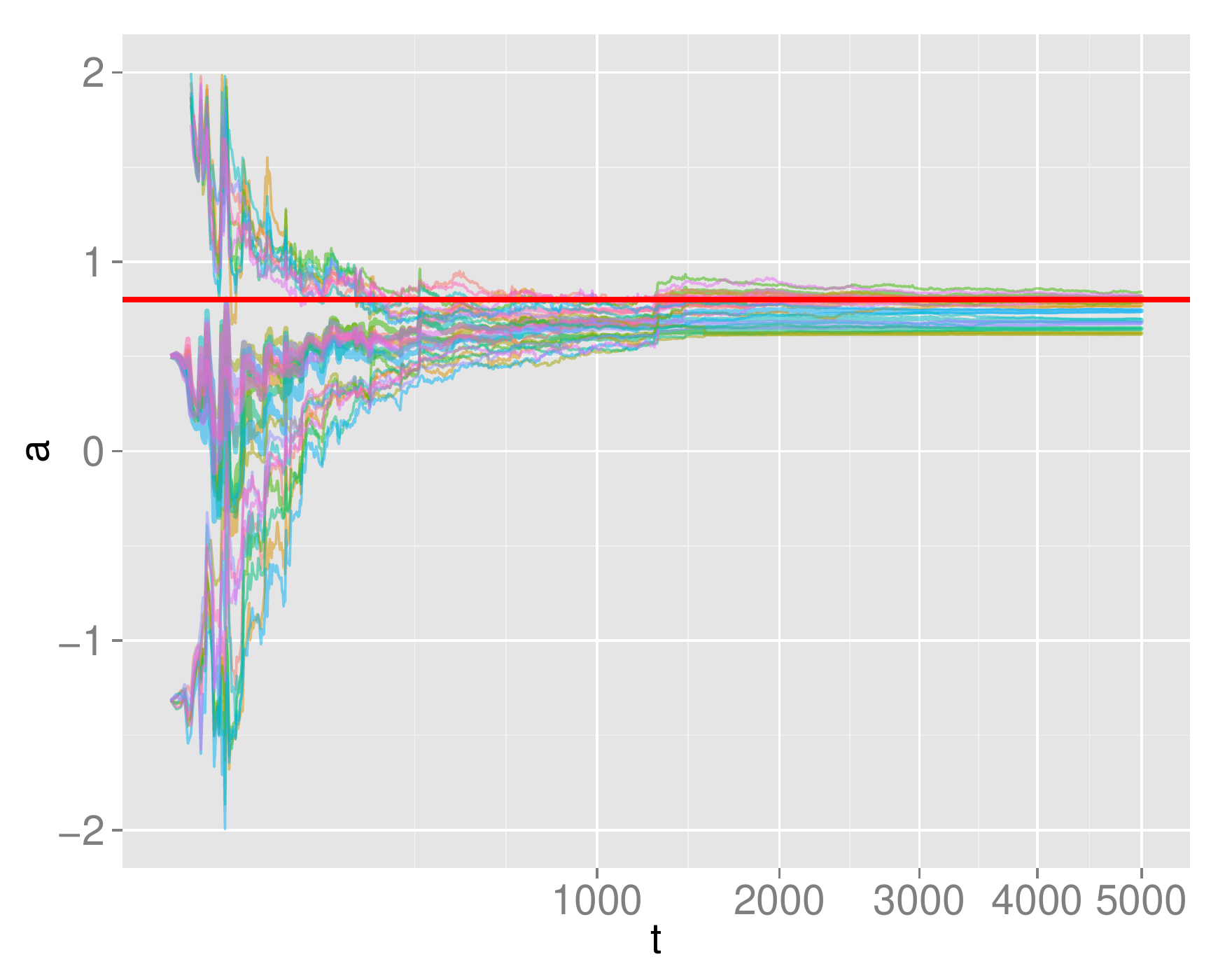}
    \subcaption{pomp-bsmc: the normal case in Section \ref{subsec:example_1}}
    \label{fig_si:ex01_bsmc_mfd_a}
\end{subfigure}
\end{adjustbox}

\begin{adjustbox}{center}
\begin{subfigure}[b]{0.55\textwidth}
    \centering
    \includegraphics[scale=0.37]{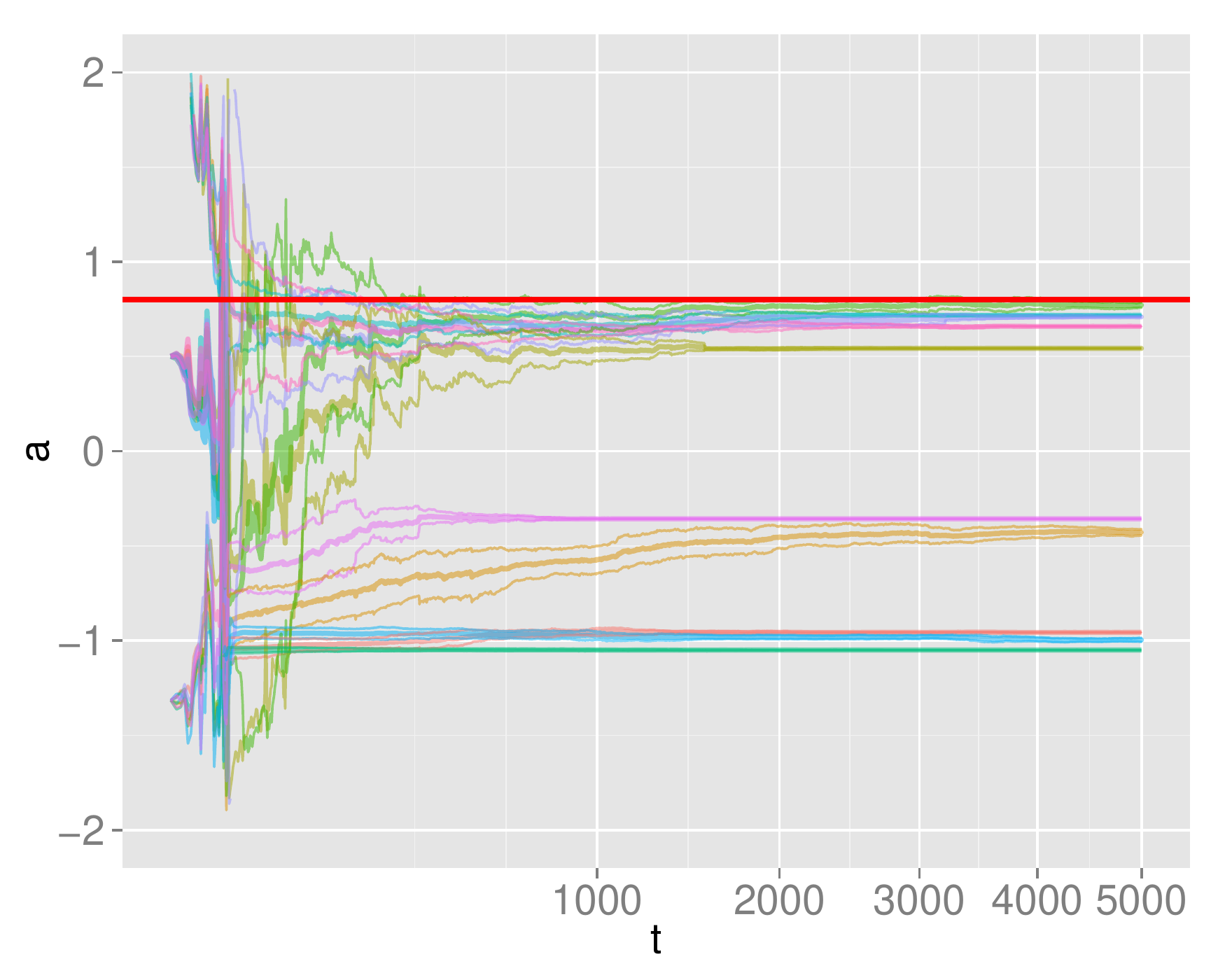}
    \subcaption{pomp-bsmc: the outlier experiment}
    \label{fig_si:ex01_ol_bsmc_mfd_a}
\end{subfigure}
\begin{subfigure}[b]{0.55\textwidth}
    \centering
    \includegraphics[scale=0.37]{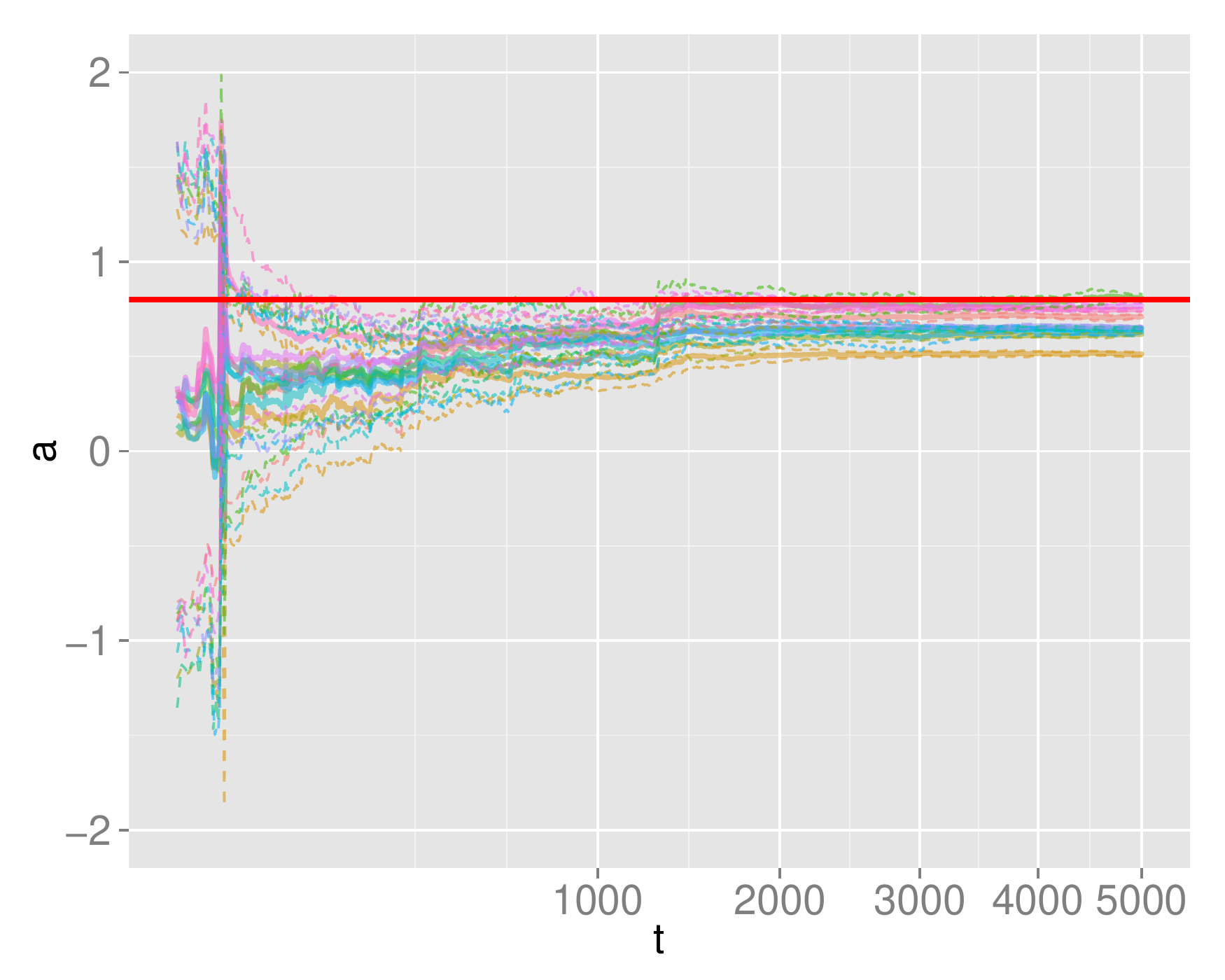}
    \subcaption{SIG: the outlier experiment}
    \label{fig_si:ex01_ol_SIG_mfd_a}
\end{subfigure}
\end{adjustbox}

\begin{adjustbox}{center}
\begin{subfigure}[b]{0.55\textwidth}
    \centering
    \includegraphics[scale=0.37]{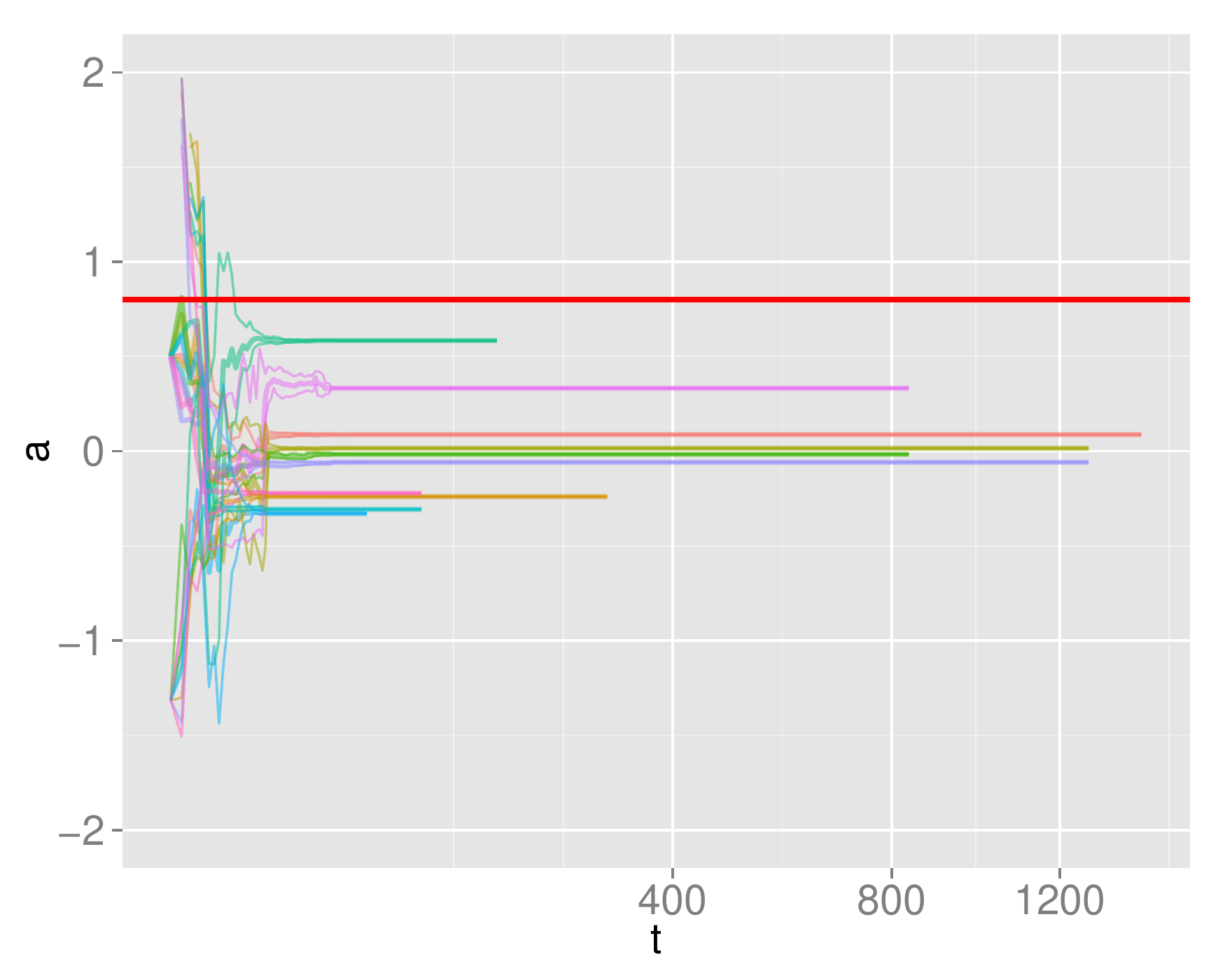}
    \subcaption{pomp-bsmc: the case with $y_{t} = \exp(\alpha_t) + v_t$}
    \label{fig_si:ex01_expalpha_bsmc_mfd_a}
\end{subfigure}
\begin{subfigure}[b]{0.55\textwidth}
    \centering
    \includegraphics[scale=0.37]{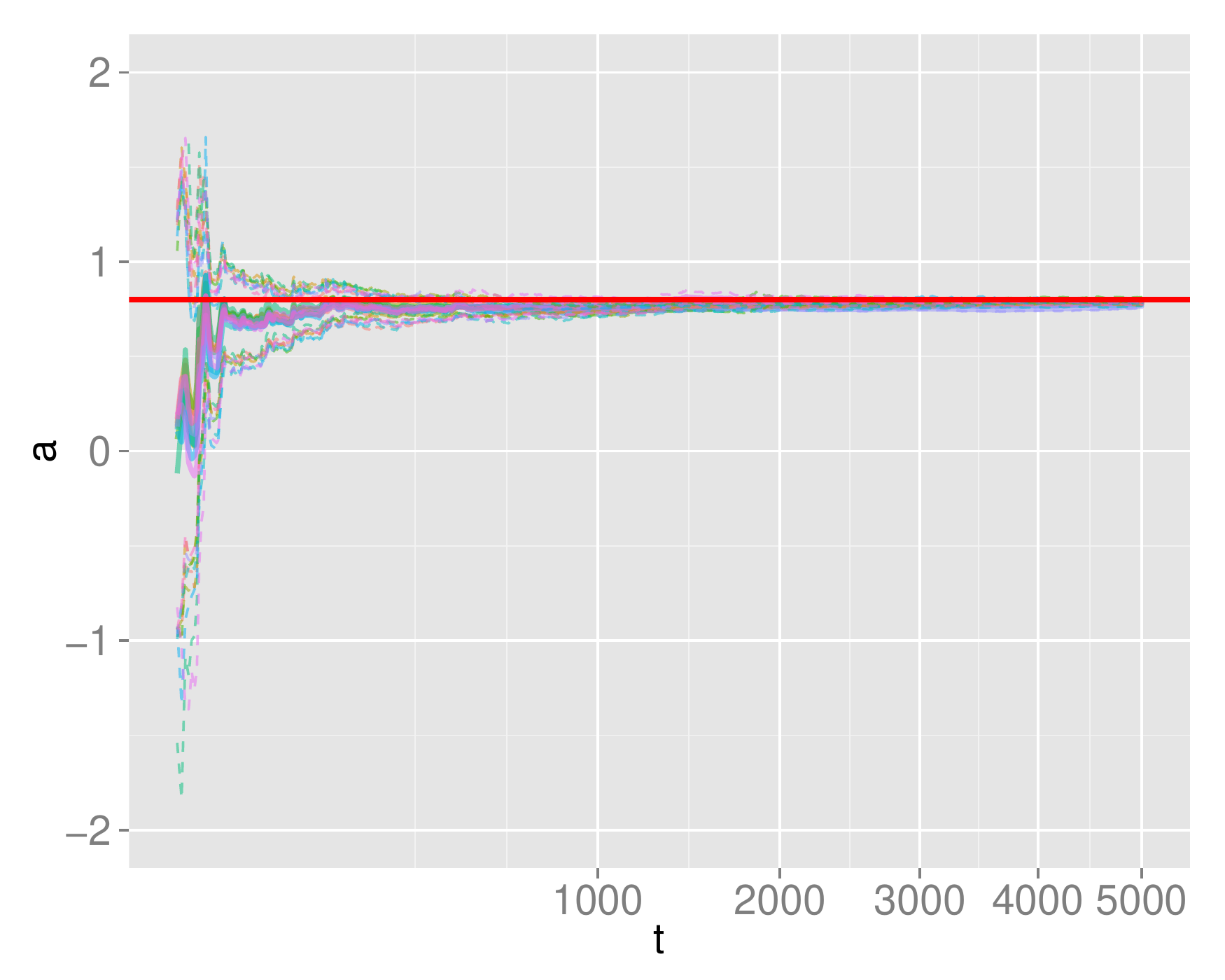}
    \subcaption{SIG: the case with $y_{t} = \exp(\alpha_t) + v_t$}
    \label{fig_si:ex01_expalpha_SIG_mfd_a}
\end{subfigure}
\end{adjustbox}
\caption{Sequential estimation of $a$ by SIG and pomp-bsmc.}
\label{fig_si:ex01_comp_pomp_SIG}
\end{figure}

We first test pomp-bsmc directly on the same data and prior of Section \ref{subsec:example_1} with $10$ parallel runs each of which has $10000$ particles. In this case, pomp-bsmc provides similar results to SIG's of Figure \ref{fig_si:ex01_SIG_mfd}; illustrative sequential traces of parameter $a$ by pomp-bsmc are given in Figure \ref{fig_si:ex01_bsmc_mfd_a}.

Then, both methods are tested for the robustness with respect to outliers. Three consecutive and independent outliers $o_{t=21:23} \sim N(0,\sigma=30)$ are added to the data $y_{21:23}$. Estimation results of parameter $a$ by SIG and pomp-bsmc are shown in Figures \ref{fig_si:ex01_ol_bsmc_mfd_a} and \ref{fig_si:ex01_ol_SIG_mfd_a}. As expected, particle-based pomp-bsmc decays quickly to few particles while SIG estimation, even also affected by outliers, is quite robust and maintains a wide standard deviation.

The final test is with a different observation equation, $y_{t} = \exp(\alpha_t) + v_t$. Again, Figures \ref{fig_si:ex01_expalpha_bsmc_mfd_a} and \ref{fig_si:ex01_expalpha_SIG_mfd_a} show that SIG provides more stable results than pomp-bsmc \footnote{It is noted that pomp-bsmc has run-time error and stops halfway during the experiments. It is reported that this error is due to the degeneracy into few particles of pomp-bsmc.}.

\section{Population-based strategies}
\label{sec:pop_based_strategies}

It is seen from the previous section that the SIG approach is capable of locally approximating the target filtering distributions but may be sensitive to the prior; this is seen especially in the example where both variances are unknown. In this section a method for implementing parallel runs of SIG is described that improves the identification of the principal mode of the target distribution and offers a pragmatic solution to the issue of specifying a prior that permits the algorithm to produce a good approximation.

It is noted that any differences between independent runs of the SIG approximation of Section \ref{sec:sequential_iterlap} is due to the importance sampling, making it reasonable to select a run that has good performance and propagate that run forward, without weighting across all runs. This is analogous to optimisation with several starting values.

In this population-based strategy, $\bar{N}$ parallel runs of the SIG method are made and are re-sampled from time to time, according to their predictive performance.  The objective is to identify runs that are performing well and favour them in the re-sampling. ``Bad'' runs, that are stuck at a local mode that has poor consistency with the data, end up being rejected. The method will focus on exploration of the space of both state and unknown parameters that give small prediction error.  At any time, each single run can be used as a proper best-effort approximation of the filtering distribution $p(x_t, \varphi \, | \, x_{1:t})$. These runs can also be combined in a weighted average for a more robust sequential approximation.  

We define a general predictive performance function based on square error. Define a set of re-sampling times $r_1 < r_2 < \cdots r_S$ and a maximum lag $L$ at which to compute predictions.  
For any run $\bar{n}$, at one of the re-sampling times $t = r_s$, the cumulative square prediction error since the last re-sampling is examined for each lag $l$ and component $i$ of $y_t$:
\begin{equation*} 
SSE_{\bar{n};l,i} = \sum_{t=r_{s-1}+1}^{r_s} \!\!\! (y_{t,i} - \hat{y}_{\bar{n};t,i}^{(t-l)})^2, \; l=0,\ldots,L; \: i=1,\ldots,\mbox{dim}(y_t), 
% \label{eq:pred_error_SSE}
\end{equation*}
where $\hat{y}_{\bar{n};t,i}^{(t-l)}$ is the $l$-step ahead prediction of $y_{t,i}$ of run $\bar{n}$ i.e.\ the prediction of $y_{t,i}$ made at time $t-l$. The prediction $\hat{y}_{\bar{n};t,i}^{(t-l)}$ can be obtained either by using Monte Carlo approximation of $\mathbb{E}_{\bar{n}}(y_{t,i} \, | \, y_{1:(t-l)})$ where the expectation is with respect to approximation $\widetilde{p}(x_{t-1},\varphi \, | \, y_{1:(t-1)})$ of run $\bar{n}$ or a point prediction of $y_{t,i}$ at the approximated mean $(\widetilde{x}_{t-1},\widetilde{\varphi})$ of $\widetilde{p}(x_{t-1},\varphi \, | \, y_{1:(t-1)})$. To save the computation cost, point prediction is used in this paper. Note that $l$ starts from zero as we also want to take into account the filtering as well as prediction error.

The predictive performance measure is a weighted sum of these squared errors over lags $0,1,\ldots,L$ and the components of $y_t$:
\begin{equation}
{\cal S}_{\bar{n}} = \sum_{l=0}^L \sum_{i=0}^{\mbox{\scriptsize{dim}}(y_t)} \omega_{l,i} \: SSE_{\bar{n};l,i}.
\label{eqn:pred_error}
\end{equation}
Notice that both the cumulative square prediction error and predictive performance measure can be updated sequentially at each time step $t$ by keeping previous predictions $\hat{y}_{t,i}^{(t-l)}$ made at time $(t-l)$.

At a re-sampling time, $\bar{N}$ runs are re-sampled with replacement from the existing $\bar{N}$ runs with probabilities:
\[ P(\mbox{sample run } \bar{n}) \: \propto \exp(-{\cal S}_{\bar{n}}), \]
where ${\cal S}_{\bar{n}}$ is the measure of Equation \ref{eqn:pred_error} derived from the approximation of the $\bar{n}$th run.

The user-defined weights $\omega_{l,i}$ should both standardise the errors across components of $y_t$ and can reflect differing importances of predictions at different lags.  In this paper, we use:
\[ \omega_{l,i} = \frac{\omega_l}{\hat{\sigma}_i^2}, \]
where $\hat{\sigma}$ is a rough estimation of sample standard deviation of $y_{t,i}$ of the first $50$ observations and $\omega_l$ are the importances of each prediction lag. For all the subsequent experiments, we use $L=1$ and set $(\omega_0=0.2,\omega_1=0.8)$ to put more importance weight to the one-step prediction error. Also,

\begin{algorithm}[!h]
\stepcounter{algorithm}
\caption{SIG-RS: Resampling parallel SIG runs}
\label{alg_pbs:SIG_RS}
\begin{enumerate}
\item Specify a number of parallel runs $\bar{N}$, a maximum prediction lag $L$, a set of re-sampling times $r_1 < r_2 < \cdots < r_S$ and a set of weights $\omega_{l,i}, \: l=0,\ldots,L; \, i=1,\ldots,\mbox{dim}(y_t)$.
\item For each $t$:
\begin{enumerate}
\item Update each run's approximation to $p(x_t, \varphi \, | \, y_{1:t})$ independently following the SIG algorithm of Section \ref{subsec:bias_correction}. 
\item Compute and store $\hat{y}_{t+l,i}^{(t)}$, the $l$-step ahead predictions from time $t$, for $l=0,\ldots,L$;
\item Sequentially update ${\cal S}_{\bar{n}}$ for each run $\bar{n}=1,\ldots,\bar{N}$, following Equation \ref{eqn:pred_error}.
\item If $t = r_s$ from some $s$, re-sample $\bar{N}$ runs with replacement using probabilities proportional to $\exp(-{\cal S}_{\bar{n}})$.
%\item If an approximation to $p(x_t, \varphi \, | \, y_{1:t})$ is required then take an average of the approximations from each run.
\end{enumerate}
\end{enumerate}
\end{algorithm}

An extension of this algorithm also permits a pragmatic solution to the problem of vague or mis-specified priors in sequential analysis.  In both cases, the algorithms of Section \ref{sec:sequential_iterlap} can perform badly because the inference lacks good regularization from the prior and so, depending on what is observed, $\tilde{p}(x_t, \varphi \, | \, y_{1:t})$ may become degenerate or be caught at a bad local approximation from which it is very unlikely to escape even after a large sequence of observations.  This is related to problems in other methods such as convergence of an MCMC algorithm or to the  global solution in an optimisation problem. Since the SIG approach suffers from being trapped at a local approximation, in addition to the resampling step above, we propose a simple strategy called SIG-RSRP (Algorithm 5).

In the SIG-RSRP algorithm, the SIG-RS algorithm is run but in addition, at each re-sampling time $r_s$, if the predictive error of the parallel runs is too small, then the variance of $\tilde{p}(x_t, \varphi \, | \, y_{1:t})$ is tempered to a larger, pre-set, value.  This is analogous to simulated tempering in MCMC \citep{ref:Neal2001}.  Specifically, a set of threshold marginal variances for $(x_t,\varphi)$ are pre-defined.  At $t = r_s$ for some $s$, if all of the marginal posterior variances of $\tilde{p}(x_t, \varphi \, | \, y_{1:t})$ are smaller than their respective threshold then the variances in $\tilde{p}(x_t, \varphi \, | \, y_{1:t})$ are re-set to these thresholds and the algorithm proceeds. 

This is equivalent to having run the algorithm with a different prior with larger variance. In this sense it is a pragmatic solution as it violates the \textit{a priori} specification of the prior. It is noted that that, unlike SIG-RS, the SIG-RSRP method does alter the target distribution by tempering the prior. However, like SIG-RS, SIG-RSRP stops the switching and prior tempering when $t>r_{S}$; hence, one can interpret SIG-RSRP as a sequential procedure to select a prior that was able to regularize the inference up to $t = r_{S}$ before continuing with the usual SIG algorithm.

Both SIG-RS and SIG-RSRP are applied on the example of Section \ref{subsec:example_3}. For SIG-RSRP, the re-sampling times are set to $r_1=10 < r_2=25 < r_3=50 < r_4=75< r_5=100< \ldots < r_S = 2000$; and $\bar{\Sigma}_{x_t,\varphi}$ is set to $\diag(0.3^2,0.3^2,0.25^2,0.25^2,0.5^2)$, which is approximately a fraction of prior variance $\Sigma_{x_1, \varphi}^{(p)}$.

\begin{figure}[!ht]
\centering
\begin{adjustbox}{center}
\begin{subfigure}[b]{0.55\textwidth}
    \centering
    \includegraphics[scale=0.37]{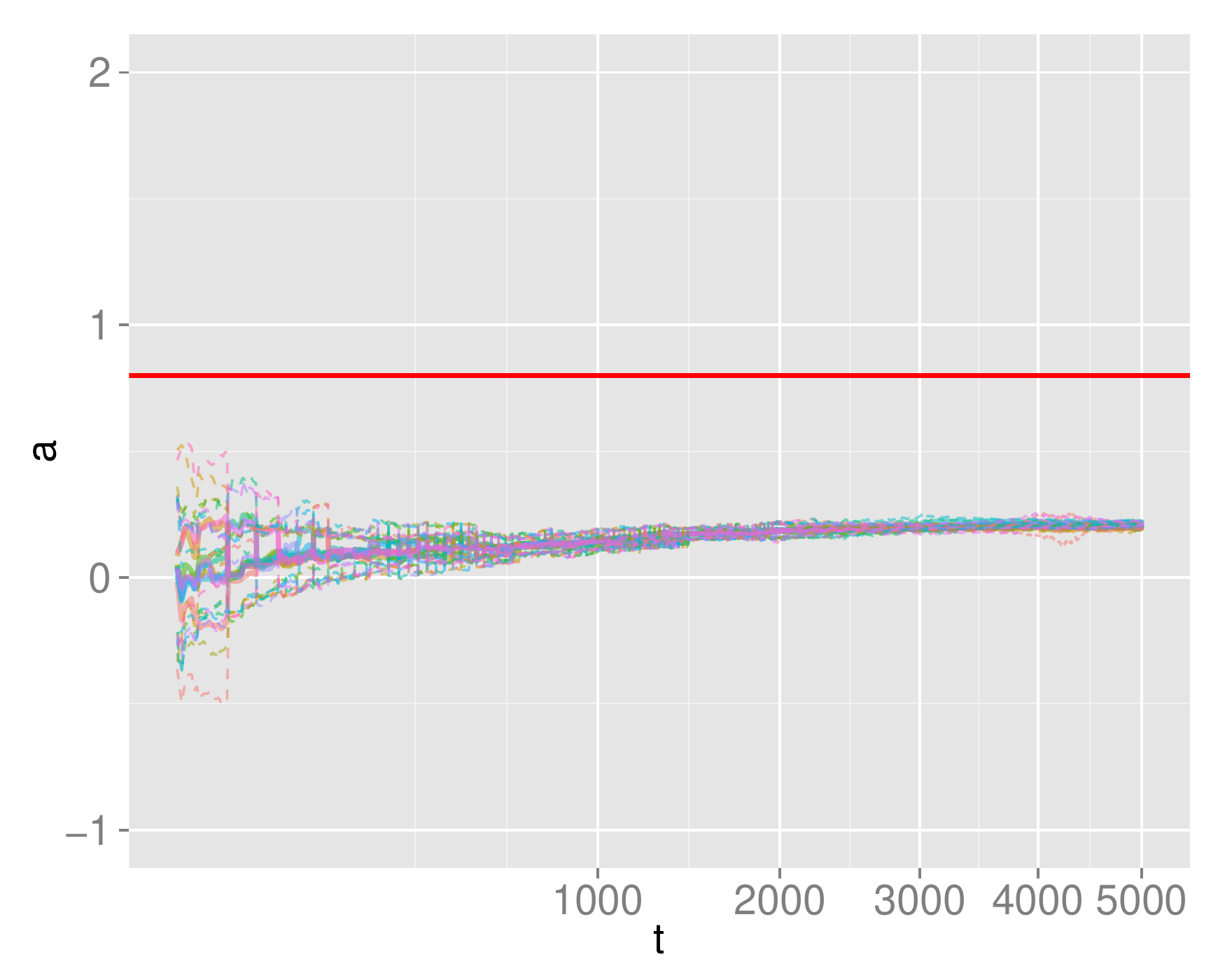}
    \subcaption{$a$}
    \label{fig_si:ex03_prior02_SIG_RS_mfd_a}
\end{subfigure}
\begin{subfigure}[b]{0.55\textwidth}
    \centering
    \includegraphics[scale=0.37]{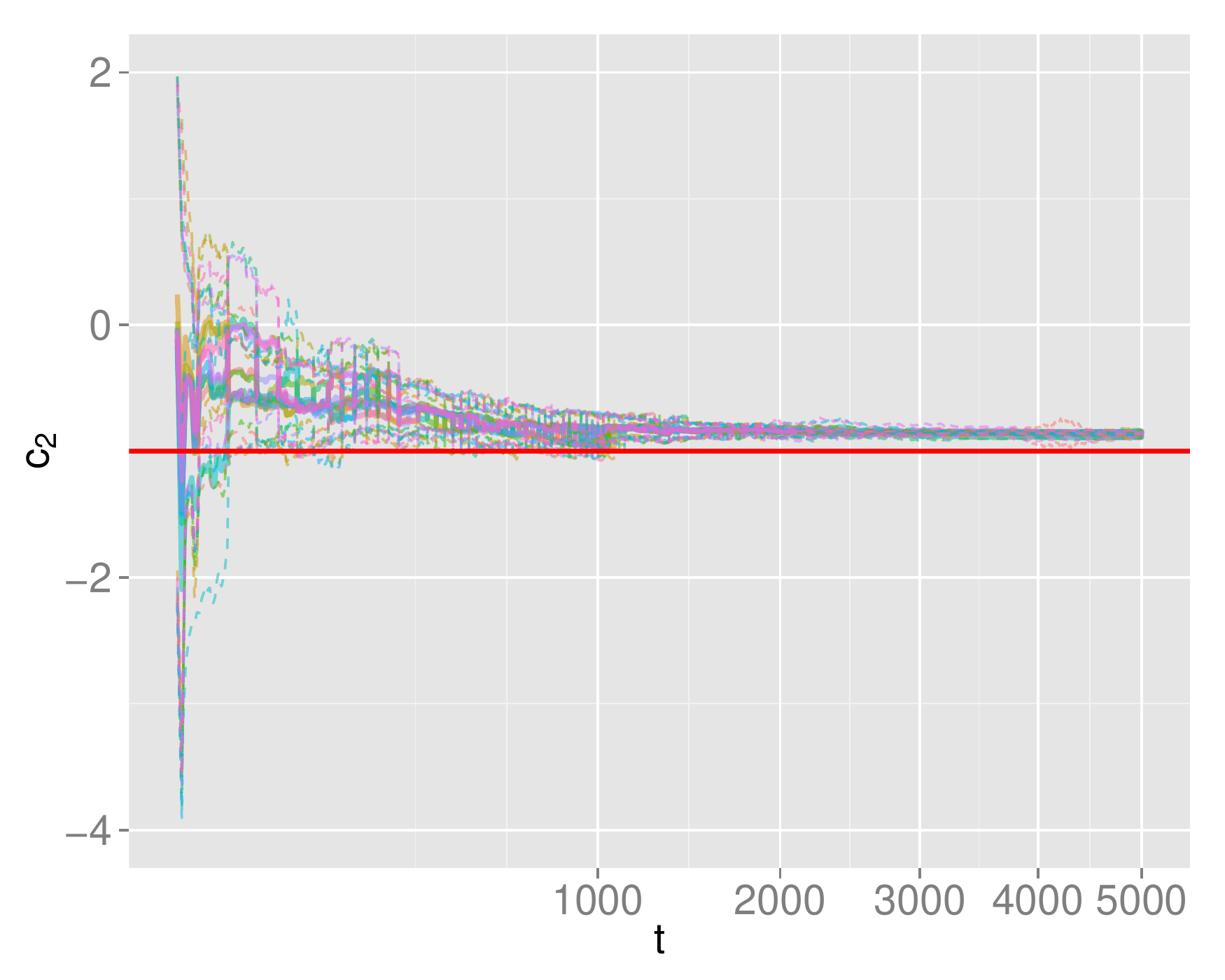}
    \subcaption{$c_2$}
    \label{fig_si:ex03_prior02_SIG_RS_mfd_b2}
\end{subfigure}
\end{adjustbox}

\begin{adjustbox}{center}
\begin{subfigure}[b]{0.55\textwidth}
    \centering
    \includegraphics[scale=0.37]{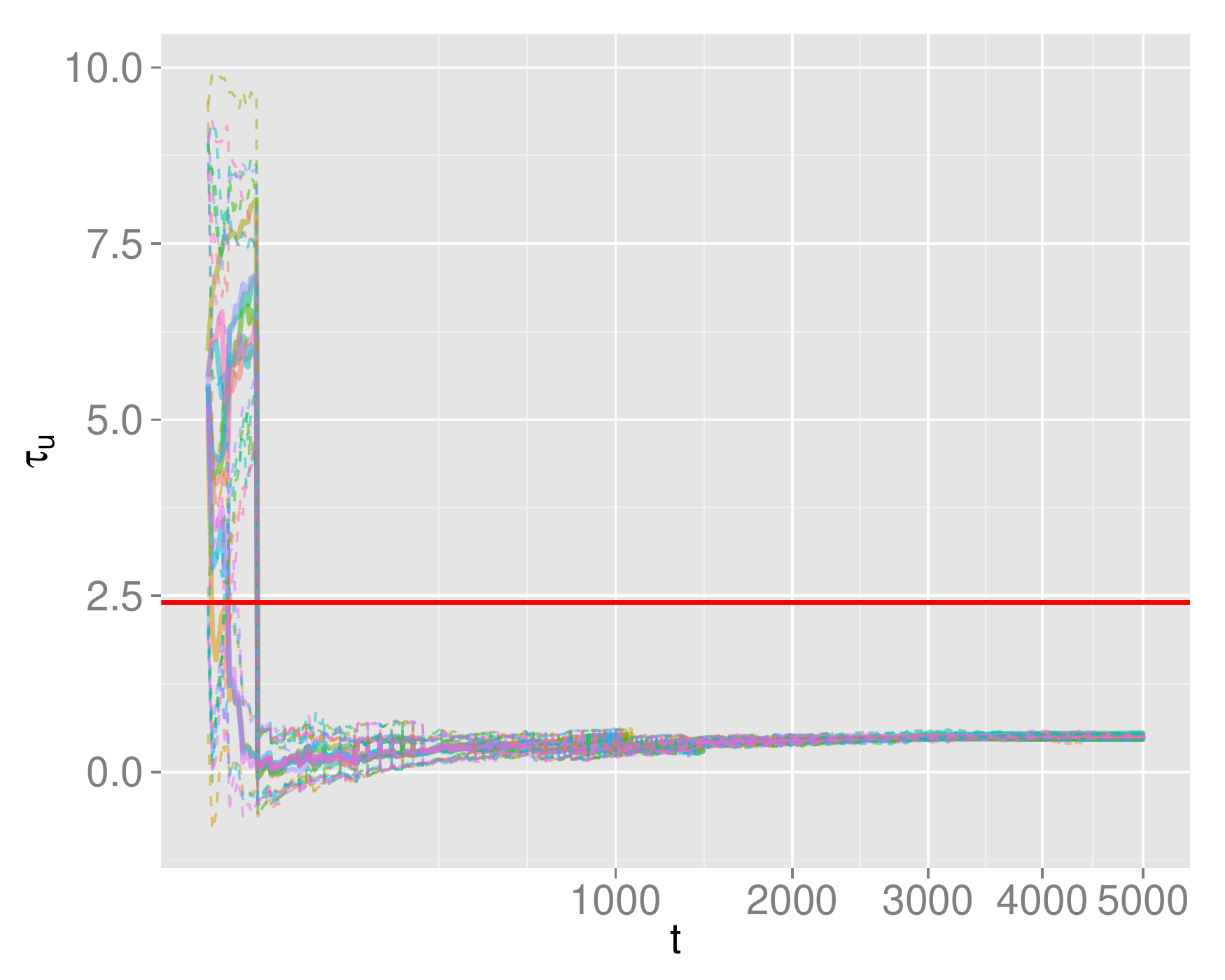}
    \subcaption{$\tau_u$}
    \label{fig_si:ex03_prior02_SIG_RS_mfd_tauu}
\end{subfigure}
\begin{subfigure}[b]{0.55\textwidth}
    \centering
    \includegraphics[scale=0.37]{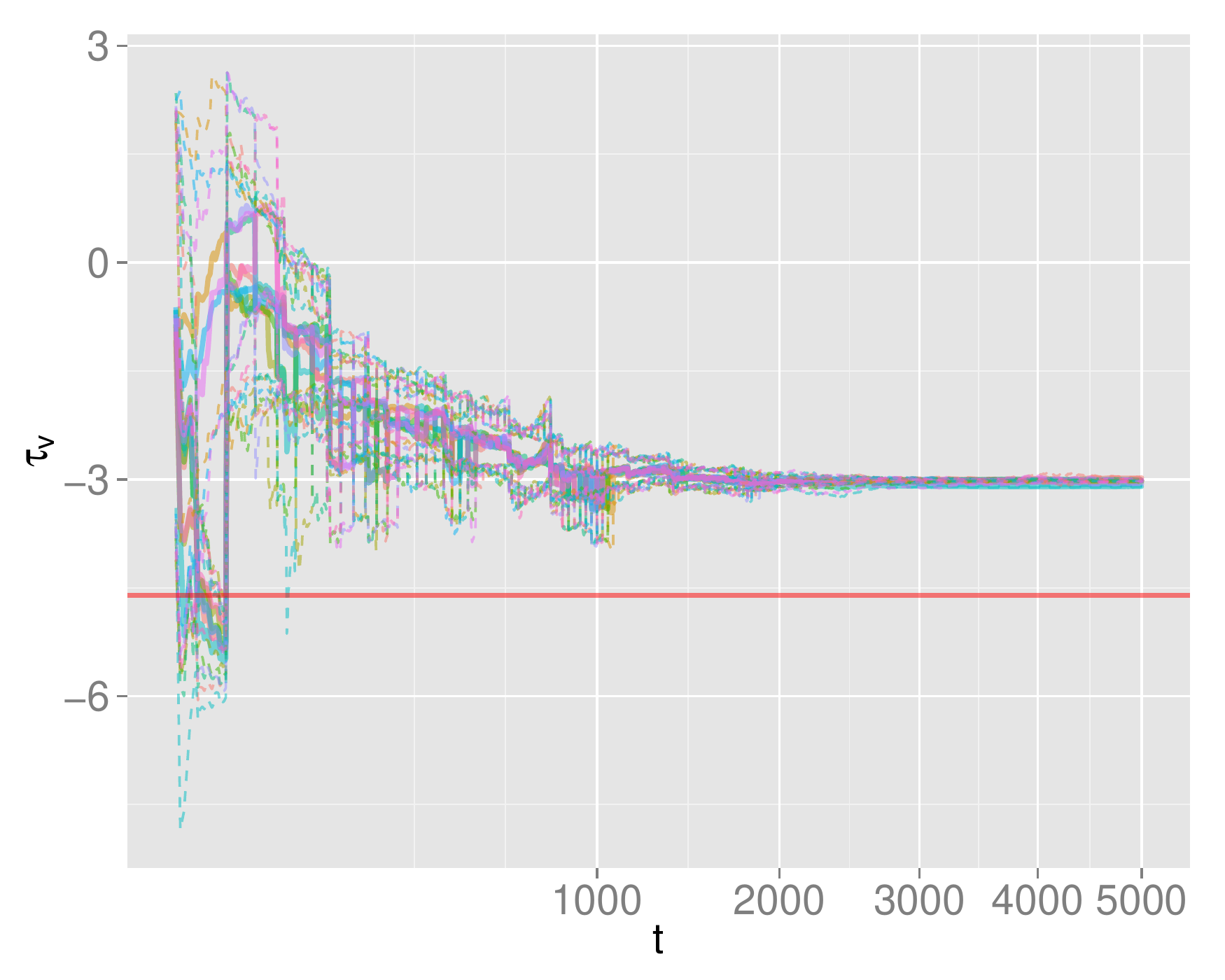}
    \subcaption{$\tau_v$}
    \label{fig_si:ex03_prior02_SIG_RS_mfd_tauv}
\end{subfigure}
\end{adjustbox}

\begin{adjustbox}{center}
\begin{subfigure}[b]{0.55\textwidth}
    \centering
    \includegraphics[scale=0.37]{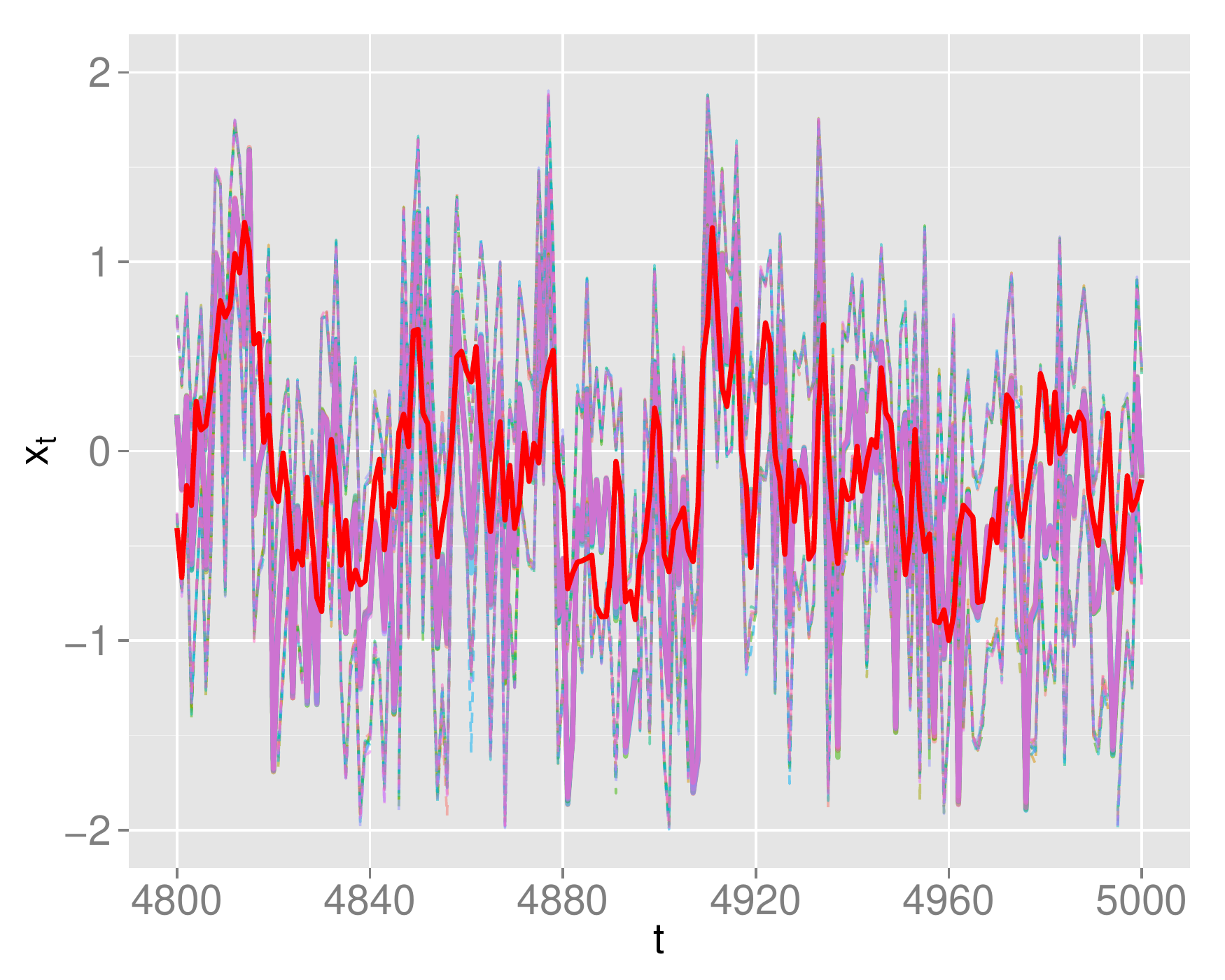}
    \subcaption{$x_t$}
    \label{fig_si:ex03_prior02_SIG_RS_mfd_xt}
\end{subfigure}
\end{adjustbox}
\caption{SIG-RS approach: the marginal filtering distributions of unknown parameters and state variable for the example in Section \ref{subsec:example_3} with prior in Equation \ref{eqn:ex03_prior02}.}
\label{fig_si:ex03_prior02_SIG_RS_mfd}
\end{figure}

\begin{figure}[!ht]
\centering
\begin{adjustbox}{center}
\begin{subfigure}[b]{0.55\textwidth}
    \centering
    \includegraphics[scale=0.37]{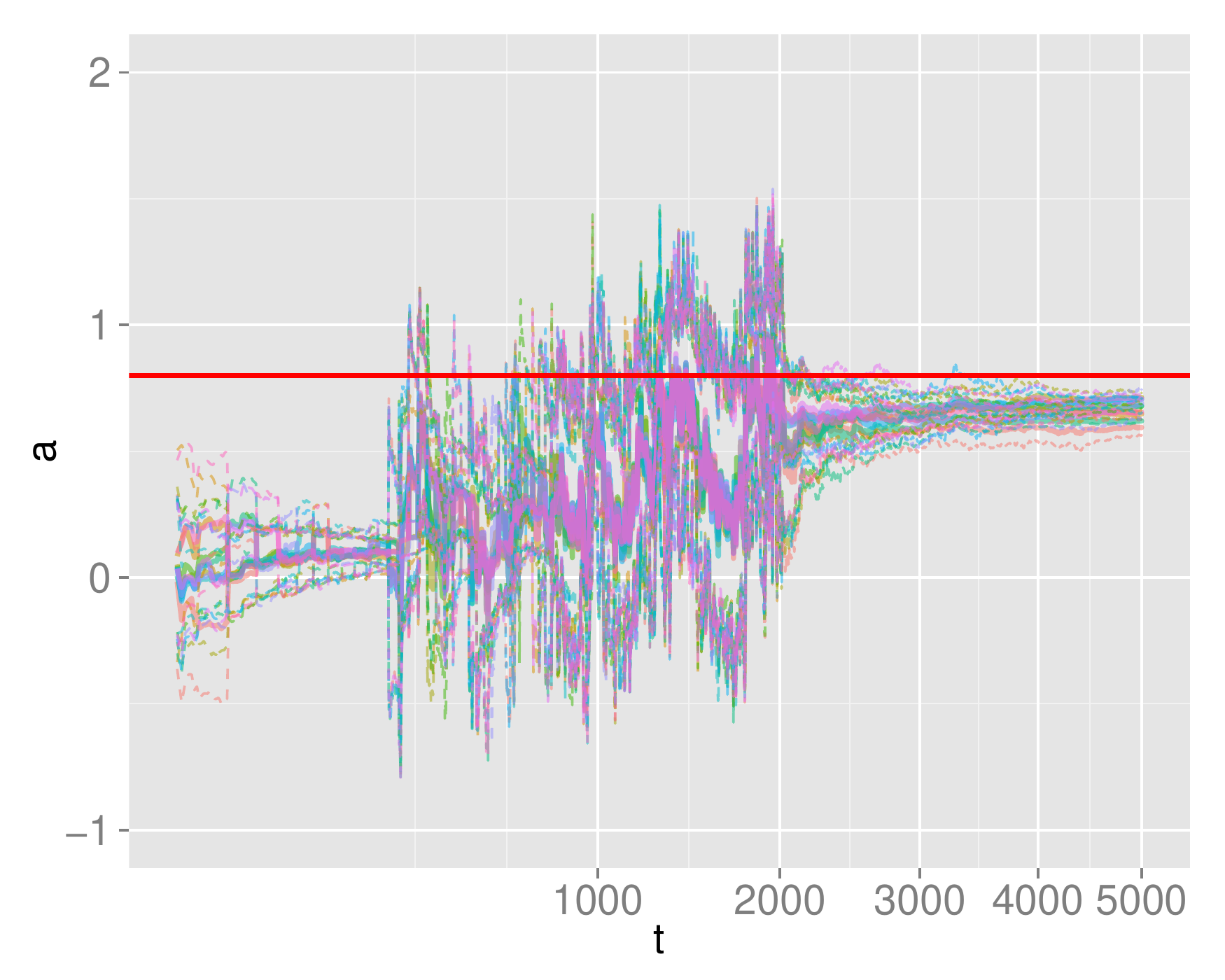}
    \subcaption{$a$}
    \label{fig_si:ex03_prior02_SIG_RSRP_mfd_a}
\end{subfigure}
\begin{subfigure}[b]{0.55\textwidth}
    \centering
    \includegraphics[scale=0.37]{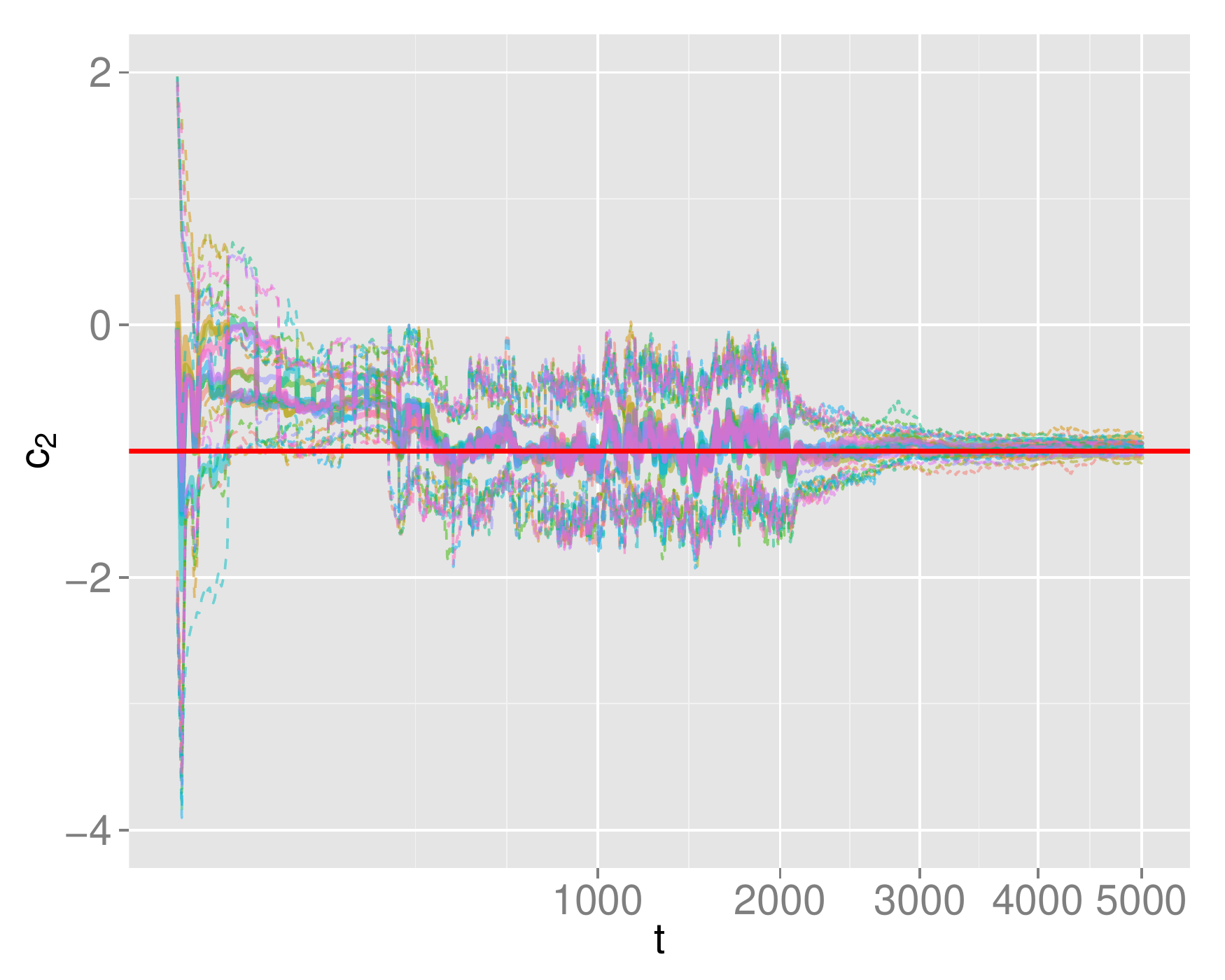}
    \subcaption{$c_2$}
    \label{fig_si:ex03_prior02_SIG_RSRP_mfd_b2}
\end{subfigure}
\end{adjustbox}

\begin{adjustbox}{center}
\begin{subfigure}[b]{0.55\textwidth}
    \centering
    \includegraphics[scale=0.37]{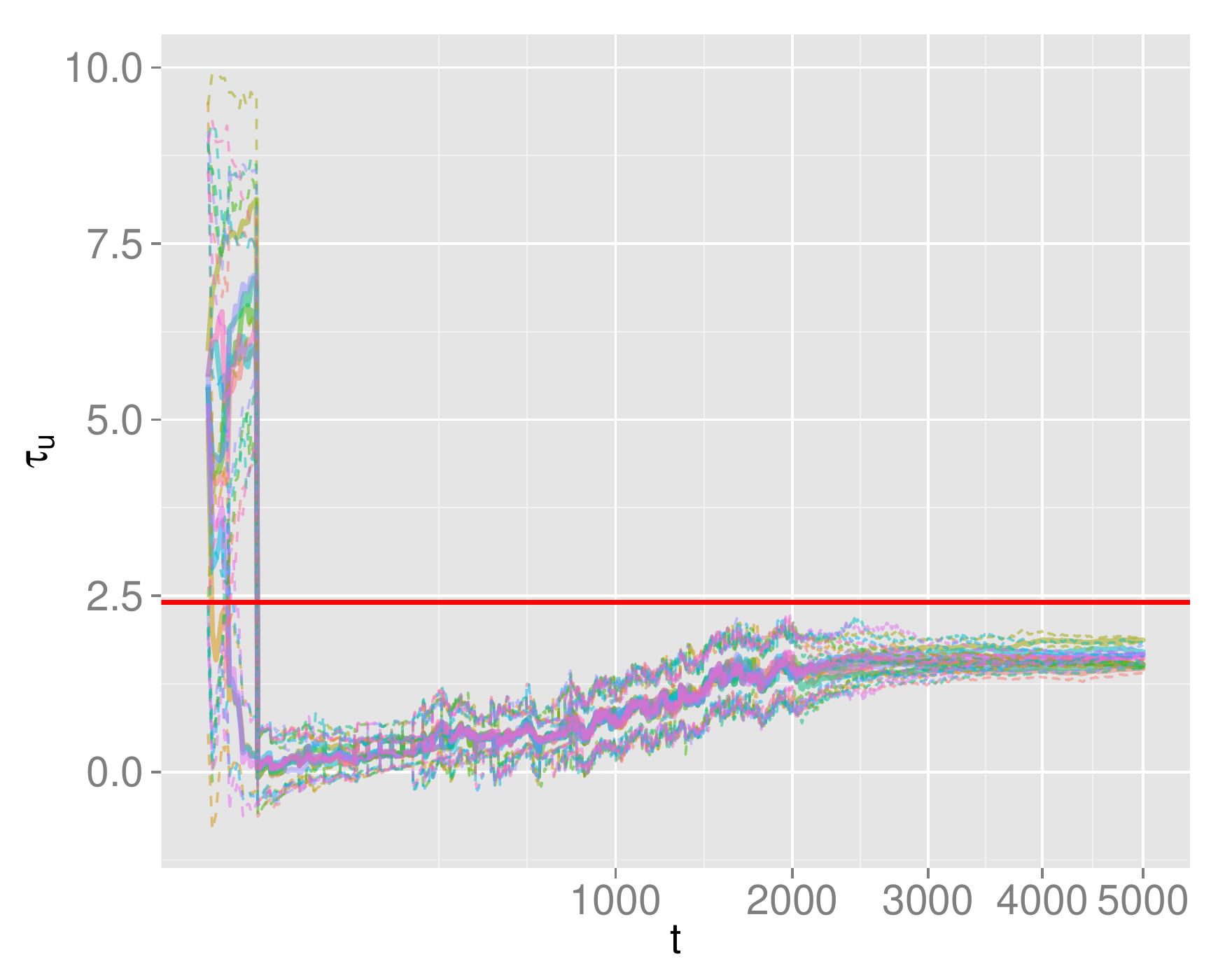}
    \subcaption{$\tau_u$}
    \label{fig_si:ex03_prior02_SIG_RSRP_mfd_tauu}
\end{subfigure}
\begin{subfigure}[b]{0.55\textwidth}
    \centering
    \includegraphics[scale=0.37]{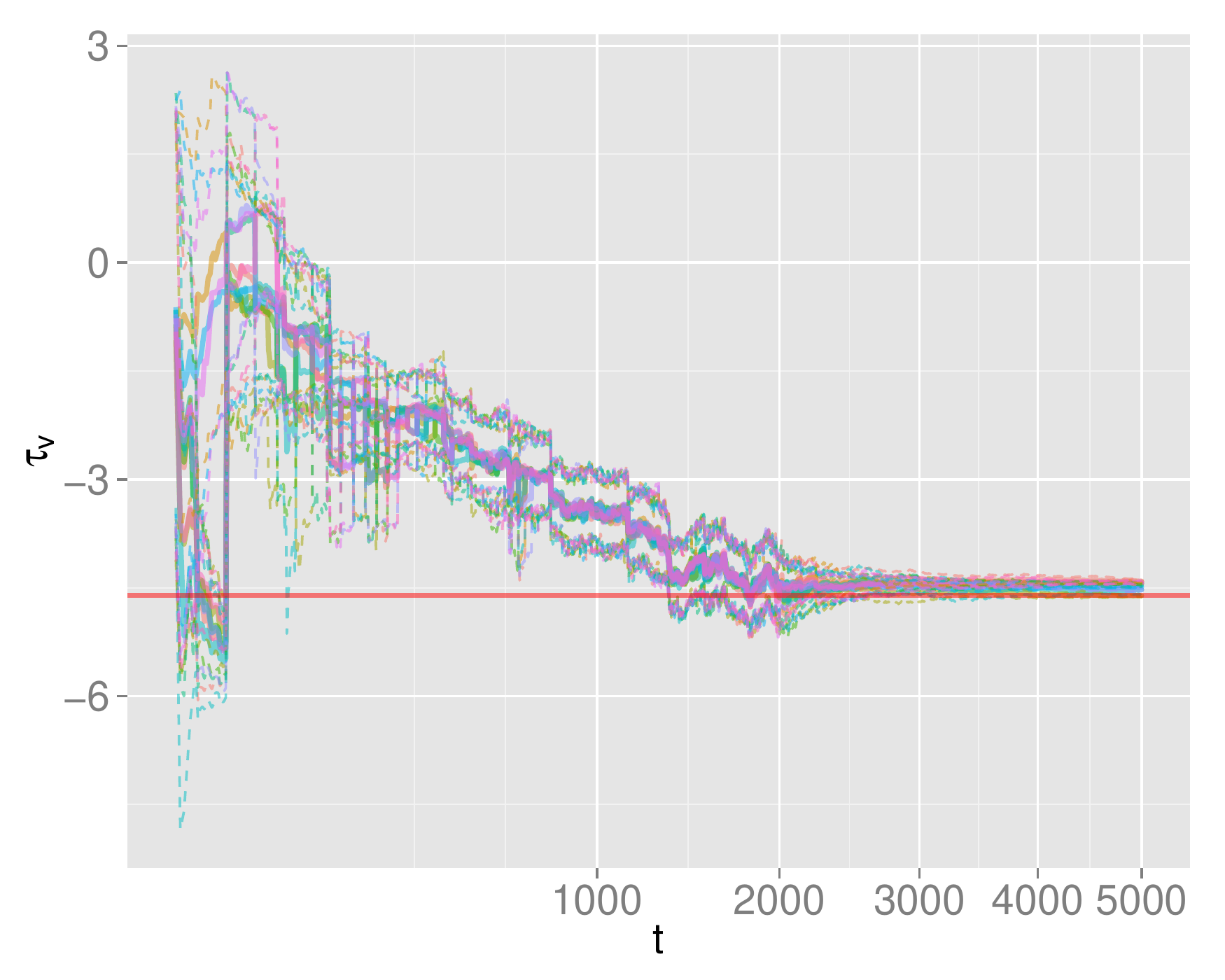}
    \subcaption{$\tau_v$}
    \label{fig_si:ex03_prior02_SIG_RSRP_mfd_tauv}
\end{subfigure}
\end{adjustbox}

\begin{adjustbox}{center}
\begin{subfigure}[b]{0.55\textwidth}
    \centering
    \includegraphics[scale=0.37]{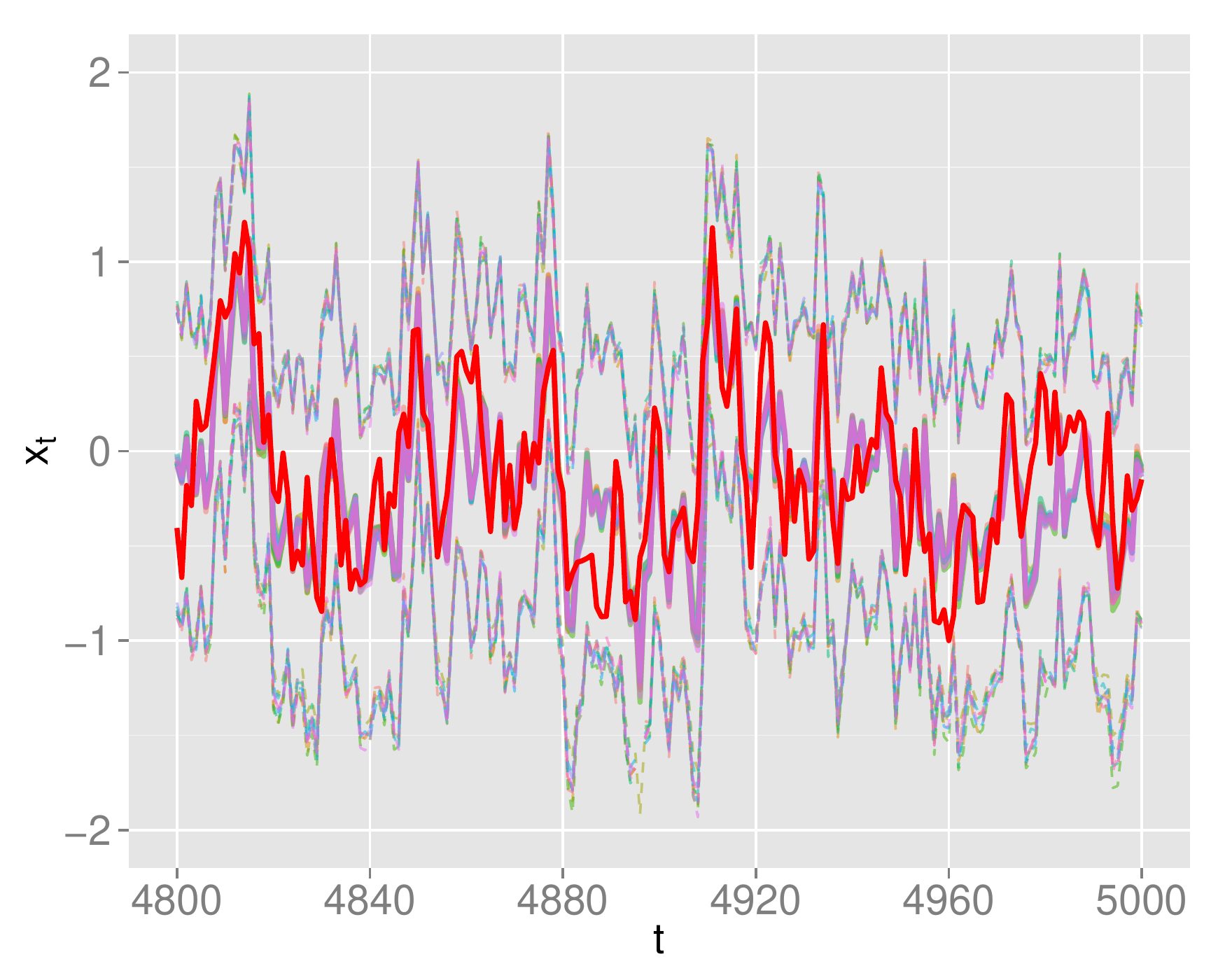}
    \subcaption{$x_t$}
    \label{fig_si:ex03_prior02_SIG_RSRP_mfd_xt}
\end{subfigure}
\end{adjustbox}
\caption{SIG-RSRP approach: the marginal filtering distributions of unknown parameters and state variable for the example in Section \ref{subsec:example_3} with prior in Equation \ref{eqn:ex03_prior02}.}
\label{fig_si:ex03_prior02_SIG_RSRP_mfd}
\end{figure}

Figures \ref{fig_si:ex03_prior02_SIG_RS_mfd} and \ref{fig_si:ex03_prior02_SIG_RSRP_mfd} show the filtering results of SIG-RS and SIG-RSRP respectively. Compared to the SIG approach, the traces of SIG-RS and SIG-RSRP favour the group which has higher state variance, making the state variable more adapted to the observation due to the prediction-error criterion. Between these two, SIG-RSRP has more room to change due to the relaxation and there are definitely trends of SIG-RSRP traces moving closer to the true values over time.

\section{Conclusion}
\label{sec:conclusion}
This paper's principal contributions are the following:
\begin{itemize}
\item It has extended the iterated Laplace approximation to a sequential setting for joint parameter and latent process inference in latent Gaussian models.  This permits approximations with skewness, multi-modality and non-linear correlations.
\item  It has developed two single Gaussian component approximations, one based on an importance sampling correction and the other based on a population idea, to this joint posterior that are robust to outlier observations and a provide good performance in terms of posterior predictions.
\item The methods also demonstrate good computational performance as they are based on Gaussian approximations.
\end{itemize}

\begin{figure}[!ht]
\centering
\includegraphics[scale=0.60]{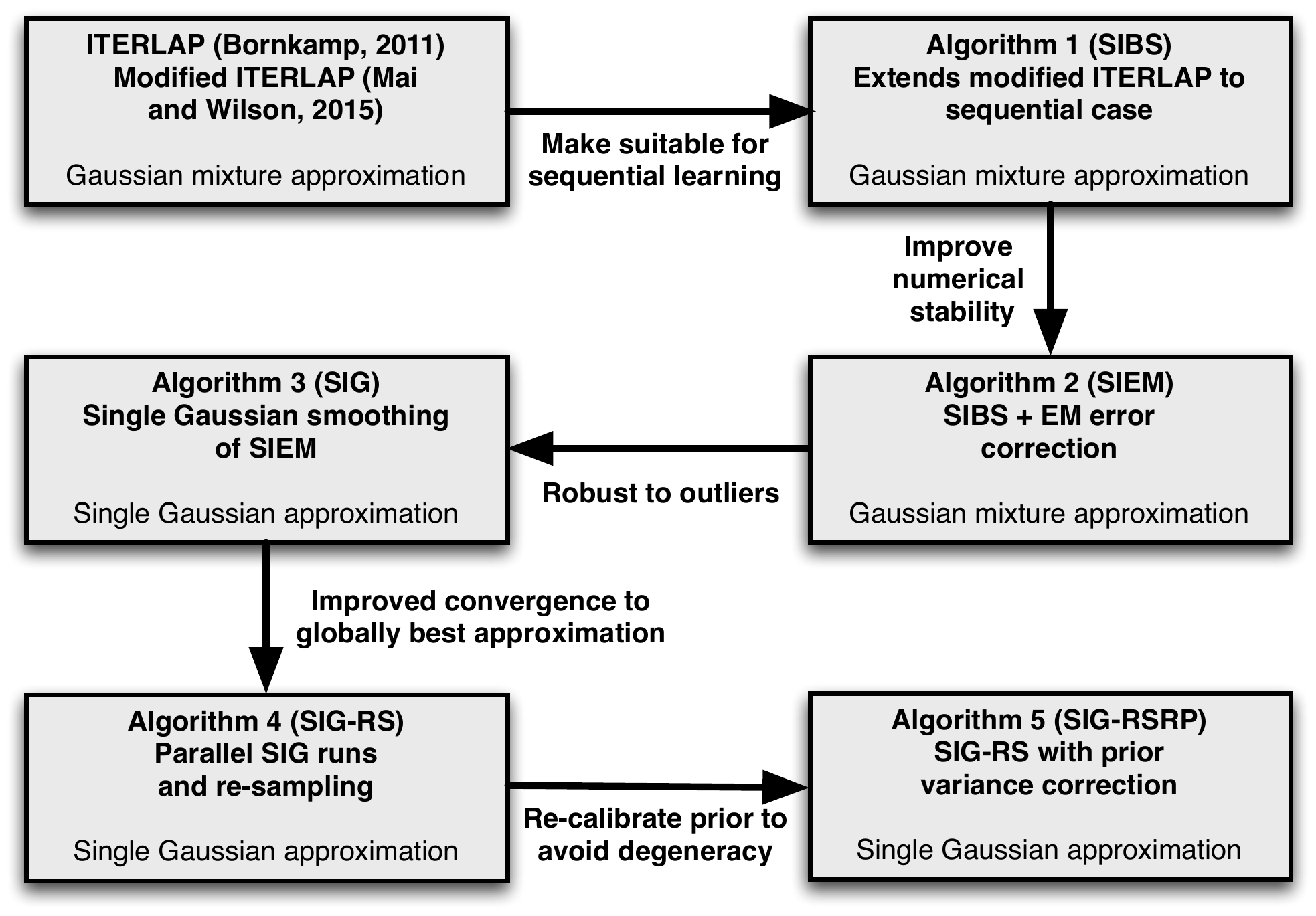}
\caption{Summary of explored algorithms.}
\label{fig_conclusion:algorithm_relationship}
\end{figure}

Figure \ref{fig_conclusion:algorithm_relationship} shows the chain of development of the 5 algorithms that are explored in this paper. Starting from Algorithm 1, which can be viewed as the natural extension of iterLap to the sequential case, subsequent algorithms are derived to address some issue that arises with the previous algorithm in the chain, following evaluation on test data.

The most interesting conclusion is that a single Gaussian approximation, obtained as a smoothed version of a Gaussian mixture, has been shown to have attractive properties in sequential analysis.  It must be emphasised that this is not the Laplace approximation but rather is derived as a smoothed version of a Gaussian mixture. The iterLap approximation is key to providing a good Gaussian mixture approximation around which the single Gaussian can be built. An important difficulty that these approximations encounter is how to react to outlier observations, also a cause of degeneracy in particle filter methods.  It is here that the single Gaussian is shown to be robust.

Non-identifiability of parameters is also an issue that causes difficulties for parameter estimation, and the population-based approach has been shown to provide a solution to this through allowing several different approximation runs to work in parallel, and then be combined in a sensible manner. 

Which of the 5 algorithms should one use?  The final algorithm SIG-RSRP addresses the most of the issues that have arisen when implementing these approximations, and produces the most consistent and accurate approximation, but a subjective Bayesian may have reservations about using it as it is implicitly altering the prior following observation of data.  Similarly, there may be some objections to the population idea that is also used in algorithm 4 (SIG-RS), as resampling the best approximations from an set of them could be seen as biasing the posterior uncertainty.  On this basis, the recommendation is that algorithm 3 (SIG) is best for fully (subjective) Bayesian analysis, while SIG-RS and SIG-RSRP provide a pragmatic solution that performs well.

%%%%%%%%%%%%%%%%%%%%%%%%%%%%%%%%%%%%%%%%%%%%%%%%%%%%%%%%%%%%%%%%%%%%%%
%%%%%%%%%%%%%%%%%%%%%%%%%%%%%%%%%%%%%%%%%%%%%%%%%%%%%%%%%%%%%%%%%%%%%%

\section*{Acknowledgement}
This work was supported by the STATICA project, funded by the Principal Investigator programme of Science Foundation Ireland, contract number 08/IN.1/I1879.

%%%%%%%%%%%%%%%%%%%%%%%%%%%%%%%%%%%%%%%%%%%%%%%%%%%%%%%%%%%%%%%%%%%%%%
%%%%%%%%%%%%%%%%%%%%%%%%%%%%%%%%%%%%%%%%%%%%%%%%%%%%%%%%%%%%%%%%%%%%%%

%\bibliographystyle{chicago}
%\bibliography{root}

\bibliography{Ref/Reference}
\bibliographystyle{agsm}

\clearpage

\appendix

\section{Non-identifiability with the model of Equations \ref{eqn_ssm:ex1_ssm_obs} to \ref{eqn_ssm:ex1_ssm_state}}
\label{app:nonident}

Assuming that all parameters are unknown, the example model suffers from some cases of non-identifiability that we know, as follows.

\begin{case}
$\mathcal{NID}(c_1,\sigma_u)$
\end{case}
Define $c_1^{\prime}=c_1/\beta$; $\sigma_u^{\prime}=\beta \sigma_u$; $x_t^{\prime}=\beta x_t$ and $u_t^{\prime}=\beta u_t$ for $t=1:n$ with any $\beta>0$. It can be easily seen that:
\begin{align*}
\alpha_t &= \mathbbm{1}(c_1^{\prime} x_t^{\prime} + c_2 z_t + 5 \geq 0) \: (c_1^{\prime} x_t^{\prime} + c_2 z_t + 5),\\
x_{t}^{\prime} &= a x_{t-1}^{\prime} + u_{t}^{\prime},
\end{align*}
and hence:
\begin{alignat}{2}
\nonumber && p(y_{1:n},x_{2:n}^\prime \, | \, x_{1}^{\prime},a,c_1^{\prime},c_{2},\sigma_u^{\prime},\sigma_v) \beta^{n-1} &= p(y_{1:n},x_{2:n} \, | \, x_{1},a,c_1,c_{2},\sigma_u,\sigma_v)\\
\nonumber \Leftrightarrow ~~~ && p(y_{1:n},x_{2:n}^\prime \, | \, x_{1}^{\prime},a,c_1^{\prime},c_{2},\sigma_u^{\prime},\sigma_v) \: \mathrm{d}x_{2:n}^{\prime} &= p(y_{1:n},x_{2:n} \, | \, x_{1},a,c_1,c_{2},\sigma_u,\sigma_v) \: \mathrm{d}x_{2:n}
\\
\nonumber \Rightarrow ~~~ && p(y_{1:n} \, | \, x_{1}^{\prime},a,c_1^{\prime},c_{2},\sigma_u^{\prime},\sigma_v) &= p(y_{1:n} \, | \, x_{1},a,c_1,c_{2},\sigma_u,\sigma_v)
\end{alignat}

\begin{case}
The non-identifiability $\mathcal{NID}(c_1)$ can be proved easily with $c_1^{\prime}= -c_1$; $x_t^{\prime}= - x_t$ for $t=1:n$.
\end{case}

Depending on the model, the non-identifiability problem can become more severe. For example, if $z_t$ is a series with identical and independent increment ($z_t-z_{t-1} \sim N(0,\cdot)$), then there exists a new non-identifiability case $\mathcal{NID}(c_2,\sigma_u)$. Or if the model is extended with $\alpha_t = \mathbbm{1}(c_1 x_t + c_2 z_t + c_3 \geq 0) \: (c_1 x_t + c_2 z_t + c_3)$ and $(x_{t}-\mu_x) = a (x_{t-1}-\mu_x) + u_{t}$, then there is $\mathcal{NID}(c_3,\mu_x)$. It is noted that all these cases can be combined, resulting in a very irregular likelihood $p(y_{1:n} \, | \, x_{1},\varphi)$ with multi-modality and narrow density ridges.

\end{document}